\documentclass[12pt]{article}
\usepackage{a4wide}
\usepackage{axodraw}
\def\plus{\!+\!}
\def\minus{\!-\!}
\usepackage[english]{babel}
\usepackage{epsfig}
\usepackage{latexsym}
\usepackage{amsmath}
\usepackage{amssymb}
\usepackage{cite}

\newcommand\Mvec{\,\mbox{\bf M}}
\newcommand{\SH}{\sqcup \!\!\! \sqcup}
\newcommand{\SHA}{\sqcup \!\! \sqcup}
\newcommand{\Li}{\rm Li}
\newcommand{\Ly}{\rm Ly}
\newcommand{\HH}{H}

\def\ul(#1){\underline{#1}}

\usepackage{pslatex}

\usepackage{color,colordvi}

\begin{document}
\sloppy

\begin{titlepage}
\begin{flushleft}
DESY 09-003 \hfill {\tt arXiv:09072557; [math-ph]} \\
NIKHEF 09-016\\
SFB/CPP-09-65\\
\end{flushleft}
\vspace{0.8cm}

\begin{center}
\Large
{\bf The Multiple Zeta Value Data Mine}

\vspace*{1cm}
\large
J. Bl\"umlein$^{\, a}$, D.J. Broadhurst$^{\, b}$, 
J.A.M. Vermaseren$^{\, a,c}$~\footnote{Alexander-von-Humboldt Awardee.}
\\
\vspace{1.2cm}
\normalsize
{\it $^a$~Deutsches Elektronen-Synchrotron, DESY,\\
          Platanenallee 6, D-15738~Zeuthen, Germany
}\\
\vspace{0.5cm}
{\it $^b$~Physics and Astronomy Department, Open University, \\
\vspace{0.1cm}
Milton Keynes MK7 6AA, UK} \\
\vspace{0.5cm}
{\it $^c$~Nikhef Theory Group \\
\vspace{0.1cm}
Science Park 105, 1098 XG Amsterdam, The Netherlands} \\
\vfill
\end{center}
\begin{abstract}
{\small
\noindent
We provide a data mine of proven results for multiple zeta values (MZVs) of 
the form $\zeta(s_1,s_2,\ldots,s_k)=\sum_{n_1>n_2>\ldots>n_k>0}^\infty 
\left\{1/(n_1^{s_1} ... n_k^{s_k})\right\}$ with weight $w=\sum_{i=1}^k 
s_i$ and depth $k$ and for Euler sums of the form 
$\sum_{n_1>n_2>\ldots>n_k>0}^\infty \left\{(\epsilon_1^{n_1} ...\epsilon_1 
^{n_k})/ (n_1^{s_1} ... n_k^{s_k}) \right\}$ with signs $\epsilon_i=\pm1$. 
Notably, we achieve explicit proven reductions of all MZVs with weights 
$w\le22$, and all Euler sums with weights $w\le12$, to bases whose 
dimensions, bigraded by weight and depth, have sizes in precise agreement 
with the Broadhurst--Kreimer and Broadhurst conjectures. Moreover, we lend 
further support to these conjectures by studying even greater weights 
($w\le30$), using modular arithmetic. To obtain these results we derive a 
new type of relation for Euler sums, the Generalized Doubling Relations. 
We elucidate the ``pushdown'' mechanism, whereby the ornate enumeration of 
primitive MZVs, by weight and depth, is reconciled with the far simpler 
enumeration of primitive Euler sums. There is some evidence that this 
pushdown mechanism finds its origin in doubling relations. We hope that our 
data mine, obtained by exploiting the unique power of the computer algebra 
language {\sc form}, will enable the study of many more such consequences 
of the double-shuffle algebra of MZVs, and their Euler cousins, which are 
already the subject of keen interest, to practitioners of quantum field 
theory, and to mathematicians alike.
}  
\end{abstract}
\vfill
\end{titlepage}
%
%
\setcounter{equation}{0}

\section{Introduction}
\renewcommand{\theequation}{\thesection.\arabic{equation}}
\setcounter{equation}{0}

\vspace{1mm}
\noindent
Multiple Zeta Values (MZVs) and Euler sums 
\cite{Euler,ZAG1,Hof1}  have been of interest to 
mathematicians~\cite{Euler,OLD,Hof2,CARTIER,ZUD} and physicists~\cite{PHYS} 
for a long time. One place in physics in which they are important is 
perturbative Quantum Field Theory. The interest became even larger when 
higher order calculations in Quantum Electrodynamics (QED) and Quantum 
Chromodynamics (QCD) started to need the multiple harmonic sums 
$S_{\vec{c}}(N)$ \cite{HSUM1,Vermaseren:1,HSUM3}.
Euler sums are obtained as the limit $N \rightarrow \infty$ of the 
related multiple sums $Z_{\vec{c}}(N)$
\begin{equation}
\label{eq:EZV}
\zeta_{\vec{c}} = \sum_{k=1}^{\infty} \frac{({\sigma}(b))^k}{k^{|b|}}
       Z_{\vec{a}}(k-1)~,
\end{equation}
with $\vec{c} = (b,\vec{a}),~~b, a_i \in {\bf Z}$ and
\begin{equation}
\label{eq:HSUM}
Z_{b,\vec{a}}(N) = \sum_{k=1}^N \frac{(\sigma(b))^k}{k^{|b|}} 
Z_{\vec{a}}(k-1)~,~~~~Z_\emptyset = 1,~~Z_{\vec{a}}(0) = 0~,
\end{equation}
with $\sigma(b) = {\rm sign(b)}$. Euler sums for which all indices are 
positive are called Multiple Zeta Values. Euler sums and MZVs with the 
first index $b=1$ diverge, but will be included symbolically in the 
following, for convenience. Their degree of divergence can be uniquely 
traced back to a polynomial in the single harmonic sum $S_1(\infty) = 
\sum_{N\rightarrow\infty}\sum_{k=1}^N\frac{1}{k}$ shown later in the text.
We call the number of indices
of the Euler sums and MZVs their depth {\sf d} and
\begin{equation}
{\sf w} = \sum_{k=1}^{{\sf d}} |c_k|
\end{equation}
their weight.

The number of Euler sums, resp. MZVs, up to a given weight {\sf w} grows 
rapidly and amounts to $2\ \cdot 3^{\sf w-1}$ and $2^{\sf w-1}$, respectively. 
A 
central question thus concerns to find all the relations between the Euler 
sums, resp. MZVs for {\it fixed weight} and {\it depth}, and even more 
importantly, {\it new relations} between MZVs at the one hand and Euler 
sums on the other hand, and the corresponding bases. Besides weight and 
depth, another degree of freedom, being discussed later, the {\it pushdown} 
{\sf p}, quantifies the relation between MZVs and Euler sums. The way to 
view MZVs, embedded into Euler sums, dates back to 
Broadhurst~\cite{Broadhurst:1}, who conjectured the counting of basis 
elements at fixed $\{w,d\}$. The corresponding conjecture for the MZVs is 
due to Broadhurst and Kreimer \cite{BK1}\footnote{Conjectures for fixed 
weight are due to Zagier~\cite{ZAG1} and probably also independently due to 
Drinfel'd, Goncharov and Kontsevich.}. For the number of basis elements for 
MZVs of a given weight, without regard to depth, an upper bound has been 
proven in \cite{W1}. This coincides with the result obtained by summing the 
numbers conjectured in \cite{BK1} over all depths at a fixed weight.

The relations between MZVs and Euler sums in Ref.~\cite{Broadhurst:1} are 
conjectured using algorithms for integer relations as {\sf PSLQ} 
\cite{PSLQ} and {\sf LLL} \cite{LLL} which use representations based on a 
large number of digits.

It is well-known that MZVs obey shuffle- and stuffle-relations. This is 
due to their representation in terms of Poincar\'{e} iterated integrals 
\cite{ITINT} at argument $x=1$, which are harmonic polylogarithms 
\cite{harmpol} on the one hand, and harmonic sums \cite{HSUM1,Vermaseren:1,
HSUM3} on the other hand. The former quantities obey a shuffle- the latter 
a quasi-shuffle algebra, i.e. shuffling with ``stuff'' from polynomials of 
harmonic sums of lower weight. Currently no other independent relation is 
known between MZVs. The Euler sums are also related by both the shuffle- 
and stuffle-relations, where now also negative indices occur to indicate 
alternating sums. However, these relations are not sufficient to obtain the 
minimal set of basis elements as being conjectured in \cite{Broadhurst:1}. 
Starting with ${\sf w = 8}$ it requires the doubling relation and with 
${\sf w = 11}$ generalized doubling relations derived in the present paper. 
Beginning with ${\sf w = 12}$ relations occur, which allow to express MZVs 
of a given depth in terms of Euler sums of a lesser depth. Part of these 
relations have been conjectured in the past using integer relations 
\cite{Broadhurst:1,BBB}. A main objective of the present paper is to prove 
these relations applying computer algebra methods and to find relations of 
this type in a more systematic way.

We investigate the Euler sums to ${\sf w = 12}$ completely, deriving 
basis-representations for all individual values in an explicit analytic 
calculation. For the MZVs the same analysis is being performed up to ${\sf 
w = 22}$. To ${\sf w = 24}$ we checked the conjectured size of the basis 
using modular arithmetic. Under the further conjecture that the basis 
elements can be chosen out of MZVs of depth ${\sf d \leq w/3}$ we confirm 
the conjecture up to ${\sf w = 26}$. Furthermore, the following runs at 
limited depth, using modular arithmetic keeping the 
highest weight terms only, were performed: 
${\sf d = 7,~~ w = 27}$;
${\sf d = 6,~~ w = 28}$;
${\sf d = 7,~~ w = 29}$;
${\sf d = 6,~~ w = 30}$.
For the Euler sums complete results were obtained for 
${\sf d \leq 3,~~ w = 29}$;
${\sf d \leq 4,~~ w = 22}$; 
${\sf d \leq 5,~~ w = 17}$ and for 
${\sf d \leq 3,~~ w = 51}$;
${\sf d \leq 4,~~ w = 30}$;
${\sf d \leq 5,~~ w = 21}$;
${\sf d \leq 6,~~ w = 17}$ using modular arithmetic neglecting products of 
lower weight. 
The conjectures on the number of basis elements 
w.r.t. $\{w,d\}$ were verified in all these cases. The results of our 
analysis are made available in the Multiple Zeta Data Mine \cite{www}, to 
allow users to search for yet un-discovered relations. 

The paper is organized as follows. In Section~\ref{BasicFormalism} we 
summarize basic notations and the well known relations between Euler sums. 
A novel type of relations, the generalized doubling relations, is derived 
in Section~\ref{GDR}. There we also discuss its impact in finding the basis 
elements at a given weight ${\sf w}$ and depth ${\sf d}$. In 
Section~\ref{ComputerProgram} an outline is given on the details of the 
computer algebra code, which allowed to derive the basis-representations of 
the MZVs and Euler sums. Details on the running for the different cases are 
reported in Section~\ref{Runningprograms}. The results are stored in the 
${\sf Multiple~Zeta~Value~Data~Mine}$~\footnote{It goes without saying that 
also the Euler sums are covered here.}, which is described in 
Section~\ref{TheDataMine}. To establish the solution of the problems dealt 
with in the current project required some new features of {\tt {\tt FORM}} 
\cite{FORM} and {\tt TFORM} \cite{Vermaseren:2}, which are described in 
Section~\ref{FORMaspects}.
In Section~\ref{FirstResults} 
we briefly review the status achieved by other groups and present first 
results of the analysis. In particular a series of conjectures made in the 
mathematical literature are confirmed within the range explored in the present 
study. Here we discuss also particular choices for the respective bases. 
An interesting aspect representing MZVs by Euler sums concerns the 
so-called pushdowns, i.e. the representation of a MZV of a given depth 
${\sf d}$ with Euler sums of depth ${\sf d'}$ with ${\sf d' < d}$. These 
are studied in Section~\ref{sec:pushdown} in which we also 
introduce a new kind of object, the $A_{\vec{a}}$--functions. They play a 
key role in representing a class of Euler sums. Some more 
special Euler sums are studied in Section~\ref{SpecialEulerSums}. 
Section~\ref{Outlook} contains the conclusions and an outlook. In the 
Appendices we provide different basis representations and discuss the 
pushdowns in more detail.


\section{Basic Formalism}
\renewcommand{\theequation}{\thesection.\arabic{equation}}
\setcounter{equation}{0}
\label{BasicFormalism}

\vspace{1mm}
\noindent
In the following we work with three types of objects, the finite nested 
harmonic $S_{\vec{a}}$-sums, $Z_{\vec{a}}$-sums, both at argument $N \in {\bf N}$, and 
the harmonic polylogarithms $\HH_{\vec{a}}$ at argument $x,~~0 \leq x \leq 
1$. They all can be used to define the MZVs and the Euler sums in the limit 
$N \rightarrow \infty$ and $x=1$, respectively. We generally consider the 
case of colored objects corresponding to $n=2$, i.e. numerator weights with 
$(\pm 1)^k$, i.e. polylogarithms of square root of unity.

The harmonic $S$-sums are defined by 
\begin{eqnarray}
   S_{\vec{a}}(0) &=& 0 \nonumber\\
   S_{b}(N) &=& \sum_{k=1}^N \frac{(\sigma(b))^k}{k^{|b|}} \nonumber \\
   S_{b,\vec{a}}(N) &=& \sum_{k=1}^N \frac{(\sigma(b))^k}{k^{|b|}} S_{\vec{a}}(k)~.
\end{eqnarray}
In this form these sums are usually used by physicists. In particular 
results in QCD \cite{MVV,HEAV,STRUCT5,STRUCT6}  
are expressed in terms of these objects\footnote{The class of Euler sums 
is known to be too small in general to represent all Feynman diagrams for 
no-scale processes in scalar field theories, but have to be extended in higher 
orders~\cite{BGK,Broadhurst:1998rz,ANDRE,BROWN}. This will apply also for 
field theories as 
QCD and QED. Feynman-integrals are periods~\cite{periods} if all ratios of 
Lorenz invariants and masses have rational values \cite{BW}.}. 

Next there 
are the $Z$-sums. They are defined in (\ref{eq:HSUM}). 
These are of course very similar to the $S$-sums and it is straightforward  
to convert from one notation to the other. The $Z$-sums are mostly used by 
mathematicians. In the limit $N \rightarrow \infty$ and when $\sigma(b)=1$ 
for all $b$ they define the Multiple Zeta Values (MZVs):
\begin{equation}
\zeta_{\vec{a}} = \lim_{N \rightarrow \infty} Z_{\vec{a}}(N)~.
\end{equation}
When we allow $\sigma(b)$ to take the values $+1$ or $-1$ and we take the 
limit $N \rightarrow \infty$ we speak of Euler sums. 

Finally, there are the harmonic polylogarithms, which we will also call 
$\HH$-functions. We consider the alphabets
\begin{eqnarray}
\mathfrak{h} &=& \{0,1,-1\}~~~{\rm and} \nonumber\\
\mathfrak{H} &=& \{1/x, 1/(1-x), 1/(1+x)\}~,
\end{eqnarray}
which define the elements of the index set of the harmonic 
polylogarithms\footnote{Special cases are the classical polylogarithms 
\cite{LEWIN} and the Nielsen polylogarithms \cite{NIELS2}. Generalizations of
harmonic polylogarithms are found in \cite{MUW,GONPL}.} 
and the functions in the iterated integrals, respectively.
Let $\vec{a} = \{m_1, \ldots, m_k\},~~m_i,b~\in~\mathfrak{h},~~k \geq 1$, then
\begin{eqnarray}
        \HH_{{\it b},\vec{a}}(x) &=& \int_0^x dz f_b(z) \HH_{\vec{a}}(z) 
\nonumber\\
	f_0(z)               &=& 1/z \nonumber \\
	f_1(z)               &=& 1/(1-z) \nonumber \\
	f_{-1}(z)            &=& 1/(1+z) \nonumber \\
	\HH_0(x)             &=& ~~~\log (x)    \nonumber\\
	\HH_1(x)             &=& -\log (1-x)  \nonumber\\
	\HH_{-1}(x)          &=& ~~~\log (1+x)~.
\end{eqnarray}
The sums to infinity and the $\HH$-functions at unity are all related and can 
be readily transformed into each other. For some applications it is most 
convenient to work with one set of objects and for others other objects may 
be more useful. For reasons being explained later our computer programs 
work mostly with $\HH$-functions at unity.

A first aspect to note is that the index fields of the sums and the 
functions are of a different nature. This can be seen by introducing the 
notation in which the index $n$ in the sums can alternatively be written as 
$n-1$ zeroes followed by a one and $-n$ is written as $n-1$ zeroes followed 
by a minus one. In the $\HH$-functions we can absorb alternatively the zeroes 
in the nonzero number to their right by raising its absolute value by one 
for each zero being absorbed. This leaves only the rightmost zeroes. Hence:
\begin{eqnarray}
	S_{-3,4}(N) & = & S_{0,0,-1,0,0,0,1}(N) \nonumber\\
	Z_{2,-5}(N) & = & Z_{0,1,0,0,0,0,-1}(N) \nonumber\\
	\HH_{0,1,-1,0,0,-1,0,0}(x) & = & \HH_{2,-1,-3,0,0}(x)~.
\end{eqnarray}
The notation in terms of the $0,\pm 1$ we call the (iterated) integral 
notation. The natural notation of the sums we call the (nested) sum 
notation.

Reference to the alphabet $\mathfrak{h}$ allows us to count the number of 
objects and to classify them. The number of indices in this integral 
notation is called the weight of the sum or the function. For a given 
weight ${\sf w}$ there are $2 \cdot 3^{{\sf w}-1}$ sums and $3^{\sf w}$ 
$\HH$-functions. When 
the sums are written in the original sum notation, the number of indices 
indicates the number of nested sums. This is also called the depth of the 
sum. When there are no trailing zeroes in the $\HH$-functions we can 
introduce the depth in the same way. Because of algebraic relations we can 
express the functions with trailing zeroes as products of powers of 
$\log(x)$ and $\HH$-functions with fewer indices~\cite{ALGEBRA,harmpol}.
In that case the concept of depth can be used in a similar way as with the sums.

For any argument $x \ne 1$ the $\HH$-functions form a shuffle algebra:
\begin{eqnarray}
	\HH_{\vec{p}}(x) H_{\vec{q}}(x) & = & \sum_{\vec{r}~\in~\vec{p}~ 
{\SHA}~ \vec{q}}
		\HH_{\vec{r}}(x)~,
\end{eqnarray}
where $\vec{p}~\SH~\vec{q}$ denotes the shuffle product, cf. e.g. 
\cite{ALGEBRA}, and $p_i, q_i \in \mathfrak{h}$.
When $x=1$ $\HH$-functions for which the first index is one are divergent. 
It is however possible to express them in terms of a single divergent 
object and other finite terms in a consistent way. The only thing that 
breaks down is that there are correction terms to the shuffle relations when 
both objects in the left hand side are divergent, see also 
Ref.~\cite{harmpol}. Because the number of non-zero indices remains the 
same during the shuffle operation, we call it depth preserving.

For general argument $N$ the sums form a stuffle algebra, \cite{ALGEBRA}. 
This is a general property of sums which we show here for a double sum:
\begin{eqnarray}
	\sum_{i=1}^N \sum_{j=1}^N & = &
		\sum_{i=1}^N \sum_{j=1}^{i}+\sum_{j=1}^N \sum_{i=1}^{j}
			-\sum_{i=j=1}^N \nonumber\\
	 & = &
		\sum_{i=1}^N \sum_{j=1}^{i-1}+\sum_{j=1}^N \sum_{i=1}^{j-1}
			+\sum_{i=j=1}^N~. 
\end{eqnarray}
The diagonal terms give extra `stuff' beyond the normal shuffling in the 
natural notation for the sums. Even though the diagonal terms add terms 
usually the stuffle relations have fewer terms because most of the time 
some of the indices will have an absolute value greater than one. We write 
in terms of $S$-- or $Z$--notation :
\begin{eqnarray}
	S_m(N) S_n(N) & = & S_{m,n}(N)+S_{n,m}(N)-S_{m \& n}(N) \\
	S_m(N) S_{n,k}(N) & = & S_{m,n,k}(N)+S_{n,m,k}(N)+S_{n,k,m}(N)
				-S_{m \& n,k}(N)-S_{n,m \& k}(N) \nonumber \\ \\
	Z_m(N) Z_n(N) & = & Z_{m,n}(N)+Z_{n,m}(N)+Z_{m \& n}(N) \\
	Z_m(N) Z_{n,k}(N) & = & Z_{m,n,k}(N)+Z_{n,m,k}(N)+Z_{n,k,m}(N)
				+Z_{m \& n,k}(N)+Z_{n,m \& k}(N)~.
\nonumber\\ 
\end{eqnarray}
Here the operator $\&$ is defined by
\begin{eqnarray}
	m \& n  &=& \sigma(m) \sigma(n) (|m| + |n|) \nonumber\\
	      & = & \sigma_n m + \sigma_m n~.
\end{eqnarray}

The above algebraic relations can be used to bring an expression with many 
harmonic polylogarithms or harmonic sums into a standard form. For 
evaluation, however, it is often useful to work it the other way and reduce 
the number of objects at the highest weight in favor of products of objects 
with a lower weight which are easier to evaluate. For this the theory of 
Lyndon words \cite{LYND} applies, but especially with the stuffles the 
extra terms which have the same weight but a lower depth have to be taken 
along and make things considerably more involved than pure shuffles.

A $k$-ary Lyndon word of length $n$ is a $n$-letter concatenation product 
over an alphabet of size $k$, which, observing lexicographical 
ordering is smaller than all its suffixes. Equivalently, it is the unique 
minimal element in the lexicographical ordering of all its cyclic 
permutations. The uniqueness implies that a Lyndon word is aperiodic. So it 
differs from any of its non-trivial rotations. In our case we will usually 
replace minimal by maximal when we form Lyndon words of indices of MZVs or 
Euler sums. That is, we will put the larger indices to the left. One could 
also say that the concept of greater than is defined in a special way 
inside the alphabet. The practical advantage is that this guarantees that 
none of the MZVs of which the index string forms a Lyndon word is
divergent.

When we use the stuffle relations to simplify the set of objects at a given 
weight, we can arrange that they are used in such a way that they never 
raise the value of the depth parameter. Some terms will have a lower value 
for the depth. Therefore we call the stuffles potentially depth lowering.

When we consider the sums to infinity there are two classes of extra 
relations worth mentioning. The first is the `rule of the triangle' which 
is based on
\begin{eqnarray}
	\lim_{N\rightarrow\infty}\sum_{i=1}^N\sum_{j=1}^N
		& = &
	\lim_{N\rightarrow\infty}\sum_{i=1}^N\sum_{j=1}^{N-i}
	+\lim_{N\rightarrow\infty}\sum_{i=1}^N\sum_{j=N-i+1}^{N}~.
\end{eqnarray}
For most sums the second term will give a limit that goes to zero with at 
least one power of $1/N$, possibly multiplied by powers of $\log(N)$. This 
system can be generalized to the product of any pair of sums and it can be 
proven that the limit of the second term vanishes when at least one of the 
sums in the left hand side is finite~\cite{Vermaseren:1}. When both are 
divergent it is possible to work out which extra terms are needed. Because 
the sums of the first term in the right hand side can be worked out, even 
in the most general case, the above gives us an extra algebraic relation 
for the sums to infinity. These relations are depth preserving.

When we consider the $\HH$-functions at unity, it is easy to see that they 
can be written as nested sums to infinity of the same variety as the $Z$-sums 
or the $S$-sums. Hence they now obey also the stuffle algebra. And it can be 
shown that the `rule of the triangle' is no more than the equivalent of the 
shuffle algebra for the $\HH$-functions, with the same restrictions for the 
double divergent terms.

The next set of relations is easy to see for finite sums:
\begin{eqnarray}
\label{doublingrelation}
	S_{m}(N) & = &
			\sum_{i=1}^N \frac{1}{i^m} 
		 = 
			\sum_{i=1}^N 2^m \frac{1}{(2i)^m} 
		 = 
			\sum_{i=1}^{2N} 2^{m-1} \frac{1+(-1)^i}{i^m} \nonumber\\
		& = & 2^{m-1} \left[ S_m(2N) + S_{-m}(2N) \right]~,
\end{eqnarray}
which generalizes into
\begin{eqnarray}
\label{doubling}
	S_{n_1,\cdots,n_p}(N) & = & 2^{n_1+\cdots+n_p-p} \sum_{\pm}
             S_{\pm n_1,\cdots,\pm n_p}(2N)~.
\end{eqnarray}
Here the sum is over all $2^p$ plus/minus combinations. These relations 
are called the `doubling relations'. For finite sums with $n_1 \neq 1$ these relations
can be used directly. In the case that divergent sums are involved 
there are again correction terms.

The equivalent formula for the $\HH$-functions is obtained by looking at 
$H_{\vec{a}}(x^2)$ and noticing that at $x=1$ this is the same as 
$H_{\vec{a}}(x)$. In that case we have
\begin{eqnarray}
\label{eq:inx2}
	H_{1,0,1}(x^2) & = & 2 \left[ H_{1,0,1}(x) - H_{-1,0,1}(x)
			- H_{1,0,-1}(x) + H_{-1,0,-1}(x) \right]~,
\end{eqnarray}
which generalizes to any number of indices. The rule is that the factor is 
identical to $2^m$ in which $m$ is the number of zeroes in the indices, and 
each one in the left hand side gives a doubling of terms in the right hand 
side: one term with a corresponding 1 and one with a corresponding $-1$ and 
an extra overall minus sign. In the left hand side one cannot have negative 
indices. Again one should be careful with the divergent functions.

Divergences are expressed in terms of the object $S_1(\infty)$.
In most cases one can use this as a regular symbol and take it along in the 
equations and expressions. Unless we mention the problems explicitly, one 
can exchange limits and sums when this object is combined with 
finite sums. The reason is that our finite sums converge faster than that 
this object diverges. A problem occurs when we use the doubling formula on 
it. We find:
\begin{eqnarray}
	S_1(\infty) & = & S_1(2\infty) + S_{-1}(2\infty) \nonumber\\
	            & = & S_1(2\infty) - \log(2)~,
\end{eqnarray}
which just shows that the divergence of $S_1(\infty)$ is logarithmic, since
\begin{equation}
S_1(N) = \ln(N) + \gamma_E + \frac{1}{2N} + \frac{1}{12 N^2} + O\left(\frac{1}{N^3}\right)~,
\end{equation}
cf.~\cite{STRUCT5}.
One 
can however use the stuffle relations on these objects. This allows one in 
principle to express the divergent sums in terms of products of 
$S_1(\infty)$ and finite sums as in
\begin{eqnarray}
	S_{1}(N) S_{m,n}(N) & = & S_{1,m,n}(N) + S_{m,1,n}(N) + S_{m,1,n}(N)
			-S_{m\& 1,n}(N) - S_{m,n\& 1}(N)~.
\nonumber\\
\end{eqnarray}
If we assume  $m\ne 1$ this allows us to express the divergent sum 
$S_{1,m,n}(\infty)$ the way we want it. Similarly one can now look at 
stuffles of $S_1 \cdot S_1$ to determine $S_{1,1}$ and then look at 
stuffles of $S_{1,1}(N)$ with finite sums. In the programs we give 
$S_1(\infty)$ the name {\tt Sinf} which, due to the above, can be treated as 
a regular symbol.

Because we have two shuffle products - the stuffle-algebra is a quasi-shuffle 
algebra \cite{HOF2} -  we can equate the result 
of the stuffle product of two objects with the result of the shuffle 
product of the same two objects. The resulting relation is called a double-shuffle 
relation and contains only objects of the same weight.
These relations have been used in a number of calculations. For our type of 
calculations they are, however, not suitable. We will use the stuffle and the 
shuffle relations individually. This will allow a better optimization of 
the algorithms.

The concept of {\it duality} is very useful and allows us to roughly half the 
number of objects that need to be computed. The duality relation is defined 
in the integral notation using harmonic polylogarithms at one. It states that 
if we have a MZV and we reverse 
the order of its indices while at the same time transforming zeroes into 
ones and ones into zeroes the new object has the same value as the 
original. An example of this duality is the 
relation
\begin{eqnarray}
	H_{0,1,0,1,1,1,1,1} & = & H_{0,0,0,0,0,1,0,1}
\end{eqnarray}
In mathematics one traditionally considers this duality in sum notation. In 
that case, for a sequence
\begin{equation}
I = (p_1+1,\{1\}_{q_1-1},p_2+1,\{1\}_{q_2-1}, \ldots,
p_k+1,\{1\}_{q_k-1})
\end{equation}
there is a dual sequence
\begin{equation}
\tau(I) = (q_k+1,\{1\}_{p_k-1},q_{k-1}+1,\{1\}_{p_{k-1}-1},\ldots,
q_1+1,
\{1\}_{p_1-1})~.
\end{equation}
The duality theorem  \cite{ZAG1}  states
\begin{equation}
\label{DUAL}
\zeta_{I} = \zeta_{\tau(I)}~.
\end{equation}
It was conjectured in \cite{HOF3} and is easily proven by the transformation
$x \to 1-t$ of the corresponding iterated integrals.

Because for even weights there are some elements that are 
self-dual this does not divide the number of terms exactly by two. 
Considering that we do not have to consider the divergent objects we have
$2^{{\sf w}-3}$ relevant objects when ${\sf w}$ is odd and 
$2^{{\sf w}-3}+2^{{\sf w}/2-2}$ relevant 
objects when $w$ is even.

For Euler sums the equivalent transformation is more complicated due to the 
three letter alphabet. It is obtained by studying the transformation
\begin{equation}
x \rightarrow \frac{1-t}{1+t}
\end{equation}
in the integral representation. Its effect is that given the alphabet
\begin{eqnarray}
	A & = \ \ 0  & \leftarrow \ \ \frac{1}{x} \nonumber \\
	B & = \ \ 1  & \leftarrow \frac{1}{1-x} \nonumber \\
	C & = -1 & \leftarrow \frac{1}{1+x}
\end{eqnarray}
and a string of letters from this alphabet as indices of an Euler sum 
$\HH$, the `dual expression' is obtained by reverting the string of letters 
and making the replacement
\begin{eqnarray}
\label{eq:pseudodual}
	A & \rightarrow & B \oplus C \nonumber \\
	B & \rightarrow & A \ominus C \nonumber \\
	C & \rightarrow & C~.
\end{eqnarray}
The addition and subtraction operators here mean that for each such 
transformation there will be a doubling of the number of terms, one with 
the first letter and the other with the other letter. The sign-operator 
$\oplus (\ominus)$ refers to 
the sign of the complete term. Because these relations can both raise and 
lower the depth of a term we call them depth mixing.

We have tested that this transformation does add something new beyond what 
the stuffles and the shuffles give us. In particular, when one derives 
equations for all sums at a given weight, they can be used to replace the 
doubling and the {\it Generalized Doubling Relations} (GDRs), see 
Section~\ref{GDR}. 
We have tested this to weight ${\sf w = 12}$. Unfortunately they cannot be 
used when 
the concept of depth of the sums is important and hence we have not used 
these equations in our programs.

A  generalization of the Riemann $\zeta$-function is Hurwitz' 
$\zeta$-function \cite{HURWITZ,CARTIER}~:
\begin{eqnarray}
\zeta(n,a) = \sum_{k=1}^\infty \frac{({\rm sign}(n))^k}{(k+a)^{|n|}}~,
\end{eqnarray}
which can be extended to generalized Euler sums analogous to (\ref{eq:EZV}).
Since $a$ is a real parameter, one may differentiate $\zeta(\vec{c},a)$ w.r.t. 
$a$ and seek for new relations. We investigated this possibility, but did not 
find new relations beyond those quoted above.

When we are discussing bases into which we write the MZVs and the Euler 
sums we recognize two types of basis~:

\vspace*{1mm}\noindent
{\bf Definition.}~A basis of a vector space of all Euler sums or MZVs
at a given weight ${\sf w}$ is called a {\it Fibonacci basis}.

\vspace*{1mm}\noindent
{\bf Definition.}~A basis of the ring of all Euler sums or MZVs
at a given weight ${\sf w}$ is called a {\it Lyndon basis} if all its 
elements have an index field that forms a Lyndon word.

\vspace*{2mm} \noindent
In a Fibonacci basis all basis elements are nested sums of the same weight. 
The name derives from the observation that the size of such a basis for the 
Euler sums seem to follow a Fibonacci rule \cite{FIBO}. Also the MZVs seem 
to follow the rule that the total number of their basis elements follow the 
Fibonacci-like Padovan numbers \cite{PADO}, see Appendix~\ref{gen:basis}.

In a Lyndon basis we write in the complete basis as many elements as 
possible as products of lower weight basis elements and what remains is the 
Lyndon basis. Simultaneously we require the index field to form a Lyndon 
word. Sometimes a Lyndon basis can be formed from a Fibonacci basis by just 
selecting the Lyndon words from it. The number of basis elements in the 
case of MZVs is counted by a Witt-type relation \cite{WITT} based on the 
Perrin numbers~\cite{Perrin}. In the case of the Euler sums the 
corresponding relation relies on the  Lucas numbers \cite{LUCAS}, see 
Appendix~\ref{gen:basis}. Any other basis we will call a mixed basis.

We will usually try to arrange the Lyndon bases in such a way that they are 
`minimal depth'. This means that if an element can be expressed in terms of 
objects with a lower depth, it cannot be a member of the basis. Details on 
a variety of bases are given in Appendix~\ref{gen:basis}. The complete 
basis we actually selected for the MZVs is presented in 
Appendix~\ref{ap:bases}.

\section{Conjectures on Bases at Fixed Weight and Depth}
\renewcommand{\theequation}{\thesection.\arabic{equation}}
\setcounter{equation}{0}
\label{ConjecturesonBases}

\vspace{1mm}\noindent
Broadhurst \cite{Broadhurst:1} and Broadhurst and Kreimer \cite{BK1} formulated
conjectures on the size of the basis for  Euler sums and MZVs, respectively, 
which we 
summarize in the following.

Euler sums $\zeta_{\vec{a}}$ at given weight and depth {\sf w,d} are 
called {\sf independent} if there exists no relation between them, cf. 
Sect. 2,4.
The elements of the basis through which all Euler sums can be represented 
in terms of polynomials are called {\sf primitive}. The numbers of 
independent and primitive sums at a given weight are fixed, while 
different basis representations may be chosen.

Let $E_{w,d}$ be the number of independent Euler sums at weight ${\sf w > 2}$ 
and depth ${\sf d}$ that cannot be reduced to primitive Euler sums of lesser 
depth
and their products. Thus we believe that $E_{3,1}=1$, since there is
no known relationship between $\zeta_3$, $\pi^2$ and $\ln(2)$. It is
rather natural to guess that $E_{w,d}$ is given by a filtration
of the coefficients of powers of $x$ and $y$ in the expansion
of $1/(1-xy-x^2)$, i.e.\ that
\begin{equation}
\prod_{w>2}\prod_{d>0}(1-x^w y^d)^{E_{w,d}}\stackrel{?}{=}
\frac{1-xy-x^2}{(1-xy)(1-x^2)}=1-\frac{x^3y}{(1-xy)(1-x^2)}.
\label{conj1}\end{equation}
It is then easy to obtain $E_{w,d}$ by M\"obius
transformation of the binomial coefficients in Pascal's triangle. Let
\begin{equation}
T(a,b)=\frac{1}{a+b}\sum_{d|a,b}\mu(d)\,\frac{(a/d+b/d)!}{(a/d)!(b/d)!}
\label{ET}
\end{equation}
where the sum is over all positive integers $d$ that divide both $a$
and $b$ and the M\"obius function is defined by
\begin{equation}
\mu(d)=\left\{\begin{array}{ll}
1&\mbox{ when $d=1$}\\
0&\mbox{ when $d$ is divisible by the square of a prime}\\
(-1)^k&\mbox{ when $d$ is the product of $k$ distinct primes}
\end{array}\right.\label{mu}
\end{equation}
When $w$ and $d$ have
the same parity, and $w>d$, one obtains from~(\ref{conj1})
\begin{equation}
E_{w,d}=T\left(\frac{w-d}{2},d\right)\,.\label{Enk}
\end{equation}
With the exception of $\ln(2)$ and $\zeta_2$, which act as the seeds $xy$ and 
$x^2$,
all elements of the basis are thereby conjecturally enumerated.
In this paper we provide extensive evidence to support conjecture~(\ref{conj1}).

Now let $D_{w,d}$ be the number of independent MZVs at weight $w>2$ 
and depth $d$ that cannot be reduced to primitive MZVs of lesser depth
and their products. Thus we believe that $D_{8,2}=1$, since there is
no known relationship between the double sum $Z_{5,3}=\sum_{m>n>0} 1/(m^5n^3)$
and single sums or their products. It is tempting to guess to that
$D_{w,d}$ is generated by filtration of the expansion of $1/(1-x^2-x^3y)$,
seeded by $\pi^2$ and $\zeta_3$. But this is not the case, since the solution
of the double-shuffle algebra at weight $w=12$ leaves one quadruple 
sum undetermined, while the obvious guess would leave none.
The conjecture \cite{BK1} in this case is rather ornate, cf. Table~\ref{BK1table}.
\begin{equation}
\prod_{w>2}\prod_{d>0}(1-x^w y^d)^{D_{w,d}}\stackrel{?}{=}
1-\frac{x^3y}{1-x^2}+\frac{x^{12}y^2(1-y^2)}{(1-x^4)(1-x^6)}\label{conj2}
\end{equation}
with a correction term whose numerator, $x^{12}y^2(1-y^2)$,
ensures that $D_{12,4}=1$ and $D_{12,2}=1$, in agreement with the solution 
of the double-shuffle algebra. The denominator $(1-x^4)(1-x^6)$ is then chosen
to give $D_{2m,2}=\lfloor(m-1)/3\rfloor$ for the number of primitive
double sums with weight $2m$.  Conjecture~(\ref{conj2}) is impressively supported
by the data mine.

Furthermore, 
\begin{equation}
\label{eq:BK2}
\prod_{w > 2} \prod_{d > 0} (1-x^w y^d)^{M_{w,d}} = \frac{1 - x^2 - x^3 
y}{1-x^2}
\end{equation}
is the conjectured generating function of the basis elements $M_{w,d}$ of the 
MZVs when 
expressed as 
Euler sums in a minimal depth representation, see~Table~\ref{BK2table}.
\section{Generalized Doubling Relations}
\renewcommand{\theequation}{\thesection.\arabic{equation}}
\setcounter{equation}{0}
\label{GDR}

\vspace{1mm}
\noindent
Up to ${\sf w = 10}$ the shuffle-, stuffle-, and doubling relations were 
sufficient to express the alternating Euler sums over a basis whose size is 
in accordance with the conjecture in Ref.~\cite{Broadhurst:1}. This is not 
the case from ${\sf w = 11}$ onwards. Therefore one has to seek a new kind 
of relations, which we derive in the following. Of course, when we derive 
all relations at a given weight we could use the relations of 
(\ref{eq:pseudodual}). The fact that they are depth mixing makes them 
useless for calculations in which the concept of depth plays a role. Hence 
we need our new (depth lowering) relations anyway. We first present the 
derivation of this class of relations which we call Generalized Doubling 
Relations (abbreviated to GDRs) and discuss then their effect on the number 
of basis elements representing the Euler sums.

\subsection{Derivation of the generalized doubling relations}

\vspace{1mm}\noindent
The only relations we could find thus far adding something new to the 
system are the depth~2 relations of Ref.~\cite{Broadhurst:1}. They 
are based on partial fractioning in two different ways. One way is:
\begin{eqnarray}
\frac{1}{(2i+j)(j)}  =  \frac{1}{(2i+2j)(2i+j)} + \frac{1}{(2i+2j)(j)}
\end{eqnarray}
We can take out the factor two and in the first term the $2i$ is taken care 
of by changing the summation over $i$ into a summation over the even 
numbers by including a factor $(1+(-1)^i)/2$, which introduces negative 
indices in some Euler sums. In the other way we use the more regular form
\begin{eqnarray}
\frac{1}{(2i+j)(j)}  =  \frac{1}{(2i)(j)}-\frac{1}{(2i)(2i+j)}~.
\end{eqnarray}
Together these partial fractions produce new types of relations.

Here we will give the new set of relations and their derivation. We will 
work with the $Z$-sums. The reason is a particularly handy representation of 
these sums to infinity~\cite{HOF3,BBBL}:

\begin{eqnarray}
Z_{m_1,\cdots,m_p}(\infty) & \! = \! &
\sum_{i_1 > i_2 > \cdots > i_p > 0}^\infty
\frac{\sigma_1^{i_1}\sigma_2^{i_2} \cdots \sigma_p^{i_p}}{i_1^{n_1} i_2^{n_2} \cdots i_p^{n_p}}
		\nonumber \\ & \! = \! &
\sum_{x_1=1}^\infty \sum_{x_2=1}^\infty \cdots \sum_{x_p=1}^\infty
\frac{\sigma_1^{x_1+x_2+\cdots+x_p}\sigma_2^{x_2+\cdots+x_p}\cdots \sigma_p^{x_p}}
{(x_1+x_2+\cdots+x_p)^{n_1}(x_2+\cdots+x_p)^{n_2}\cdots (x_p)^{n_p}}~,
\nonumber\\
\end{eqnarray}
in which we take $n_i = |m_i|$ and $\sigma_i$ to be the sign of $m_i$.

Let us start with the re-derivation of the equation for depth ${\sf d=2}$. 
Actually we do not reproduce it exactly, but we obtain a similar equation. 
Here we write for brevity $Z(a,b) = Z_{a,b}(\infty)$. Throughout this 
Section we assume that $a, b, c$ and $d$ are positive integers. We consider 
the 
following combination of $Z$-sums~:

\begin{align}
E(a,b) & = \frac{1}{2}(Z(a,b)+Z(-a,-b)) 
\nonumber\\
       & = 
	\sum_{x_1=1}^\infty \sum_{x_2=1}^\infty \frac{1}{(x_1+x_2)^a\ x_2^b}
						\frac{1+(-1)^{x_1}}{2}         
       = 
	\sum_{x_1=1}^\infty \sum_{x_2=1}^\infty \frac{1}{(2x_1+x_2)^a\ x_2^b} 
\nonumber\\ 
       & = 
	\sum_{x_1=1}^\infty \sum_{x_2=1}^\infty \left[ \sum_{i=1}^a
		A^{(a,b)}_i\frac{1}{(2x_1\plus 2x_2)^{a+b-i}\ (2x_1\plus x_2)^i} 
         + 
	\sum_{i=1}^b
		B^{(a,b)}_i\frac{1}{(2x_1\plus 2x_2)^{a+b-i}\ x_2^i} \right]
\nonumber
\end{align}\begin{align}
       & = 
	\sum_{i=1}^a A^{(a,b)}_i 2^{i-a-b} \sum_{x_1=1}^\infty \sum_{x_2=1}^\infty
		\frac{1}{(x_1\plus x_2)^{a+b-i}\ (2x_1\plus x_2)^i} 
\nonumber \\
       & +
	\sum_{i=1}^b B^{(a,b)}_i 2^{i-a-b} Z(a\plus b\minus i,i)
\nonumber\\
	    & = 
	\sum_{i=1}^b B^{(a,b)}_i 2^{i-a-b} Z(a+b-i,i)
+ 
	\sum_{i=1}^a A^{(a,b)}_i 2^{i-a-b}
		\sum_{x_1=1}^\infty \sum_{x_2=x_1+1}^\infty \frac{1}{(x_1+x_2)^i x_2^{a+b-i}}
\nonumber\\
	    & = 
	\sum_{i=1}^b B^{(a,b)}_i 2^{i-a-b} Z(a+b-i,i)
 + 
	\sum_{i=1}^a A^{(a,b)}_i 2^{i-a-b}
		\sum_{x_1=1}^\infty \sum_{x_2=1}^\infty \frac{1}{(x_1+x_2)^i x_2^{a+b-i}}
\nonumber\\ 
		&  - 
	\sum_{i=1}^a A^{(a,b)}_i 2^{i-a-b}
		\sum_{x_1=1}^\infty \sum_{x_2=1}^{x_1} \frac{1}{(x_1+x_2)^i x_2^{a+b-i}}
\nonumber\\
	   & = 
	\sum_{i=1}^b B^{(a,b)}_i 2^{i-a-b} Z(a+b-i,i)
		  + 
	\sum_{i=1}^a A^{(a,b)}_i 2^{i-a-b} Z(i,a+b-i)
\nonumber\\ &- 
	\sum_{i=1}^a A^{(a,b)}_i 2^{i-a-b}
		\sum_{x_2=1}^\infty \sum_{x_1=x_2}^\infty \frac{1}{(x_1+x_2)^i x_2^{a+b-i}}
\nonumber\\
& = 
	\sum_{i=1}^b B^{(a,b)}_i 2^{i-a-b} Z(a+b-i,i)
		+ 
	\sum_{i=1}^a A^{(a,b)}_i 2^{i-a-b} Z(i,a+b-i)
		\nonumber \\ & - 
	\sum_{i=1}^a A^{(a,b)}_i 2^{i-a-b}
		\sum_{x_2=1}^\infty \sum_{x_1=1}^\infty \frac{1}{(x_1+2x_2)^i x_2^{a+b-i}}
		- 
	\sum_{i=1}^a A^{(a,b)}_i 2^{i-a-b}
		\sum_{x_2=1}^\infty \frac{1}{(2x_2)^i x_2^{a+b-i}}
\nonumber \\ 
& = 
	\sum_{i=1}^b B^{(a,b)}_i 2^{i-a-b} Z(a+b-i,i)
		+ 
	\sum_{i=1}^a A^{(a,b)}_i 2^{i-a-b} Z(i,a+b-i)
\nonumber \\ 
& - 
	\sum_{i=1}^a A^{(a,b)}_i
		\sum_{x_2=1}^\infty \sum_{x_1=1}^\infty \frac{1+(-1)^{x_2}}{2}
			\frac{1}{(x_1+x_2)^i x_2^{a+b-i}}
		- 
	\sum_{i=1}^a A^{(a,b)}_i 2^{-a-b}
		\sum_{x_2=1}^\infty \frac{1}{x_2^{a+b}}
\nonumber \\ & = 
	\sum_{i=1}^b B^{(a,b)}_i 2^{i-a-b} Z(a+b-i,i)
		 + 
	\sum_{i=1}^a A^{(a,b)}_i 2^{i-a-b} Z(i,a+b-i)
\nonumber \\ & - 
	\sum_{i=1}^a A^{(a,b)}_i \frac{1}{2}(Z(i,a+b-i)+Z(i,-(a+b-i)))
		- 
	\frac{(a+b-1)!}{(a-1)!\ b!} 2^{-a-b} Z(a+b)~,
\end{align} 

\vspace*{-5mm}
with
\begin{eqnarray}
A^{a,b}_i & = & \frac{(a+b-i-1)!}{(a-i)!(b-1)!} \\
B^{a,b}_i & = & \frac{(a+b-i-1)!}{(b-i)!(a-1)!}~.
\end{eqnarray}
Actually there is a slight problem with the above derivation. At two points 
we changed the summation range. Once from $\infty$ to $\infty/2$ and once 
from $\infty$ to $2\infty$. This causes no problems if the sum is finite, 
but for the divergent sums this needs a correction term. The second case 
is harmless as it concerns only an inner sum, the step in which 
$(-1)^{x_2}$ is introduced. But the first case, in the very first step of 
the derivation, needs a correction term. Hence the full formula becomes~: 
\begin{eqnarray}
E(a,\sigma_b b) & = &
	\frac{1}{2}(Z(a,\sigma_b b)+Z(-a,-\sigma_b b))
		\nonumber \\ & = &
	\frac{1}{2}\delta(a-1) Z(-1) Z(\sigma_b b)
	-\frac{1}{2}\delta(a-1)\delta(\sigma_b b - 1) Z(-2)
		\nonumber \\ & + &
	\sum_{i=1}^b B^{(a,b)}_i 2^{i-a-b} Z(a+b-i,\sigma_b i)
		\nonumber \\ & + &
	\sum_{i=1}^a A^{(a,b)}_i 2^{i-a-b} Z(\sigma_b i,a+b-i)
		\nonumber \\ & - &
	\sum_{i=1}^a A^{(a,b)}_i \frac{1}{2}(
		Z(\sigma_b i,\sigma_b (a+b-i))
		+Z(\sigma_b i,-\sigma_b (a+b-i)))
		\nonumber \\ & - &
	\frac{(a+b-1)!}{(a-1)!\ b!} 2^{-a-b} Z(a+b)~.
\end{eqnarray}
Here also the signs on the indices $a$ and $b$ are included which is only 
a very mild complication in the derivation.
The function $\delta(m)$ is one when $m$ is zero and zero otherwise. 
The $\sigma$-variables have a value that is either $+1$ or $-1$ and 
indicate non-alternating and alternating sums. Due to the symmetry of the 
starting formula a sign on the first variable is not necessary. If we put it 
anyway in the form of $\sigma_a$, $\sigma_b$ will have to be replaced by 
$\sigma_a \sigma_b$ in the right hand side.

It is quite relevant to take these $\sigma$ factors along. Although they 
are usually not needed to get a complete coverage of depth ${\sf d=2}$ sums, 
in the 
case of greater depth sums they are necessary.

The above derivation shows basically all techniques we need for the 
derivation of the greater depth formulas. In the sequel we will only carry 
the $\sigma$ factors that survive conditions posed during the derivation.

The derivation of the depth 3 formula follows a similar but slightly more 
complicated path. Again, we first omit the signs of the indices and the 
correction terms for divergent integrals when we double or half the 
summation range. Then we present the complete formula. In the derivation we 
will be a bit shorter this time as the techniques are all similar to what 
we have shown above.

\begin{eqnarray}
E(a,b,c) & = & \frac{1}{2}(Z(a,b,c)+Z(-a,-b,c))
		\nonumber \\ & = &
	\sum_{x_1=1}^\infty \sum_{x_2=1}^\infty \sum_{x_3=1}^\infty
		 \frac{1}{(x_1+x_2+x_3)^a\ (x_2+x_3)^b\ x_3^c}
						\frac{1+(-1)^{x_1}}{2}
		\nonumber \\ & = &
	\sum_{x_1=1}^\infty \sum_{x_2=1}^\infty \sum_{x_3=1}^\infty
		\frac{1}{(2x_1+x_2+x_3)^a\ (x_2+x_3)^b\ x_3^c}
		\nonumber 
\end{eqnarray}\begin{eqnarray}
& = &
	\sum_{x_1=1}^\infty \sum_{x_2=1}^\infty \sum_{x_3=1}^\infty
		\sum_{i=1}^a A^{(a,b)}_i
		\frac{1}{(2x_1+2x_2+2x_3)^{a+b-i} (2x_1+x_2+x_3)^i\ x_3^c}
		\nonumber \\ & + &
	\sum_{x_1=1}^\infty \sum_{x_2=1}^\infty \sum_{x_3=1}^\infty
		\sum_{i=1}^b B^{(a,b)}_i
		\frac{1}{(2x_1+2x_2+2x_3)^{a+b-i} (x_2+x_3)^i\ x_3^c}
		\nonumber \\ & = &
	\sum_{i=1}^b B^{(a,b)}_i 2^{-a-b+i} Z(a+b-i,i,c)
		\nonumber \\ & + &
	\sum_{i=1}^a A^{(a,b)}_i 2^{-a-b+i} Z(i,a+b-i,c)
		\nonumber \\ & - &
	\sum_{i=1}^a A^{(a,b)}_i 2^{-a-b+i} K^{(1)}_1(a+b-i,i,c)
		\nonumber \\ & - &
	\sum_{i=1}^a A^{(a,b)}_i 2^{-a-b+i} K^{(1)}_2(i,a+b-i,c)~,
\end{eqnarray}
with the $K$ functions given below. The full formula becomes
\begin{eqnarray}
E(a,\sigma_b\ b,\sigma_c\ c) & = &
		\frac{1}{2}(Z(a,\sigma_b\ b,\sigma_c\ c)+Z(-a,-\sigma_b\ b,\sigma_c\ c))
		\nonumber \\ & = &
		\frac{1}{2} Z(-1) Z(\sigma_b b,\sigma_c c) \delta(a-1)
		\nonumber \\ & - &
		\frac{1}{2} Z(-2) Z(\sigma_c c) \delta_(a-1) \delta(\sigma_b b-1)
		\nonumber \\ & + &
		\frac{1}{2} Z(-3) \delta(a-1) \delta(\sigma_b b-1) \delta(\sigma_c c-1)
		\nonumber \\ & + &
	\sum_{i=1}^b B^{(a,b)}_i 2^{-a-b+i} Z(a+b-i,\sigma_b i,\sigma_c c)
		\nonumber \\ & + &
	\sum_{i=1}^a A^{(a,b)}_i 2^{-a-b+i} Z(\sigma_b i,a+b-i,\sigma_c c)
		\nonumber \\ & - &
	\sum_{i=1}^a A^{(a,b)}_i 2^{-a-b+i} K^{(1)}_1(a+b-i,\sigma_b i,\sigma_c c)
		\nonumber \\ & - &
	\sum_{i=1}^a A^{(a,b)}_i 2^{-a-b+i} K^{(1)}_2(\sigma_b i,a+b-i,\sigma_c c)~.
\end{eqnarray}
The correction terms with the $\delta$--functions are due to the halving of 
the summation range in the first step. The $K$--functions are given by
\begin{eqnarray}
K^{(1)}_1(a,\sigma_b b,\sigma_c c) & = &
	\sum_{x_1=1}^\infty \sum_{x_2=1}^\infty
		\frac{\sigma_b^{2x_1+x_2}\sigma_c^{x_2}}{(x_1+x_2)^a\ (2x_1+x_2)^b\ x_2^c}
		\nonumber \\ & = &
	(-1)^b \sum_{i=1}^a A^{(a,b)}_i 2^{a-i} Z(i,\sigma_b\sigma_c(a+b+c-i))
		\nonumber 
\end{eqnarray}\begin{eqnarray}
& + &
	(-1)^b \sum_{i=1}^b B^{(a,b)}_i 2^{a-1}
			(Z(i,\sigma_b\sigma_c(a+b+c-i))
		\nonumber \\ & & \ \ \ \ \ \ \ \ \ \ \ \ \ \ \ 
					+Z(-i,-\sigma_b\sigma_c(a+b+c-i)))
		\nonumber \\ & + &
	(-1)^b B^{(a,b)}_1 2^{a-1} Z(-1) Z(\sigma_b\sigma_c(a+b+c-1))
		\\
K^{(1)}_2(\sigma_a a,b,\sigma_c c) & = &
	\sum_{x_1=1}^\infty \sum_{x_2=1}^\infty \sum_{x_3=1}^\infty
		\frac{\sigma_a^{x_1+2x_2+x_3}\sigma_c^{x_3}}{(x_1+2x_2+x_3)^a\ (x_2+x_3)^b\ x_3^c}
		\nonumber \\ & = &
	(-1)^c 2^{b-1} \sum_{i=1}^c B^{(b,c)}_i (-1)^i (
        Z(\sigma_a a,(b+c-i),\sigma_c i)
		\nonumber \\ & & \ \ \ \ \ \ \ \ \ \ \ \ \ \ \ 
			+Z(\sigma_a a,-(b+c-i),-\sigma_c i) )
		\nonumber \\ & - &
	(-1)^c 2^{b-1} \sum_{i=1}^b A^{(b,c)}_i (
        Z(\sigma_a a,\sigma_c (b+c-i),i)
		\\ & & \ \ \ \ \ \ \ \ \ \ \ \ \ \ \ 
			+Z(\sigma_a a,\sigma_c (b+c-i),-i) )
		\nonumber \\ & - &
	(-1)^c 2^{b-1} \frac{(b+c-1)!}{(b-1)!\ c!} (
        Z(\sigma_a a,(b+c))+Z(\sigma_a a,-(b+c)) )
		\nonumber
\end{eqnarray}
The last term in the function $K^{(1)}_1$ is also a correction term because 
we have to double the summation range on the $Z$-function of which the 
first index is one. Because the second index cannot be one in that case, we 
only need one correction term.

At depth 4 the relation becomes yet a bit more complicated but the 
derivation follows exactly the same path. We start with applying the 
non-trivial partial fractioning and then we have to try to rewrite the 
results in terms of $Z$--functions by percolating the factors two to the 
right. As there is one more sum this takes another step and we get two 
layers of $K$--functions:
\begin{eqnarray}
\label{starteq4}
E(a,\sigma_b\ b,\sigma_c\ c,\sigma_d\ d) & = &
		\frac{1}{2}(Z(a,\sigma_b\ b,\sigma_c\ c,\sigma_d\ d)
				+Z(-a,-\sigma_b\ b,\sigma_c\ c,\sigma_d\ d))
		\nonumber \\ & = &
		\frac{1}{2} Z(-1) Z(\sigma_b b,\sigma_c c,\sigma_d\ d)) \delta(a-1)
		\nonumber \\ & - &
		\frac{1}{2} Z(-2) Z(\sigma_c c,\sigma_d\ d)) \delta_(a-1) \delta(\sigma_b b-1)
		\nonumber \\
& + &
		\frac{1}{2} Z(-3) Z(\sigma_d\ d)) \delta_(a-1) \delta(\sigma_b b-1) \delta(\sigma_c c-1)
		\nonumber \\ & - &
		\frac{1}{2} Z(-4) \delta(a-1) \delta(\sigma_b b-1)
				 \delta(\sigma_c c-1) \delta(\sigma_d d-1)
		\nonumber \\ & + &
	\sum_{i=1}^b B^{(a,b)}_i 2^{-a-b+i} Z(a+b-i,\sigma_b i,\sigma_c c,\sigma_d\ d))
		\nonumber \\ & + &
	\sum_{i=1}^a A^{(a,b)}_i 2^{-a-b+i} Z(\sigma_b i,a+b-i,\sigma_c c,\sigma_d\ d))
		\nonumber \\ & - &
	\sum_{i=1}^a A^{(a,b)}_i 2^{-a-b+i} K^{(1)}_1(a+b-i,\sigma_b i,\sigma_c c,\sigma_d\ d))
		\nonumber \\ & - &
	\sum_{i=1}^a A^{(a,b)}_i 2^{-a-b+i} K^{(1)}_2(\sigma_b i,a+b-i,\sigma_c c,\sigma_d\ d))
\end{eqnarray} \begin{eqnarray}
\label{diveq4}
K^{(1)}_1(a,\sigma_b b,\sigma_c c,\sigma_d d) & = &
	\sum_{x_1=1}^\infty \sum_{x_2=1}^\infty \sum_{x_3=1}^\infty
		\frac{\sigma_b^{2x_1+x_2+x_3}\sigma_c^{x_2+x_3}\sigma_d^{x_3}}{
				(x_1+x_2+x_3)^a\ (2x_1+x_2+x_3)^b\ (x_2+x_3)^c\ x_3^d}
		\nonumber \\ & = &
	(-1)^b \sum_{i=1}^a A^{(a,b)}_i 2^{a-i} Z(i,\sigma_b\sigma_c(a+b+c-i),\sigma_d d)
		\nonumber \\ & + &
	(-1)^b \sum_{i=1}^b B^{(a,b)}_i 2^{a-1}
			(Z(i,\sigma_b\sigma_c(a+b+c-i),\sigma_d d)
		\nonumber \\ & & \ \ \ \ \ \ \ \ \ \ \ \ \ \ \ 
					+Z(-i,-\sigma_b\sigma_c(a+b+c-i),\sigma_d d))
		\nonumber \\ & + &
	(-1)^b B^{(a,b)}_1 2^{a-1} Z(-1) Z(\sigma_b\sigma_c(a+b+c-1),\sigma_d d)
				\\
K^{(1)}_2(\sigma_a a,b,\sigma_c c,\sigma_d d) & = &
	\sum_{x_1=1}^\infty \cdots \sum_{x_4=1}^\infty
		\frac{\sigma_a^{x_1+2x_2+x_3+x_4}\sigma_c^{x_3+x_4}\sigma_d^{x_4}}{
				(x_1\plus 2x_2\plus x_3\plus x_4)^a\ (
				x_2\plus x_3\plus x_4)^b\ (x_3\plus x_4)^c\ x_4^d}
		\nonumber \\ & = &
	(-1)^c 2^{b-1} \sum_{i=1}^c B^{(b,c)}_i (-1)^i (
        Z(\sigma_a a,(b+c-i),\sigma_c i,\sigma_d d)
		\nonumber \\ & & \ \ \ \ \ \ \ \ \ \ \ \ \ \ \ 
			+Z(\sigma_a a,-(b+c-i),-\sigma_c i,\sigma_d d) )
		\nonumber \\ & + &
	(-1)^c \sum_{i=1}^b A^{(b,c)}_i 2^{b-i}
		Z(\sigma_a a,\sigma_c (b+c-i),i,\sigma_d d)
		\nonumber \\ & - &
	(-1)^c \sum_{i=1}^b A^{(b,c)}_i 2^{b-i}
			 K^{(2)}_1(\sigma_a a,\sigma_c\ (b+c-i),i,\sigma_d\ d))
		\nonumber \\ & - &
	(-1)^c \sum_{i=1}^b A^{(b,c)}_i 2^{b-i}
			 K^{(2)}_2(\sigma_a a,\sigma_c\ (b+c-i),i,\sigma_d\ d))
				\\
\label{spliteq4}
K^{(2)}_1(\sigma_a\ a,\sigma_b b,c,\sigma_d d) & = &
	\sum_{x_1=1}^\infty \sum_{x_2=1}^\infty \sum_{x_3=1}^\infty
		\frac{\sigma_a^{x_1+2x_2+x_3}\sigma_b^{2x_2+x_3}\sigma_d^{x_3}}{
				(x_1+2x_2+x_3)^a\ (2x_2+x_3)^b\ (x_2+x_3)^c\ x_3^d}
		\nonumber \\ & = &
	(-1)^d \sum_{i=1}^c A^{(c,d)}_i 2^{c-i}
				Z(\sigma_a\ a,\sigma_b\sigma_d\ (b\plus c\plus d\minus i),i)
		\nonumber \\ & + &
	(-1)^d 2^{c-1} \sum_{i=1}^d B^{(c,d)}_i (-1)^i (
			Z(\sigma_a\ a,(b\plus c\plus d\minus i),\sigma_b\sigma_d\ i)
		\nonumber \\ & & \ \ \ \ \ \ \ \ \ \ \ \ \ \ \ 
			+Z(\sigma_a\ a,-(b\plus c\plus d\minus i),-\sigma_b\sigma_d\ i))
		\nonumber \\ & - &
	(-1)^d 2^{c-1} \sum_{i=1}^c A^{(c,d)}_i (
		Z(\sigma_a\ a,\sigma_b\sigma_d\ (b\plus c\plus d\minus i),i)
		\nonumber \\ & & \ \ \ \ \ \ \ \ \ \ \ \ \ \ \ 
		+Z(\sigma_a\ a,\sigma_b\sigma_d\ (b\plus c\plus d\minus i),-i))
		\nonumber \\ & - &
	(-1)^d 2^{c-1} \frac{(c+d-1)!}{(c-1)!\ d!} (
		Z(\sigma_a\ a,(b\plus c\plus d))
		\nonumber \\ & & \ \ \ \ \ \ \ \ \ \ \ \ \ \ \ 
		+Z(\sigma_a\ a,-(b\plus c\plus d)))
				\\
K^{(2)}_2(\sigma_a a,\sigma_b b,c,\sigma_d d) & = &
	\sum_{x_1=1}^\infty \cdots \sum_{x_4=1}^\infty
		\frac{\sigma_a^{x_1+x_2+2x_3+x_4}\sigma_b^{x_2+2x_3+x_4}\sigma_d^{x_4}}{
			(x_1\plus x_2\plus 2x_3\plus x_4)^a\ (x_2\plus 2x_3
			\plus x_4)^b\ (x_3\plus x_4)^c\ x_4^d}
		\nonumber 
\end{eqnarray}\begin{eqnarray}
& = &
	(-1)^d 2^{c-1} \sum_{i=1}^d B^{(c,d)}_i (-1)^i (
        Z(\sigma_a a,\sigma_b b,(c\plus d\minus i),\sigma_d i)
		\nonumber \\ & & \ \ \ \ \ \ \ \ \ \ \ \ \ \ \ 
        +Z(\sigma_a a,\sigma_b b,-(c\plus d\minus i),-\sigma_d i) )
		\nonumber \\ & + &
	(-1)^d \sum_{i=1}^c A^{(c,d)}_i 2^{c-i}
        Z(\sigma_a a,\sigma_b b,\sigma_d (c\plus d\minus i),i)
		\nonumber \\ & - &
	(-1)^d 2^{c-1} \sum_{i=1}^c A^{(c,d)}_i (
        Z(\sigma_a a,\sigma_b b,\sigma_d (c\plus d\minus i),i)
		\nonumber \\ & & \ \ \ \ \ \ \ \ \ \ \ \ \ \ \ 
        +Z(\sigma_a a,\sigma_b b,\sigma_d (c\plus d\minus i),-i)
		\nonumber \\ & - &
	(-1)^d 2^{c-1} \frac{(c+d-1)!}{(c-1)!\ d!} (
        Z(\sigma_a a,\sigma_b b,(b+c))
		\nonumber \\ & & \ \ \ \ \ \ \ \ \ \ \ \ \ \ \ 
        +Z(\sigma_a a,\sigma_b b,-(b+c)) )~.
\end{eqnarray}
When we do depth 5 we see that, like  $K^{(1)}_2$, also the $K^{(1)}_1$ 
splits off two new functions. Hence to produce a generic routine for 
any depth we have to look at a few very general steps.

In the general case the equations (\ref{starteq4}, \ref{diveq4}) and 
(\ref{spliteq4}) stay more or less the same. They just get more indices to 
the right. The difference comes with the equations for $K^{(2)}$. We have 
to make a distinction whether there are still many indices to the right or 
whether we are terminating. The terminating equations are also more or less 
the same as the equations for $K^{(2)}$ above, but now with more indices to 
the left. This leaves the `intermediary' objects:
\begin{eqnarray}
&&
K^{(i)}_1(M,\sigma_a a,b,\sigma_c c,N) =
	\nonumber \\ && \ \ \ \ \ \ 
	\sum_{x_1=1}^\infty \sum_{x_2=1}^\infty
	\frac{\sigma_a^{x_M\plus 2x_1\plus x_2\plus x_N}\sigma_c^{x_2\plus x_N}
	}{(x_M\plus 2x_1\plus x_2\plus x_N)^a (x_1\plus x_2\plus x_N)^b (x_2\plus x_N)^c}
		\\
&&
K^{(i)}_2(M,\sigma_a a,\sigma_b b,c,\sigma_d d,N) =
	\nonumber \\ && \ \ \ \ \ \ 
	\sum_{x_1=1}^\infty \sum_{x_2=1}^\infty
	\frac{\sigma_a^{x_M\plus 2x_1\plus x_2\plus x_N}\sigma_b^{2x_1\plus x_2\plus x_N}\sigma_d^{x_2\plus x_N}
	}{(x_M\plus 2x_1\plus x_2\plus x_N)^a 
(2x_1\plus x_2\plus x_N)^b (x_1\plus x_2\plus x_N)^c (x_2\plus x_N)^d}~.
\nonumber\\
\end{eqnarray}
In these formulas $M$ and $N$ indicate a range of indices. There are more 
sums and factors in the numerator and denominator, but we just omit them as 
they do not take part in the `action'. We use the same techniques  
applied before to move the factor 2 that multiplies $x_1$ to $x_2$, to the 
right. When $N$ is empty we run into a termination condition and  switch 
to the equations for $K^{(2)}$,
\begin{eqnarray}
&&
K^{(i)}_1((M),\sigma_a a,b,\sigma_c c,(n_1,N)) =
	\nonumber \\ && \ \ \ \ \ \ \ \ 
	(-1)^c \sum_{i=1}^b A^{(b,c)}_i 2^{b-i}
		 Z(M,\sigma_a a,\sigma_c (b+c-i),i,n_1,N)
	\nonumber \\ && \ \ \ \ \ \ +
	(-1)^c \sum_{i=1}^c B^{(b,c)}_i 2^{b-1} (-1)^i (
		 Z(M,\sigma_a a,(b+c-i),\sigma_c i,n_1,N)
		\nonumber \\ & & \ \ \ \ \ \ \ \ \ \ \ \ \ \ \ 
		 +Z(M,\sigma_a a,-(b+c-i),-\sigma_c i,n_1,N) )
	\nonumber
\end{eqnarray}\begin{eqnarray} 
&& \ \ \ \ \ \ -
	(-1)^c \sum_{i=1}^b A^{(b,c)}_i 2^{b-i}
		K^{(i)}_1((M,\sigma_a a),\sigma_c(b+c-i),i,n_1,(N))
	\nonumber \\ && \ \ \ \ \ \ -
	(-1)^c \sum_{i=1}^b A^{(b,c)}_i 2^{b-i}
		K^{(i)}_2((M),\sigma_a a,\sigma_c(b+c-i),i,n_1,(N))
		\\
&&
K^{(i)}_2((M),\sigma_a a,\sigma_b b,c,\sigma_d d,(n_1,N)) =
	\nonumber \\ && \ \ \ \ \ \ \ \ 
	(-1)^d \sum_{i=1}^c A^{(c,d)}_i 2^{c-i}
		 Z(M,\sigma_a a,\sigma_b\sigma_d (c+d-i),i,n_1,N)
	\nonumber \\ && \ \ \ \ \ \ +
	(-1)^d \sum_{i=1}^d B^{(c,d)}_i 2^{c-1} (-1)^i (
		 Z(M,\sigma_a a,\sigma_b (b+c+d-i),\sigma_d i,n_1,N)
		\nonumber \\ & & \ \ \ \ \ \ \ \ \ \ \ \ \ \ \ 
		 +Z(M,\sigma_a a,-\sigma_b(b+c+d-i),-\sigma_d i,n_1,N) )
	\nonumber \\ && \ \ \ \ \ \ -
	(-1)^d \sum_{i=1}^c A^{(c,d)}_i 2^{c-i}
		K^{(i)}_1((M,\sigma_a a),\sigma_b \sigma_d(b+c+d-i),i,n_1,(N))
	\nonumber \\ && \ \ \ \ \ \ -
	(-1)^d \sum_{i=1}^c A^{(c,d)}_i 2^{c-i}
		K^{(i)}_2((M,\sigma_a a,(b+c+d-i),\sigma_b \sigma_d i,n_1,(N))~.
\end{eqnarray}
As one can see, each step of the iteration diminishes $N$ by one unit ($n_1$ 
is an index with its sign) and $M$ may or may not get one more index.

The above formulas can be programmed rather easily and compactly in a 
language like {\tt FORM}. We have first programmed and tested the cases 2, 
3, 4, 5 and after that we have made a generic routine that can handle any 
depth. Also this routine has been tested exhaustively. It can be found in 
the library.


\subsection{The Role of the Generalized Doubling Relations}
\renewcommand{\theequation}{\thesection.\arabic{equation}}
\label{RoleofGDR}
\vspace{1mm}
\noindent
Let us start with a modification of the program for expressing Euler sums 
into a minimal set that was used for testing {\tt 
TFORM}~\cite{Vermaseren:2}. It was modified, so as to allow running only 
with sums/functions up to a given depth. We use the same relations, up to 
that depth, as in the complete program, i.e. we use the 
stuffles, the shuffles and the doubling relations, but not the GDRs. 
This should generate new information because one is often interested in 
sums of limited depth but large weight.

When we compare the number of remaining variables with the conjectures 
\cite{BK1,Broadhurst:1}, we note that in many cases we have more variables 
left. However, if we increase the depth these remaining variables are 
eliminated after all. We set up the program in such a way that these objects 
may be recognized easily. In Table~\ref{dewe} we present how many 
of these constants are left and at which depth.

\begin{table}[t]\centering
\begin{tabular}[htb]{|c|c|c|l|} \hline
 weight & depth & number & type \\ \hline
   6    &   2   &   1    & $d=2$  \\
   6    &   3   &   1    & $d=2$  \\
   6    &   4   &   0    &        \\
   6    &   5   &   0    &        \\
   6    &   6   &   0    &        \\ \hline
   7    &   3   &   1    & $d=3$  \\
   7    &   5   &   0    &        \\ \hline
   8    &   2   &   1    & $d=2$  \\
   8    &   4   &   2    & $d=2$,$d=4$  \\
   8    &   6   &   0    &        \\ \hline
   9    &   3   &   3    & $3\times(d=3)$ \\
   9    &   5   &   2    & $d=3$,$d=5$  \\
   9    &   7   &   0    &        \\ \hline
  10    &   2   &   2    & $2\times(d=2)$ \\
  10    &   3   &   2    & $2\times(d=2)$ \\
  10    &   4   &   6    & $2\times(d=2)$,$4\times(d=4)$  \\
  10    &   5   &   6    & $2\times(d=2)$,$4\times(d=4)$  \\
  10    &   6   &   3    & $d=2$,$d=4$,$d=6$  \\
  10    &   8   &   0    &      \\ \hline
\end{tabular}
\caption{Number of constants remaining when running at fixed depth for a 
given weight. With fixed depth we mean all depths up to the given value.}
\label{dewe}
\end{table}

Table~1 indicates that there must be a significant `leaking' of relations 
at greater depths that create nontrivial results at lower depth. As an 
example we derived the {\sf d = 2} relation at weight 6 without 
substituting the lower weight constants and keeping track of all products 
of lower weight objects that combined in shuffles and in stuffles. The 
relation we refer to is given as Eq.~(27) in Ref.~\cite{Broadhurst:1}~:
\begin{eqnarray}
	Z_{-4,-2}(\infty) & = & -\HH_{-4,2}(1) = \frac{97}{420}\zeta_{2}^3 - 
         \frac{3}{4}\zeta^2_{3}~.
\label{eq:depthrel}
\end{eqnarray}
The results are shown in Table \ref{shufstufdiv}.

\begin{table}\centering
\begin{tabular}{|c|c|c|}
	\hline
	depth & shuffles & stuffles \\ \hline
      2   &    11    &     8    \\
      3   &    52    &    19    \\
      4   &    72    &    41    \\ \hline
\end{tabular}
\caption{Number of shuffles and stuffles separated by depth contributing to 
equation (\ref{eq:depthrel}).}
\label{shufstufdiv}
\end{table}
We see that a total of 203 equations make contributions to the final 
result. Considering this, it should not come as a great surprise that 
attempts to derive this equation by hand using shuffle and stuffle 
relations have failed thus far.

It is of course possible to obtain this result by different means as was 
shown in ref \cite{STRUCT6} where the finite harmonic sum $S_{-4,-2}(N)$ was 
calculated in terms of the following one-dimensional integral 
representation:
\begin{eqnarray}
S_{-4,-2}(N) &=& - \Mvec\left[\left(\frac{4 \Li_5(-x) - \ln(x)
\Li_4(-x)}{x-1}\right)_+\right](N) \\ & &
+ \frac{1}{2} \zeta_2 \left[S_4(N) -S_{-4}(N)\right] - \frac{3}{2} \zeta_3
S_3(N) + \frac{21}{8} \zeta_4 S_2(N) - \frac{15}{4} \zeta_5 S_1(N)~,
\nonumber
\end{eqnarray}
where 
\begin{eqnarray}
\Mvec[f(x)](N) = \int_0^1 dx~x^N~f(x)~.
\end{eqnarray}
Since
\begin{eqnarray}
\nonumber\\ 
\int_0^1 dx \frac{4\left[\Li_5(-x) + (15/16) \zeta_5\right] 
- \ln(x) \Li_4(-x)}{x-1} = - \frac{811}{840} \zeta_2^3 + \frac{3}{4} \zeta_3^2
\end{eqnarray}
one obtains with
\begin{eqnarray}
Z_{-4,-2} &=& \lim_{N \rightarrow \infty} S_{-4,-2}(N) -  \zeta_6
\end{eqnarray}
the above result. It should, however, be clear that if such methods are 
needed to replace the phenomenon of leakage, it will be a near 
impossibility to go to much greater values of the weight parameter.

Using the GDRs at depth ${\sf d=2}$ resolves the problem completely. Only 
the depth ${\sf d=2}$ shuffles and stuffles in combination with these GDRs 
give already the desired formula.

To study the problem at depth ${\sf d=3}$, we recreated an old program by 
one of us\footnote{The program had an error and hence gave rise to a wrong 
conjecture.} that only determines relations at leading depth for objects of 
which the index field is a Lyndon word. The {\tt FORM} version of the 
program is rather fast when applied at depth {\sf d = 3}, see 
Table~\ref{missing3}.

\begin{table}\centering
\begin{tabular}[htb]{|c|c|c|} \hline
  weight &  constants &  expected \\ \hline
    5    &     1      &     1     \\
    7    &     3      &     2     \\
    9    &     6      &     3     \\
   11    &    11      &     5     \\
   13    &    17      &     7     \\
   15    &    23      &     9     \\
   17    &    32      &    12     \\
   19    &    41      &    15     \\
   21    &    51      &    18     \\
   23    &    63      &    22     \\
   25    &    76      &    26     \\
   27    &    89      &    30     \\
   29    &   105      &    35     \\
   31    &   121      &    40     \\
   33    &   138      &    45     \\
   35    &   157      &    51     \\
   37    &   177      &    57     \\
   39    &   197      &    63     \\
   41    &   220      &    70     \\
   43    &   243      &    77     \\
   45    &   267      &    84     \\
   47    &   293      &    92     \\
   49    &   320      &   100     \\
   51    &   347      &   108     \\  \hline
\end{tabular}
\caption{Remaining constants at depth {\sf d = 3} compared to the number of
expected constants.}
\label{missing3}
\end{table}
We see a steady increase in the number of undetermined constants. 
In Tables~\ref{missing3},~\ref{missing4} we list  
under `expected' the number of undetermined constants  
according to conjecture \cite{Broadhurst:1}. The results for the weights 7 and 9 are 
in agreement with the numbers in Table~\ref{dewe}.

To see whether we could improve the situation, we tried programming 
generalizations of the formulas $D_0$ and $D_1$ of Ref.~\cite{Girgensohn}. 
They made no difference. Close inspection reveals that the formula $D_0$ is 
another form of the shuffle formulas with the combinatorics included 
properly. The formula $D_1$, or Markett formula~\cite{Markett:1}, also does 
not add anything new. It seems to be a combination of shuffles and 
stuffles. Next we applied the GDRs at depth ${\sf d=3}$ and these reduce 
the number of undetermined constants to their expected value. This means 
that if we include the GDRs we can run the program at maximum depth ${\sf 
d=3}$ and get a complete set of expressions for all depth $d = 1, 2$ and 
$3$ objects. At the moment we have verified this for all weights up to ${\sf w 
= 51}$. The run for the highest weight took about 20 hours of CPU time on a 
single Xeon processor at 2.33 GHz.

We have made a similar program for depth ${\sf d=4}$. This is of course much 
slower 
and hence we cannot go to such large values for the weight. The results 
are given in Table~\ref{missing4}.
\begin{table}\centering
\begin{tabular}{|c|c|c|} \hline
  weight &  constants &  expected \\ \hline
    6    &     1      &     1     \\
    8    &     3      &     2     \\
   10    &     9      &     5     \\
   12    &    21      &     8     \\
   14    &    39      &    14     \\
   16    &    66      &    20     \\
   18    &   102      &    30     \\
   20    &   149      &    40     \\
   22    &   209      &    55     \\  \hline
\end{tabular}
\caption{Remaining constants at depth {\sf d = 4} compared to the number of
expected constants.}
\label{missing4}
\end{table}
Again we see an increase in the number of extra undetermined objects and 
again application of the GDRs resolved the issue.

The phenomenon of leakage is rather messy. Basically equations that are in 
nature of a greater depth have to combine first to eliminate most objects 
of this depth. After this a few equations remain between lower depth 
objects. Such leakage is impossible without the stuffle relations. The 
shuffle relations by themselves do not give terms with a lower depth and 
neither do the relations based on the doubling formula. But whether these 
extra relations come from the stuffles alone or materialize only after 
combining stuffles and shuffles, and maybe doublings, is currently not 
clear. What is clear is that they involve a very large number of equations. 
In all cases which we studied the leakage goes over at least two units of 
depth. This makes it very difficult to investigate. Fortunately the GDRs 
seem to resolve these problems. We formulate

\vspace{1mm}\noindent
{\bf Conjecture~1}: The stuffle, shuffle, doubling and Generalized Doubling 
Relations are sufficient to reduce the Euler sums of a given weight and 
depth to a minimal set that is in agreement with the conjecture 
\cite{Broadhurst:1}, both in weight and in depth. $\Box$

Even if we could dispense with
the GDRs up to weight ${\sf w=10}$, the 
whole situation changes at weight ${\sf w=11}$, see Table~\ref{nodoubling}. 
Running only stuffles, shuffles and doubling relations leaves one variable 
in excess of the conjecture \cite{Broadhurst:1}. The GDRs provide the 
missing equation by which this variable is expressed in terms of the other 
remaining variables and agreement with conjecture \cite{Broadhurst:1} is 
reached. The same effect occurs at weight ${\sf w=12}$.  Again there is one 
variable too many if the GDRs are not used. We cannot check this beyond 
weight ${\sf w=12}$, because leakage forces us to run all depths for a given 
weight if we exclude the GDRs. This becomes excessive in terms of current 
computer resources.
Alternatively one could have used the relations of equation 
(\ref{eq:pseudodual}) to resolve this issue, but these relations do not help 
with the problem of running at a limited depth. Hence we have to add the 
GDRs anyway.
 
\vspace*{2mm}\noindent
\begin{table}\centering
\begin{tabular}{|c|c|c|}
	\hline
	weight & no doubling & no GDRs  \\ \hline
      8    &      1      &     0    \\
      10   &      1      &     0    \\
      11   &      2      &     1    \\
      12   &      3      &     1    \\ \hline
\end{tabular}
\caption{Number of excess elements when no doubling relations (also no 
GDRs) are used, and when only no GDRs are used.}
\label{nodoubling}
\end{table}


\section{The Computer Program}
\renewcommand{\theequation}{\thesection.\arabic{equation}}
\setcounter{equation}{0}
\label{ComputerProgram}

\vspace{1mm}
\noindent
We have combined the above relations into a new computer program to resolve 
all relations between MZVs and reduce them to a minimal set. In principle 
this is done by writing down all equations for the MZVs of a given weight 
and then solving the system. A few variables at the given weight may remain 
and there will be products of objects of lower weight.

Considering the size of the problem and its sparsity it did not look to us 
like a typical problem to solve by matrix techniques even though other 
people have done so \cite{ENR,KNT}. Typically there would be many 
thousands of zeroes for each non-zero element. The advantage of computer 
algebra is that in a sparse polynomial representation those zeroes will not 
be present and need no attention. Hence we have selected a rather special 
method the essence of which has already been used in 
references~\cite{Vermaseren:1,harmpol,Vermaseren:2}, although not described 
there in detail. We select the {\tt FORM} system, because it is by far the best 
suited for this kind of problems. Since we go to much greater weights than 
previously investigated, we take the opportunity to give here a better 
description of the completely renewed version of the program.

We start generating a master expression which contains one term for each 
sum that we want to compute. For the MZVs of weight 
${\sf w=4}$ this expression looks in computer terms like
\begin{verbatim}
   FF =
      +E(0,0,0,1)*(H(0,0,0,1))
      +E(0,0,1,1)*(H(0,0,1,1))
      +E(0,1,0,1)*(H(0,1,0,1));
\end{verbatim}
We have used already that we will only compute the finite elements and that 
there is a duality that allows us to eliminate all elements with a depth 
greater than half the weight. When the depth is exactly half the weight we 
choose from a sum and its dual the element that comes first 
lexicographically. We work in terms of the $\HH$-functions because for the 
Euler sums the basis of reference ~\cite{Broadhurst:1} turns out to be 
ideal. This basis consists of all Lyndon words of negative odd integers 
that add up in absolute value to the weight. For the MZVs these 
$\HH$-functions and the $Z$-functions are identical anyway and hence we 
could keep a single program for most procedures.

We pull the function E outside brackets. The contents of a bracket is what 
we know about the object indicated by the indices of the function E. In the 
beginning this is all trivial knowledge.

Assume now that we generate the stuffle relation
\begin{eqnarray}
	H_{0,1} H_{0,1} & = & H_{0,0,0,1} + 2 H_{0,1,0,1}
\end{eqnarray}
The left hand side can be substituted from the tables for the lower weight 
MZVs. Hence it becomes $\zeta_2^2$. In the program $\zeta_2$ is called 
\verb:z2:. The right hand side objects are replaced by the contents of the 
corresponding E brackets in the master expression. These are for now 
trivial substitutions. From the result we generate the substitution
\begin{verbatim}
   id H(0,1,0,1) = z2^2/2-H(0,0,0,1)/2;
\end{verbatim}
which we apply to the master expression. Hence the master expression becomes
\begin{verbatim}
   FF =
      +E(0,0,0,1)*(H(0,0,0,1))
      +E(0,0,1,1)*(H(0,0,1,1))
      +E(0,1,0,1)*(z2^2/2-H(0,0,0,1)/2);
\end{verbatim}

Let us now generate the corresponding shuffle relation:
\begin{eqnarray}
	\HH_{0,1} \HH_{0,1} & = & 4 \HH_{0,0,1,1} + 2 \HH_{0,1,0,1}
\end{eqnarray}
and replace the right hand side objects by the contents of the 
corresponding E brackets in the master expression. This gives
\begin{eqnarray}
	\zeta_2^2 & = & 4 \HH_{0,0,1,1} + \zeta_2^2 - \HH_{0,0,0,1}
\end{eqnarray}
which leads to the substitution
\begin{verbatim}
   id H(0,0,1,1) = H(0,0,0,1)/4;
\end{verbatim}
and we obtain
\begin{verbatim}
   FF =
      +E(0,0,0,1)*(H(0,0,0,1))
      +E(0,0,1,1)*(H(0,0,0,1)/4)
      +E(0,1,0,1)*(z2^2/2-H(0,0,0,1)/2);
\end{verbatim}

We also need the divergent shuffles and stuffles. This is done by including 
the shuffles involving the basic divergent object and breaking down the 
multiple divergent sums with the stuffle relations as in:
\begin{eqnarray}
	H_1 H_{0,0,1} & = & 2 H_{0,0,1,1} + H_{0,1,0,1} + H_{1,0,0,1} \nonumber
		\\ & = & - H_{0,0,0,1} + H_{0,0,1,1} + H_{0,1,0,1} + H_1 H_{0,0,1};
\end{eqnarray}
In the case we use $H_1$ as the only divergent object, this is equivalent 
to using Hoffmann's~\cite{HOF4} relation. We can use any combination 
involving divergent objects, provided not both are divergent 
simultaneously. Substituting from the master expression we get the relation
\begin{eqnarray}
	0 & = & -\frac{5}{4} H_{0,0,0,1} + \frac{1}{2} \zeta_2^2
\end{eqnarray}
and hence the substitution
\begin{verbatim}
   id H(0,0,0,1) = z2^2*2/5;
\end{verbatim}
and finally the master expression becomes
\begin{verbatim}
   FF =
      +E(0,0,0,1)*(z2^2*2/5)
      +E(0,0,1,1)*(z2^2/10)
      +E(0,1,0,1)*(z2^2*3/10);
\end{verbatim}
Now we can read off the values of all MZVs of weight 4 that we set out to 
compute. All other elements can be obtained from these by trivial 
operations that involve the use of one or two relations only.

The method should be clear now: we generate the master expression 
that contains all nontrivial objects that we need to compute. Then we 
generate all known equations one by one, putting in the knowledge that is 
contained in the master expression. After that we incorporate the new 
knowledge in the master expression (provided the equation does not become 
trivial which will happen frequently, because we have more equations than 
variables).

With this method we do not need all equations to be in memory 
simultaneously. But there is a very important observation: the order in 
which the equations are generated will determine the size the master 
expression can have during the calculation. This intermediate expression 
swell should be controlled as much as possible, because it can make many 
orders of magnitude difference in the execution time and the space needed. 
And there is another problem: substituting a new equation in the master 
expression can be rather costly when this expression becomes rather big. To 
have to do this each time is wasteful because the master expression will 
have to be brought to normal order again. Therefore we have adopted a 
scheme in which we generate the equations in groups. Then we apply first a 
Gaussian elimination scheme among the equations in the group, eliminating 
both above and below the diagonal. If we have $G$ equations left we can 
substitute $G$ variables in the master expression simultaneously. Again, this 
is not optimal yet as that would give $G$ substitution statements and hence 
each term needs $G$ pattern matchings. To improve upon this we enter these $G$ 
objects in a temporary table and the substitution in the master expression 
is by a single table lookup. This is a binary search inside {\tt FORM} and hence 
when we have grouped for instance 512 equations, the lookup takes only 9 
compares, each of which is anyway much faster than a full pattern 
matching. The difference shows in a run we made on a machine with a single 
Opteron processor. When running the equations for MZVs one 
by one at weight 18, the run took 26761 sec, while with groups of 256 
equations the same program ran in 2974 sec. Over the range in weights that 
we experimented with, the optimal group size we found for the MZVs 
was close to $2^{({\sf w}-1)/2}$. This is the value we use in the program. For 
the Euler sums the best value obeys a more involved relation because the 
number of variables goes with a power of three. We have measured the effect and it 
is shown in Table~\ref{optimumgroup}.
From this Table it looks like a decent value for the size of the groups is 
$2^{3{\sf w}/2-7}$ in which the exponent is rounded down to the nearest 
integer. 
We see, however, that the exact value is not very critical.

\begin{table}\centering
\begin{tabular}{|c|c|c|c|c|c|c|c|}
\hline
{\sf w/g} &   64  &  128  &   256  &   512  &  1024  &  2048  &  4096 \\ 
\hline
  9 &    62 &    56 &     61 &        &        &        &       \\
 10 &       &   477 &    406 &    442 &        &        &       \\
 11 &       &  5826 &   4651 &   3799 &   3623 &   5157 &       \\
 12 &       &       &        &        &  65591 &  50926 & 62867 \\
\hline
\end{tabular}
\caption{Execution times in seconds for Euler sums at any depth as a 
function of weight and the size of the groups in the Gaussian elimination 
scheme. All runs were with {\tt TFORM} on an 8 Xeon-cores machine at 3 GHz.}
\label{optimumgroup}
\end{table}

If it would be of great importance to improve over this scheme, one could 
set up a tree structure in the Gaussian scheme. This would change its 
quadratic (in the size of the groups) nature to a $G\log(G)$ behaviour. It 
would, however, make the code much more complicated and anyway, this is not 
where currently most computer time is used. As a consequence we decided to 
stay with the simple grouping.

This leaves determining a good order in which to generate the equations. 
It requires much trial and error and we are not claiming that we have the 
best scheme possible. The scheme for the stuffles is rather good, but for 
the shuffles it could probably be better. 
Once we could run what we wanted to run, we have 
stopped searching intensively. Anyway, the intermediate expression swell is 
rather moderate as is shown in the Tables containing the results below.

Before we discuss the order of the equations, we make several
observations:
\begin{itemize}
\item Shuffles preserve depth.
\item Stuffles either preserve depth or lower it.
\item The number of indices that are one in sum notation is either 
preserved or lowered by stuffles.
\item The shuffle relations can contain many more terms than the stuffle 
relations.
\item The shuffles (which are executed in integral notation) can contain 
large combinatoric factors when there are long sequences of zeroes or 
ones. This lowers the number of terms in the equation.
\end{itemize}

Based on the above observations we start with the equations with the lowest 
depth, and then do the ones with the next depth, etc. In the case that we 
only look at the MZVs, we only need to go up to half the 
weight (rounded down), because the duality relation takes care of the other 
sums. In the case of the Euler sums we have to go `all the way'.

For each depth we do first the stuffles and then the shuffles. There are 
actually conjectures about that one does not need all stuffles but only a 
limited subset. We do not use these conjectures because they would make it 
necessary to apply more shuffle relations and those are more complicated 
than the stuffles that we would omit. We have verified experimentally that 
this would make the program significantly slower.

In the case of Euler sums we have two more categories of equations: 
the equations due to the doubling relation and the equations due to the 
GDRs. It looks like we do not need all equations 
from the latter category, but because they are not extremely costly, we 
have not been motivated enough to run many programs testing what can be 
done here. We just run them all and this way there is no risk that we omit 
something essential. They are, however, more costly than the shuffle 
equations and hence we put them after the shuffles. But more ordering 
within the group of (generalized) doubling equations is not relevant as 
there are only comparatively few substitutions generated by them.

To deal with the stuffles at a given weight and depth we generate an 
expression that contains one term for each stuffle relation that we will 
use. Then we apply several operations that multiply each term with a 
function with arguments based on the equation to be generated. 
The effect of this is that at the next sorting the equations will be 
ordered according to these arguments. This can be done in a rather flexible 
way. The ordering is in sum notation according to:
\begin{itemize}
\item The number of indices that are one.
\item Next comes the number of indices that are two, then three etc.
\item The number of indices in the sum with the smallest depth.
\item The largest first index in either of the two sums.
\end{itemize}
This relatively simple ordering is amazingly effective. When we compute the 
size of the basis, using arithmetic over a 31-bit prime number, it gives a 
nearly monotonically increasing size in the master expression, indicating 
that it will be very hard to improve upon it. Once we have this expression 
we use a feature of {\tt FORM} that allows one to define a loop in which the loop 
variable takes a value which is (sequentially) each time a term from a 
given expression. This way we can now create expressions for each equation 
and each time we have enough equations to fill a group we call the routine 
that will expand the equations and process them. We do not consider stuffle 
equations that contain a divergent sum. Those are taken into consideration 
anyway when we have to extract the divergences in the shuffle equations, 
and for the Euler sums the GDRs.

For the shuffles things are more complicated. Again we generate an 
expression for all shuffles for the given depth. In this case we generate 
however only those objects that correspond to shuffles in which one of the 
objects is only of depth one. This seems to be sufficient. We have never 
run across a case where the other shuffles had any additional effect. It is 
actually possible to restrict the number of shuffle equations even more, 
although this is only based on conjectures and experimentation. A formal 
proof is missing. The ordering is now done according to
\begin{itemize}
\item The weight of the object of depth one.
\item The number of indices that are one in sum notation.
\item For each sum we compute the sum of the squares of the indices in sum 
notation. We order by the maximum of either of the two. The biggest comes 
first.
\item We select which of the two sums has the smallest first index. The 
larger values for this number come first.
\item We add the first indices of the two sums. The larger values come 
first.
\end{itemize}
The complicating factor here is that we have to keep divergent sums. We 
only keep those equations in which at most one object is divergent, and 
there is only a single divergence. Hence sums that have the first two 
indices equal to one are not considered.

According to observation the shuffle equations that fulfill all following 
requirements always reduce to trivial ($0=0$) equations:
\begin{itemize}
\item The combined depth is at least three.
\item There is at least one index that is equal to one.
\item The depth one object has at least weight two.
\item If the depth one object has weight ${\sf w=2}$, there are at least two 
indices 
equal to one in the other object.
\end{itemize}
Harmonic sums with all the same index decompose algebraically into a polynomial
of single harmonic sums. It is easily shown that the algebraic relations 
\cite{ALGEBRA} always allow to write any harmonic sum in terms of 
polynomials of $S_1(N)$ and sums, which converge in the limit $N \rightarrow 
\infty$. All the 
above greatly reduces the number of shuffle equations that have to be 
evaluated. Because this evaluation is one of the expensive steps, it speeds 
up the program significantly. On the other hand, it is only an observation 
made in runs that do not involve the greatest weights. For the more critical 
runs\footnote{With this we mean the programs that determine the size of the 
basis when using arithmetic over a prime number. Once we have established 
this, any further runs to for instance determine all values over the 
rational numbers, we can safely drop these equations.} we have left these 
equations active and spent the extra computer time.

The above describes the basic program. At this point we split it in several 
varieties. To first determine whether shuffles and stuffles are sufficient 
to reduce all MZVs to a basis of the conjectured size, we have made the 
simplifications:
\begin{itemize}
\item All products of lower weight objects are set to zero. This means we 
will only determine whether reduction to a Lyndon basis takes place.
\item We work modulus a 31-bit prime.
\end{itemize}
We have also made runs over the rational numbers. This becomes only 
problematic for the very highest values of the weight.

\begin{figure}[t]
\begin{center}
\begin{picture}(410,210)(0,0)
\SetPFont{Helvetica}{24}
\SetScale{0.5}
\BBox(0.0,0.0)(400,400)
\PText(20,360)(0)[l]{MZVs}
\PText(20,335)(0)[l]{Rational arithmetic}
\PText(20,310)(0)[l]{Complete results}
\PText(20,285)(0)[l]{W = 21}
\Vertex(0.582242,0.000309){1}
\Vertex(1.164483,4.318050){1}
\Vertex(1.746725,2.311175){1}
\Vertex(2.328967,3.022624){1}
\Vertex(2.911208,3.022748){1}
\Vertex(3.493450,3.023530){1}
\Vertex(4.075691,3.031869){1}
\Vertex(4.657933,3.051924){1}
\Vertex(5.240175,3.224057){1}
\Vertex(5.822416,3.272835){1}
\Vertex(6.404658,3.503300){1}
\Vertex(6.986900,3.752687){1}
\Vertex(7.569141,3.996741){1}
\Vertex(8.151383,4.277343){1}
\Vertex(8.733624,4.622598){1}
\Vertex(9.315866,4.962375){1}
\Vertex(9.898108,5.353072){1}
\Vertex(10.480349,15.570778){1}
\Vertex(11.062591,6.918577){1}
\Vertex(11.644833,7.285678){1}
\Vertex(12.227074,7.682264){1}
\Vertex(12.809316,8.010305){1}
\Vertex(13.391557,8.405037){1}
\Vertex(13.973799,8.763840){1}
\Vertex(14.556041,9.174757){1}
\Vertex(15.138282,9.696489){1}
\Vertex(15.720524,10.045285){1}
\Vertex(16.302766,10.546056){1}
\Vertex(16.885007,10.925676){1}
\Vertex(17.467249,11.448396){1}
\Vertex(18.049491,11.916676){1}
\Vertex(18.631732,12.344127){1}
\Vertex(19.213974,12.730829){1}
\Vertex(19.796215,13.215705){1}
\Vertex(20.378457,13.739784){1}
\Vertex(20.960699,14.328805){1}
\Vertex(21.542940,14.900468){1}
\Vertex(22.125182,15.406511){1}
\Vertex(22.707424,15.944221){1}
\Vertex(23.289665,16.597668){1}
\Vertex(23.871907,28.489554){1}
\Vertex(24.454148,33.160373){1}
\Vertex(25.036390,33.973146){1}
\Vertex(25.618632,28.385203){1}
\Vertex(26.200873,20.697957){1}
\Vertex(26.783115,19.774985){1}
\Vertex(27.365357,20.133315){1}
\Vertex(27.947598,20.698677){1}
\Vertex(28.529840,20.840275){1}
\Vertex(29.112082,21.257060){1}
\Vertex(29.694323,21.490057){1}
\Vertex(30.276565,21.996759){1}
\Vertex(30.858806,22.688997){1}
\Vertex(31.441048,22.764707){1}
\Vertex(32.023290,23.399293){1}
\Vertex(32.605531,24.083687){1}
\Vertex(33.187773,24.103927){1}
\Vertex(33.770015,24.660868){1}
\Vertex(34.352256,25.139732){1}
\Vertex(34.934498,25.772301){1}
\Vertex(35.516739,25.790749){1}
\Vertex(36.098981,26.158632){1}
\Vertex(36.681223,26.558842){1}
\Vertex(37.263464,27.061075){1}
\Vertex(37.845706,27.601503){1}
\Vertex(38.427948,28.046125){1}
\Vertex(39.010189,28.361380){1}
\Vertex(39.592431,28.724672){1}
\Vertex(40.174672,29.090310){1}
\Vertex(40.756914,29.463547){1}
\Vertex(41.339156,29.998601){1}
\Vertex(41.921397,30.499311){1}
\Vertex(42.503639,31.233038){1}
\Vertex(43.085881,31.647599){1}
\Vertex(43.668122,32.405829){1}
\Vertex(44.250364,32.536349){1}
\Vertex(44.832606,33.362177){1}
\Vertex(45.414847,33.683320){1}
\Vertex(45.997089,34.288010){1}
\Vertex(46.579330,34.607053){1}
\Vertex(47.161572,35.329827){1}
\Vertex(47.743814,35.637113){1}
\Vertex(48.326055,36.077123){1}
\Vertex(48.908297,36.615616){1}
\Vertex(49.490539,37.376275){1}
\Vertex(50.072780,37.787480){1}
\Vertex(50.655022,38.308615){1}
\Vertex(51.237263,38.725564){1}
\Vertex(51.819505,39.147373){1}
\Vertex(52.401747,39.634967){1}
\Vertex(52.983988,40.109424){1}
\Vertex(53.566230,40.763077){1}
\Vertex(54.148472,41.426469){1}
\Vertex(54.730713,41.991955){1}
\Vertex(55.312955,42.636692){1}
\Vertex(55.895197,43.349336){1}
\Vertex(56.477438,43.947931){1}
\Vertex(57.059680,44.506560){1}
\Vertex(57.641921,45.152327){1}
\Vertex(58.224163,45.889287){1}
\Vertex(58.806405,46.198962){1}
\Vertex(59.388646,64.078013){1}
\Vertex(59.970888,68.745744){1}
\Vertex(60.553130,88.701613){1}
\Vertex(61.135371,99.828042){1}
\Vertex(61.717613,91.034037){1}
\Vertex(62.299854,97.694538){1}
\Vertex(62.882096,110.617864){1}
\Vertex(63.464338,91.260878){1}
\Vertex(64.046579,69.321010){1}
\Vertex(64.628821,69.578777){1}
\Vertex(65.211063,66.293547){1}
\Vertex(65.793304,54.878548){1}
\Vertex(66.375546,55.269306){1}
\Vertex(66.957787,55.647237){1}
\Vertex(67.540029,56.053459){1}
\Vertex(68.122271,56.669144){1}
\Vertex(68.704512,56.713124){1}
\Vertex(69.286754,57.843705){1}
\Vertex(69.868996,58.672600){1}
\Vertex(70.451237,58.747116){1}
\Vertex(71.033479,58.615236){1}
\Vertex(71.615721,59.396280){1}
\Vertex(72.197962,59.778638){1}
\Vertex(72.780204,59.779318){1}
\Vertex(73.362445,60.218339){1}
\Vertex(73.944687,61.329669){1}
\Vertex(74.526929,61.791978){1}
\Vertex(75.109170,61.865814){1}
\Vertex(75.691412,61.820927){1}
\Vertex(76.273654,62.634400){1}
\Vertex(76.855895,62.697097){1}
\Vertex(77.438137,63.103587){1}
\Vertex(78.020378,63.512609){1}
\Vertex(78.602620,63.787054){1}
\Vertex(79.184862,65.057730){1}
\Vertex(79.767103,65.458640){1}
\Vertex(80.349345,65.343664){1}
\Vertex(80.931587,66.057008){1}
\Vertex(81.513828,66.619982){1}
\Vertex(82.096070,67.701044){1}
\Vertex(82.678311,67.854646){1}
\Vertex(83.260553,68.197924){1}
\Vertex(83.842795,69.443665){1}
\Vertex(84.425036,70.422518){1}
\Vertex(85.007278,71.957735){1}
\Vertex(85.589520,72.312091){1}
\Vertex(86.171761,72.433696){1}
\Vertex(86.754003,72.515356){1}
\Vertex(87.336245,72.505287){1}
\Vertex(87.918486,72.161803){1}
\Vertex(88.500728,72.324280){1}
\Vertex(89.082969,73.421031){1}
\Vertex(89.665211,73.675175){1}
\Vertex(90.247453,73.686046){1}
\Vertex(90.829694,74.891637){1}
\Vertex(91.411936,75.133550){1}
\Vertex(91.994178,75.095108){1}
\Vertex(92.576419,76.340911){1}
\Vertex(93.158661,76.857311){1}
\Vertex(93.740902,76.735047){1}
\Vertex(94.323144,77.902093){1}
\Vertex(94.905386,79.162597){1}
\Vertex(95.487627,79.628365){1}
\Vertex(96.069869,79.811329){1}
\Vertex(96.652111,79.827080){1}
\Vertex(97.234352,79.684041){1}
\Vertex(97.816594,80.567376){1}
\Vertex(98.398836,81.082664){1}
\Vertex(98.981077,81.151270){1}
\Vertex(99.563319,81.798046){1}
\Vertex(100.145560,82.044880){1}
\Vertex(100.727802,82.597373){1}
\Vertex(101.310044,83.230580){1}
\Vertex(101.892285,83.383647){1}
\Vertex(102.474527,83.951933){1}
\Vertex(103.056769,84.644790){1}
\Vertex(103.639010,85.044793){1}
\Vertex(104.221252,86.277295){1}
\Vertex(104.803493,87.226436){1}
\Vertex(105.385735,87.491348){1}
\Vertex(105.967977,87.394101){1}
\Vertex(106.550218,88.484614){1}
\Vertex(107.132460,88.698874){1}
\Vertex(107.714702,89.784466){1}
\Vertex(108.296943,90.805467){1}
\Vertex(108.879185,91.135670){1}
\Vertex(109.461426,91.052754){1}
\Vertex(110.043668,91.734697){1}
\Vertex(110.625910,92.338934){1}
\Vertex(111.208151,93.220375){1}
\Vertex(111.790393,93.608168){1}
\Vertex(112.372635,94.325918){1}
\Vertex(112.954876,94.714824){1}
\Vertex(113.537118,96.009487){1}
\Vertex(114.119360,96.555289){1}
\Vertex(114.701601,96.757628){1}
\Vertex(115.283843,96.934476){1}
\Vertex(115.866084,98.087830){1}
\Vertex(116.448326,98.905771){1}
\Vertex(117.030568,99.116099){1}
\Vertex(117.612809,99.148281){1}
\Vertex(118.195051,100.092336){1}
\Vertex(118.777293,100.246988){1}
\Vertex(119.359534,101.051525){1}
\Vertex(119.941776,101.399045){1}
\Vertex(120.524017,102.171955){1}
\Vertex(121.106259,102.826370){1}
\Vertex(121.688501,103.135098){1}
\Vertex(122.270742,103.855792){1}
\Vertex(122.852984,104.597550){1}
\Vertex(123.435226,105.224950){1}
\Vertex(124.017467,106.172300){1}
\Vertex(124.599709,106.641775){1}
\Vertex(125.181951,107.550250){1}
\Vertex(125.764192,108.157225){1}
\Vertex(126.346434,108.835957){1}
\Vertex(126.928675,109.599705){1}
\Vertex(127.510917,110.218067){1}
\Vertex(128.093159,110.989968){1}
\Vertex(128.675400,111.735803){1}
\Vertex(129.257642,112.442393){1}
\Vertex(129.839884,120.500693){1}
\Vertex(130.422125,130.793450){1}
\Vertex(131.004367,135.157086){1}
\Vertex(131.586608,144.961653){1}
\Vertex(132.168850,146.548304){1}
\Vertex(132.751092,154.476290){1}
\Vertex(133.333333,151.606464){1}
\Vertex(133.915575,157.629785){1}
\Vertex(134.497817,169.626145){1}
\Vertex(135.080058,178.489106){1}
\Vertex(135.662300,178.146384){1}
\Vertex(136.244541,184.100357){1}
\Vertex(136.826783,175.174349){1}
\Vertex(137.409025,181.077629){1}
\Vertex(137.991266,178.818403){1}
\Vertex(138.573508,175.246147){1}
\Vertex(139.155750,177.448565){1}
\Vertex(139.737991,160.623110){1}
\Vertex(140.320233,138.992402){1}
\Vertex(140.902475,133.075861){1}
\Vertex(141.484716,132.408372){1}
\Vertex(142.066958,121.854450){1}
\Vertex(142.649199,122.455475){1}
\Vertex(143.231441,122.768485){1}
\Vertex(143.813683,123.354129){1}
\Vertex(144.395924,124.325363){1}
\Vertex(144.978166,124.449727){1}
\Vertex(145.560408,124.455246){1}
\Vertex(146.142649,124.376941){1}
\Vertex(146.724891,124.756314){1}
\Vertex(147.307132,125.166633){1}
\Vertex(147.889374,125.704755){1}
\Vertex(148.471616,126.038664){1}
\Vertex(149.053857,126.885740){1}
\Vertex(149.636099,128.235050){1}
\Vertex(150.218341,129.166607){1}
\Vertex(150.800582,130.780808){1}
\Vertex(151.382824,131.149185){1}
\Vertex(151.965066,131.328113){1}
\Vertex(152.547307,131.830161){1}
\Vertex(153.129549,131.803126){1}
\Vertex(153.711790,130.890677){1}
\Vertex(154.294032,130.883861){1}
\Vertex(154.876274,131.060009){1}
\Vertex(155.458515,132.398921){1}
\Vertex(156.040757,133.428260){1}
\Vertex(156.622999,135.095687){1}
\Vertex(157.205240,135.407277){1}
\Vertex(157.787482,136.307022){1}
\Vertex(158.369723,135.401450){1}
\Vertex(158.951965,136.111787){1}
\Vertex(159.534207,137.273809){1}
\Vertex(160.116448,138.752795){1}
\Vertex(160.698690,139.071879){1}
\Vertex(161.280932,139.503777){1}
\Vertex(161.863173,139.478019){1}
\Vertex(162.445415,139.344430){1}
\Vertex(163.027656,139.015112){1}
\Vertex(163.609898,140.163627){1}
\Vertex(164.192140,140.462184){1}
\Vertex(164.774381,140.432349){1}
\Vertex(165.356623,141.014513){1}
\Vertex(165.938865,141.742619){1}
\Vertex(166.521106,141.884794){1}
\Vertex(167.103348,141.857780){1}
\Vertex(167.685590,142.563361){1}
\Vertex(168.267831,143.784579){1}
\Vertex(168.850073,144.289139){1}
\Vertex(169.432314,144.191192){1}
\Vertex(170.014556,144.438706){1}
\Vertex(170.596798,145.765263){1}
\Vertex(171.179039,147.161435){1}
\Vertex(171.761281,148.484987){1}
\Vertex(172.343523,152.106659){1}
\Vertex(172.925764,153.301749){1}
\Vertex(173.508006,154.132518){1}
\Vertex(174.090247,156.515552){1}
\Vertex(174.672489,156.275637){1}
\Vertex(175.254731,152.934319){1}
\Vertex(175.836972,152.924415){1}
\Vertex(176.419214,154.123314){1}
\Vertex(177.001456,155.265405){1}
\Vertex(177.583697,156.755200){1}
\Vertex(178.165939,157.111965){1}
\Vertex(178.748180,157.739489){1}
\Vertex(179.330422,157.722172){1}
\Vertex(179.912664,157.036770){1}
\Vertex(180.494905,157.026372){1}
\Vertex(181.077147,157.870153){1}
\Vertex(181.659389,158.733702){1}
\Vertex(182.241630,158.629804){1}
\Vertex(182.823872,159.947158){1}
\Vertex(183.406114,161.364291){1}
\Vertex(183.988355,162.430281){1}
\Vertex(184.570597,163.609495){1}
\Vertex(185.152838,166.308506){1}
\Vertex(185.735080,166.903127){1}
\Vertex(186.317322,167.314949){1}
\Vertex(186.899563,167.585008){1}
\Vertex(187.481805,169.289517){1}
\Vertex(188.064047,168.959500){1}
\Vertex(188.646288,167.109048){1}
\Vertex(189.228530,166.919908){1}
\Vertex(189.810771,167.676078){1}
\Vertex(190.393013,168.886920){1}
\Vertex(190.975255,170.546439){1}
\Vertex(191.557496,171.046778){1}
\Vertex(192.139738,171.734405){1}
\Vertex(192.721980,171.685421){1}
\Vertex(193.304221,171.100951){1}
\Vertex(193.886463,171.031212){1}
\Vertex(194.468705,171.495416){1}
\Vertex(195.050946,172.186110){1}
\Vertex(195.633188,172.571865){1}
\Vertex(196.215429,173.812747){1}
\Vertex(196.797671,174.274830){1}
\Vertex(197.379913,174.170685){1}
\Vertex(197.962154,174.430100){1}
\Vertex(198.544396,175.727646){1}
\Vertex(199.126638,176.267580){1}
\Vertex(199.708879,176.216002){1}
\Vertex(200.291121,176.308904){1}
\Vertex(200.873362,176.936407){1}
\Vertex(201.455604,177.483115){1}
\Vertex(202.037846,178.196438){1}
\Vertex(202.620087,179.699597){1}
\Vertex(203.202329,179.915772){1}
\Vertex(203.784571,180.593207){1}
\Vertex(204.366812,181.888508){1}
\Vertex(204.949054,183.591432){1}
\Vertex(205.531295,184.516524){1}
\Vertex(206.113537,185.730845){1}
\Vertex(206.695779,185.620090){1}
\Vertex(207.278020,184.974138){1}
\Vertex(207.860262,184.825087){1}
\Vertex(208.442504,185.223237){1}
\Vertex(209.024745,186.222495){1}
\Vertex(209.606987,186.442747){1}
\Vertex(210.189229,187.235362){1}
\Vertex(210.771470,188.510032){1}
\Vertex(211.353712,189.119067){1}
\Vertex(211.935953,188.904559){1}
\Vertex(212.518195,189.532639){1}
\Vertex(213.100437,191.121494){1}
\Vertex(213.682678,192.502985){1}
\Vertex(214.264920,192.136523){1}
\Vertex(214.847162,192.612978){1}
\Vertex(215.429403,193.932452){1}
\Vertex(216.011645,195.611491){1}
\Vertex(216.593886,196.626006){1}
\Vertex(217.176128,197.942886){1}
\Vertex(217.758370,197.828302){1}
\Vertex(218.340611,197.367784){1}
\Vertex(218.922853,197.054609){1}
\Vertex(219.505095,197.254889){1}
\Vertex(220.087336,198.634960){1}
\Vertex(220.669578,199.516792){1}
\Vertex(221.251820,199.672000){1}
\Vertex(221.834061,199.556860){1}
\Vertex(222.416303,200.798381){1}
\Vertex(222.998544,201.185907){1}
\Vertex(223.580786,201.571703){1}
\Vertex(224.163028,202.511270){1}
\Vertex(224.745269,202.761748){1}
\Vertex(225.327511,203.508509){1}
\Vertex(225.909753,204.415606){1}
\Vertex(226.491994,204.358447){1}
\Vertex(227.074236,204.910962){1}
\Vertex(227.656477,205.650125){1}
\Vertex(228.238719,206.599761){1}
\Vertex(228.820961,207.280016){1}
\Vertex(229.403202,208.105822){1}
\Vertex(229.985444,208.842309){1}
\Vertex(230.567686,209.852170){1}
\Vertex(231.149927,210.341123){1}
\Vertex(231.732169,211.304121){1}
\Vertex(232.314410,211.804522){1}
\Vertex(232.896652,212.405670){1}
\Vertex(233.478894,213.136124){1}
\Vertex(234.061135,213.848150){1}
\Vertex(234.643377,214.556140){1}
\Vertex(235.225619,215.410340){1}
\Vertex(235.807860,216.263120){1}
\Vertex(236.390102,218.829695){1}
\Vertex(236.972344,220.865024){1}
\Vertex(237.554585,221.548759){1}
\Vertex(238.136827,226.421382){1}
\Vertex(238.719068,228.575620){1}
\Vertex(239.301310,231.491979){1}
\Vertex(239.883552,243.375259){1}
\Vertex(240.465793,237.665175){1}
\Vertex(241.048035,239.082123){1}
\Vertex(241.630277,241.125442){1}
\Vertex(242.212518,247.452507){1}
\Vertex(242.794760,243.429472){1}
\Vertex(243.377001,249.259987){1}
\Vertex(243.959243,276.358289){1}
\Vertex(244.541485,261.034057){1}
\Vertex(245.123726,246.821174){1}
\Vertex(245.705968,261.966932){1}
\Vertex(246.288210,243.561661){1}
\Vertex(246.870451,241.488610){1}
\Vertex(247.452693,241.620757){1}
\Vertex(248.034934,242.971260){1}
\Vertex(248.617176,245.708774){1}
\Vertex(249.199418,245.494122){1}
\Vertex(249.781659,246.810652){1}
\Vertex(250.363901,246.483414){1}
\Vertex(250.946143,245.617910){1}
\Vertex(251.528384,246.671278){1}
\Vertex(252.110626,247.735106){1}
\Vertex(252.692868,248.968390){1}
\Vertex(253.275109,243.671756){1}
\Vertex(253.857351,245.799103){1}
\Vertex(254.439592,233.957807){1}
\Vertex(255.021834,223.774218){1}
\Vertex(255.604076,224.143975){1}
\Vertex(256.186317,224.750971){1}
\Vertex(256.768559,225.930556){1}
\Vertex(257.350801,227.199091){1}
\Vertex(257.933042,227.821549){1}
\Vertex(258.515284,227.822270){1}
\Vertex(259.097525,227.761509){1}
\Vertex(259.679767,227.359570){1}
\Vertex(260.262009,228.028789){1}
\Vertex(260.844250,229.198450){1}
\Vertex(261.426492,229.731197){1}
\Vertex(262.008734,230.034407){1}
\Vertex(262.590975,230.033028){1}
\Vertex(263.173217,230.042911){1}
\Vertex(263.755459,229.858197){1}
\Vertex(264.337700,230.187886){1}
\Vertex(264.919942,231.074762){1}
\Vertex(265.502183,232.322769){1}
\Vertex(266.084425,234.993508){1}
\Vertex(266.666667,235.831669){1}
\Vertex(267.248908,236.877625){1}
\Vertex(267.831150,236.905421){1}
\Vertex(268.413392,236.654511){1}
\Vertex(268.995633,235.803749){1}
\Vertex(269.577875,237.221417){1}
\Vertex(270.160116,238.703883){1}
\Vertex(270.742358,240.045451){1}
\Vertex(271.324600,241.043637){1}
\Vertex(271.906841,242.485437){1}
\Vertex(272.489083,244.425723){1}
\Vertex(273.071325,247.314904){1}
\Vertex(273.653566,247.834165){1}
\Vertex(274.235808,248.633575){1}
\Vertex(274.818049,249.404756){1}
\Vertex(275.400291,253.175562){1}
\Vertex(275.982533,252.891522){1}
\Vertex(276.564774,252.659122){1}
\Vertex(277.147016,252.647303){1}
\Vertex(277.729258,252.640735){1}
\Vertex(278.311499,252.643659){1}
\Vertex(278.893741,251.073953){1}
\Vertex(279.475983,250.854833){1}
\Vertex(280.058224,250.839761){1}
\Vertex(280.640466,250.572749){1}
\Vertex(281.222707,247.374409){1}
\Vertex(281.804949,247.597605){1}
\Vertex(282.387191,248.858707){1}
\Vertex(282.969432,249.720155){1}
\Vertex(283.551674,250.337240){1}
\Vertex(284.133916,250.285909){1}
\Vertex(284.716157,249.997071){1}
\Vertex(285.298399,251.344157){1}
\Vertex(285.880640,252.640138){1}
\Vertex(286.462882,253.962042){1}
\Vertex(287.045124,254.275361){1}
\Vertex(287.627365,255.199135){1}
\Vertex(288.209607,255.234139){1}
\Vertex(288.791849,255.134174){1}
\Vertex(289.374090,255.136500){1}
\Vertex(289.956332,254.507247){1}
\Vertex(290.538574,255.090379){1}
\Vertex(291.120815,255.224420){1}
\Vertex(291.703057,256.260348){1}
\Vertex(292.285298,256.462831){1}
\Vertex(292.867540,257.930740){1}
\Vertex(293.449782,259.285588){1}
\Vertex(294.032023,260.614596){1}
\Vertex(294.614265,261.734161){1}
\Vertex(295.196507,264.214648){1}
\Vertex(295.778748,266.132512){1}
\Vertex(296.360990,266.513017){1}
\Vertex(296.943231,267.271988){1}
\Vertex(297.525473,270.219150){1}
\Vertex(298.107715,270.315408){1}
\Vertex(298.689956,270.208463){1}
\Vertex(299.272198,270.273116){1}
\Vertex(299.854440,269.195472){1}
\Vertex(300.436681,269.066208){1}
\Vertex(301.018923,269.030360){1}
\Vertex(301.601164,266.082540){1}
\Vertex(302.183406,266.014737){1}
\Vertex(302.765648,266.932993){1}
\Vertex(303.347889,268.380846){1}
\Vertex(303.930131,269.774816){1}
\Vertex(304.512373,270.989754){1}
\Vertex(305.094614,274.145349){1}
\Vertex(305.676856,277.558259){1}
\Vertex(306.259098,280.267193){1}
\Vertex(306.841339,281.494713){1}
\Vertex(307.423581,290.978547){1}
\Vertex(308.005822,291.343444){1}
\Vertex(308.588064,291.205244){1}
\Vertex(309.170306,291.302182){1}
\Vertex(309.752547,291.814978){1}
\Vertex(310.334789,290.961992){1}
\Vertex(310.917031,290.174813){1}
\Vertex(311.499272,284.514063){1}
\Vertex(312.081514,283.308555){1}
\Vertex(312.663755,283.555101){1}
\Vertex(313.245997,284.983805){1}
\Vertex(313.828239,286.448563){1}
\Vertex(314.410480,287.548980){1}
\Vertex(314.992722,289.026216){1}
\Vertex(315.574964,292.021270){1}
\Vertex(316.157205,292.647456){1}
\Vertex(316.739447,293.790741){1}
\Vertex(317.321689,294.393742){1}
\Vertex(317.903930,297.312407){1}
\Vertex(318.486172,297.051263){1}
\Vertex(319.068413,297.050748){1}
\Vertex(319.650655,297.098147){1}
\Vertex(320.232897,295.932460){1}
\Vertex(320.815138,295.773978){1}
\Vertex(321.397380,292.993822){1}
\Vertex(321.979622,294.048467){1}
\Vertex(322.561863,295.348463){1}
\Vertex(323.144105,296.886913){1}
\Vertex(323.726346,297.324658){1}
\Vertex(324.308588,298.648992){1}
\Vertex(324.890830,298.670220){1}
\Vertex(325.473071,298.546824){1}
\Vertex(326.055313,298.536899){1}
\Vertex(326.637555,297.769445){1}
\Vertex(327.219796,298.473235){1}
\Vertex(327.802038,299.840602){1}
\Vertex(328.384279,301.399436){1}
\Vertex(328.966521,302.088504){1}
\Vertex(329.548763,303.402975){1}
\Vertex(330.131004,303.438122){1}
\Vertex(330.713246,303.330560){1}
\Vertex(331.295488,303.305934){1}
\Vertex(331.877729,302.580854){1}
\Vertex(332.459971,302.902142){1}
\Vertex(333.042213,303.472672){1}
\Vertex(333.624454,303.969573){1}
\Vertex(334.206696,304.462314){1}
\Vertex(334.788937,305.852495){1}
\Vertex(335.371179,308.304259){1}
\Vertex(335.953421,309.814583){1}
\Vertex(336.535662,312.716550){1}
\Vertex(337.117904,313.556872){1}
\Vertex(337.700146,312.973864){1}
\Vertex(338.282387,311.125554){1}
\Vertex(338.864629,312.264556){1}
\Vertex(339.446870,313.629123){1}
\Vertex(340.029112,314.452397){1}
\Vertex(340.611354,314.564551){1}
\Vertex(341.193595,314.148693){1}
\Vertex(341.775837,315.541674){1}
\Vertex(342.358079,316.352429){1}
\Vertex(342.940320,316.819618){1}
\Vertex(343.522562,318.225695){1}
\Vertex(344.104803,319.613281){1}
\Vertex(344.687045,321.111889){1}
\Vertex(345.269287,323.762719){1}
\Vertex(345.851528,324.584675){1}
\Vertex(346.433770,325.266968){1}
\Vertex(347.016012,328.649549){1}
\Vertex(347.598253,328.436689){1}
\Vertex(348.180495,328.543098){1}
\Vertex(348.762737,327.324927){1}
\Vertex(349.344978,327.248682){1}
\Vertex(349.927220,327.242978){1}
\Vertex(350.509461,325.633163){1}
\Vertex(351.091703,325.417256){1}
\Vertex(351.673945,326.339774){1}
\Vertex(352.256186,327.628198){1}
\Vertex(352.838428,329.181123){1}
\Vertex(353.420670,330.820341){1}
\Vertex(354.002911,330.931795){1}
\Vertex(354.585153,330.780849){1}
\Vertex(355.167394,330.444695){1}
\Vertex(355.749636,330.367359){1}
\Vertex(356.331878,331.865638){1}
\Vertex(356.914119,331.940586){1}
\Vertex(357.496361,333.008408){1}
\Vertex(358.078603,334.342357){1}
\Vertex(358.660844,335.214450){1}
\Vertex(359.243086,335.264567){1}
\Vertex(359.825328,334.916203){1}
\Vertex(360.407569,336.144587){1}
\Vertex(360.989811,337.394549){1}
\Vertex(361.572052,338.239113){1}
\Vertex(362.154294,338.285935){1}
\Vertex(362.736536,337.939466){1}
\Vertex(363.318777,338.699590){1}
\Vertex(363.901019,339.405892){1}
\Vertex(364.483261,340.216029){1}
\Vertex(365.065502,340.931988){1}
\Vertex(365.647744,341.701377){1}
\Vertex(366.229985,342.895890){1}
\Vertex(366.812227,343.417170){1}
\Vertex(367.394469,344.705512){1}
\Vertex(367.976710,345.306907){1}
\Vertex(368.558952,345.724886){1}
\Vertex(369.141194,346.307626){1}
\Vertex(369.723435,347.351152){1}
\Vertex(370.305677,348.023851){1}
\Vertex(370.887918,348.521513){1}
\Vertex(371.470160,349.560489){1}
\Vertex(372.052402,349.971900){1}
\Vertex(372.634643,350.955241){1}
\Vertex(373.216885,351.560013){1}
\Vertex(373.799127,352.356190){1}
\Vertex(374.381368,353.646839){1}
\Vertex(374.963610,353.710071){1}
\Vertex(375.545852,362.654795){1}
\Vertex(376.128093,355.599623){1}
\Vertex(376.710335,355.403894){1}
\Vertex(377.292576,353.070790){1}
\Vertex(377.874818,359.911907){1}
\Vertex(378.457060,356.716409){1}
\Vertex(379.039301,350.227443){1}
\Vertex(379.621543,358.662357){1}
\Vertex(380.203785,351.416191){1}
\Vertex(380.786026,351.118232){1}
\Vertex(381.368268,351.376473){1}
\Vertex(381.950509,345.580076){1}
\Vertex(382.532751,344.796747){1}
\Vertex(383.114993,345.186023){1}
\Vertex(383.697234,363.636345){1}
\Vertex(384.279476,346.027869){1}
\Vertex(384.861718,346.018459){1}
\Vertex(385.443959,346.734912){1}
\Vertex(386.026201,345.782867){1}
\Vertex(386.608443,346.410371){1}
\Vertex(387.190684,347.664492){1}
\Vertex(387.772926,346.966200){1}
\Vertex(388.355167,346.947237){1}
\Vertex(388.937409,348.987632){1}
\Vertex(389.519651,349.134027){1}
\Vertex(390.101892,348.523017){1}
\Vertex(390.684134,348.503477){1}
\Vertex(391.266376,349.061736){1}
\Vertex(391.848617,348.793303){1}
\Vertex(392.430859,356.983112){1}
\Vertex(393.013100,347.265004){1}
\Vertex(393.595342,347.131518){1}
\Vertex(394.177584,347.801108){1}
\Vertex(394.759825,348.112389){1}
\Vertex(395.342067,349.042093){1}
\Vertex(395.924309,349.058668){1}
\Vertex(396.506550,349.499748){1}
\Vertex(397.088792,349.942826){1}
\Vertex(397.671033,349.546838){1}
\Vertex(398.253275,345.490076){1}
\Vertex(398.835517,346.505847){1}
\SetOffset(210.0,0.0)
\BBox(0.0,0.0)(400,400)
\PText(20,360)(0)[l]{MZVs}
\PText(20,335)(0)[l]{Modular arithmetic}
\PText(20,310)(0)[l]{Basis only}
\PText(20,285)(0)[l]{W = 26, D = 8}
\Vertex(1.028278,0.000010){1}
\Vertex(2.056555,0.341186){1}
\Vertex(3.084833,0.341186){1}
\Vertex(4.113111,0.366793){1}
\Vertex(5.141388,0.366793){1}
\Vertex(6.169666,0.366803){1}
\Vertex(7.197943,0.367050){1}
\Vertex(8.226221,0.367078){1}
\Vertex(9.254499,0.369891){1}
\Vertex(10.282776,0.383022){1}
\Vertex(11.311054,0.399375){1}
\Vertex(12.339332,0.415898){1}
\Vertex(13.367609,0.435073){1}
\Vertex(14.395887,0.455596){1}
\Vertex(15.424165,0.463350){1}
\Vertex(16.452442,0.672529){1}
\Vertex(17.480720,0.474552){1}
\Vertex(18.508997,0.489147){1}
\Vertex(19.537275,0.504121){1}
\Vertex(20.565553,0.519324){1}
\Vertex(21.593830,0.533909){1}
\Vertex(22.622108,0.549520){1}
\Vertex(23.650386,0.562984){1}
\Vertex(24.678663,0.576914){1}
\Vertex(25.706941,0.594701){1}
\Vertex(26.735219,0.613372){1}
\Vertex(27.763496,0.631814){1}
\Vertex(28.791774,0.648376){1}
\Vertex(29.820051,0.667655){1}
\Vertex(30.848329,0.687504){1}
\Vertex(31.876607,0.703115){1}
\Vertex(32.904884,0.721653){1}
\Vertex(33.933162,0.738908){1}
\Vertex(34.961440,0.758016){1}
\Vertex(35.989717,0.780221){1}
\Vertex(37.017995,0.802854){1}
\Vertex(38.046272,0.824870){1}
\Vertex(39.074550,0.838856){1}
\Vertex(40.102828,12.185471){1}
\Vertex(41.131105,18.063097){1}
\Vertex(42.159383,4.657088){1}
\Vertex(43.187661,3.949146){1}
\Vertex(44.215938,1.585356){1}
\Vertex(45.244216,1.266166){1}
\Vertex(46.272494,1.295450){1}
\Vertex(47.300771,1.319679){1}
\Vertex(48.329049,1.347358){1}
\Vertex(49.357326,1.375730){1}
\Vertex(50.385604,1.400424){1}
\Vertex(51.413882,1.441833){1}
\Vertex(52.442159,1.453662){1}
\Vertex(53.470437,1.480486){1}
\Vertex(54.498715,1.503841){1}
\Vertex(55.526992,1.533904){1}
\Vertex(56.555270,1.555273){1}
\Vertex(57.583548,1.580016){1}
\Vertex(58.611825,1.605167){1}
\Vertex(59.640103,1.633852){1}
\Vertex(60.668380,1.663688){1}
\Vertex(61.696658,1.696221){1}
\Vertex(62.724936,1.730266){1}
\Vertex(63.753213,1.793186){1}
\Vertex(64.781491,1.806992){1}
\Vertex(65.809769,1.857712){1}
\Vertex(66.838046,1.854234){1}
\Vertex(67.866324,1.899481){1}
\Vertex(68.894602,1.957451){1}
\Vertex(69.922879,1.960739){1}
\Vertex(70.951157,1.981576){1}
\Vertex(71.979434,2.013559){1}
\Vertex(73.007712,2.050121){1}
\Vertex(74.035990,2.089211){1}
\Vertex(75.064267,2.110086){1}
\Vertex(76.092545,2.148634){1}
\Vertex(77.120823,2.183620){1}
\Vertex(78.149100,2.198955){1}
\Vertex(79.177378,2.249058){1}
\Vertex(80.205656,2.300918){1}
\Vertex(81.233933,2.314382){1}
\Vertex(82.262211,2.329109){1}
\Vertex(83.290488,2.369872){1}
\Vertex(84.318766,2.391155){1}
\Vertex(85.347044,2.430169){1}
\Vertex(86.375321,2.453182){1}
\Vertex(87.403599,2.478086){1}
\Vertex(88.431877,2.511152){1}
\Vertex(89.460154,2.546470){1}
\Vertex(90.488432,2.570528){1}
\Vertex(91.516710,2.597665){1}
\Vertex(92.544987,2.640308){1}
\Vertex(93.573265,2.676966){1}
\Vertex(94.601542,2.712730){1}
\Vertex(95.629820,2.752523){1}
\Vertex(96.658098,2.785998){1}
\Vertex(97.686375,2.836537){1}
\Vertex(98.714653,2.861897){1}
\Vertex(99.742931,2.906090){1}
\Vertex(100.771208,2.947384){1}
\Vertex(101.799486,2.977552){1}
\Vertex(102.827763,3.037545){1}
\Vertex(103.856041,3.060549){1}
\Vertex(104.884319,3.091667){1}
\Vertex(105.912596,3.135698){1}
\Vertex(106.940874,3.165229){1}
\Vertex(107.969152,3.204252){1}
\Vertex(108.997429,3.236226){1}
\Vertex(110.025707,3.279648){1}
\Vertex(111.053985,3.311726){1}
\Vertex(112.082262,3.353467){1}
\Vertex(113.110540,3.393478){1}
\Vertex(114.138817,3.421812){1}
\Vertex(115.167095,3.462413){1}
\Vertex(116.195373,3.508525){1}
\Vertex(117.223650,3.552888){1}
\Vertex(118.251928,3.599627){1}
\Vertex(119.280206,3.646803){1}
\Vertex(120.308483,3.688468){1}
\Vertex(121.336761,3.734665){1}
\Vertex(122.365039,18.318635){1}
\Vertex(123.393316,44.999379){1}
\Vertex(124.421594,59.252577){1}
\Vertex(125.449871,74.636789){1}
\Vertex(126.478149,87.788656){1}
\Vertex(127.506427,99.067942){1}
\Vertex(128.534704,109.525893){1}
\Vertex(129.562982,115.025723){1}
\Vertex(130.591260,23.100984){1}
\Vertex(131.619537,20.899036){1}
\Vertex(132.647815,18.305694){1}
\Vertex(133.676093,15.398645){1}
\Vertex(134.704370,14.997807){1}
\Vertex(135.732648,14.943172){1}
\Vertex(136.760925,4.204713){1}
\Vertex(137.789203,3.815847){1}
\Vertex(138.817481,3.842680){1}
\Vertex(139.845758,3.885114){1}
\Vertex(140.874036,3.903586){1}
\Vertex(141.902314,3.934922){1}
\Vertex(142.930591,3.958876){1}
\Vertex(143.958869,3.977870){1}
\Vertex(144.987147,4.023630){1}
\Vertex(146.015424,4.043336){1}
\Vertex(147.043702,4.065998){1}
\Vertex(148.071979,4.072317){1}
\Vertex(149.100257,4.132491){1}
\Vertex(150.128535,4.124101){1}
\Vertex(151.156812,4.143465){1}
\Vertex(152.185090,4.181358){1}
\Vertex(153.213368,4.191335){1}
\Vertex(154.241645,4.213131){1}
\Vertex(155.269923,4.246463){1}
\Vertex(156.298201,4.266901){1}
\Vertex(157.326478,4.358222){1}
\Vertex(158.354756,4.319921){1}
\Vertex(159.383033,4.468214){1}
\Vertex(160.411311,4.419346){1}
\Vertex(161.439589,4.515979){1}
\Vertex(162.467866,4.505593){1}
\Vertex(163.496144,4.517480){1}
\Vertex(164.524422,4.547040){1}
\Vertex(165.552699,4.559962){1}
\Vertex(166.580977,4.621495){1}
\Vertex(167.609254,4.872710){1}
\Vertex(168.637532,4.845734){1}
\Vertex(169.665810,4.756381){1}
\Vertex(170.694087,4.804070){1}
\Vertex(171.722365,4.812403){1}
\Vertex(172.750643,4.850105){1}
\Vertex(173.778920,4.941559){1}
\Vertex(174.807198,5.227911){1}
\Vertex(175.835476,5.246553){1}
\Vertex(176.863753,5.145494){1}
\Vertex(177.892031,5.141959){1}
\Vertex(178.920308,5.015596){1}
\Vertex(179.948586,5.089215){1}
\Vertex(180.976864,5.190684){1}
\Vertex(182.005141,5.129369){1}
\Vertex(183.033419,5.170093){1}
\Vertex(184.061697,5.195035){1}
\Vertex(185.089974,5.234980){1}
\Vertex(186.118252,5.238572){1}
\Vertex(187.146530,5.296143){1}
\Vertex(188.174807,5.285482){1}
\Vertex(189.203085,5.311697){1}
\Vertex(190.231362,5.410353){1}
\Vertex(191.259640,5.372698){1}
\Vertex(192.287918,5.396519){1}
\Vertex(193.316195,5.491991){1}
\Vertex(194.344473,5.451894){1}
\Vertex(195.372751,5.600092){1}
\Vertex(196.401028,5.542683){1}
\Vertex(197.429306,5.609480){1}
\Vertex(198.457584,5.780282){1}
\Vertex(199.485861,5.734303){1}
\Vertex(200.514139,5.673768){1}
\Vertex(201.542416,5.702416){1}
\Vertex(202.570694,5.752138){1}
\Vertex(203.598972,5.751501){1}
\Vertex(204.627249,5.814326){1}
\Vertex(205.655527,5.798022){1}
\Vertex(206.683805,5.885903){1}
\Vertex(207.712082,5.846528){1}
\Vertex(208.740360,5.945526){1}
\Vertex(209.768638,5.906930){1}
\Vertex(210.796915,5.935245){1}
\Vertex(211.825193,5.961270){1}
\Vertex(212.853470,5.987020){1}
\Vertex(213.881748,6.016009){1}
\Vertex(214.910026,6.051660){1}
\Vertex(215.938303,6.083680){1}
\Vertex(216.966581,6.112660){1}
\Vertex(217.994859,6.141773){1}
\Vertex(219.023136,6.151855){1}
\Vertex(220.051414,6.176198){1}
\Vertex(221.079692,6.273628){1}
\Vertex(222.107969,6.277001){1}
\Vertex(223.136247,6.308918){1}
\Vertex(224.164524,6.332396){1}
\Vertex(225.192802,6.358336){1}
\Vertex(226.221080,6.387629){1}
\Vertex(227.249357,6.436335){1}
\Vertex(228.277635,6.581815){1}
\Vertex(229.305913,6.557672){1}
\Vertex(230.334190,6.652916){1}
\Vertex(231.362468,6.849782){1}
\Vertex(232.390746,6.792829){1}
\Vertex(233.419023,6.686134){1}
\Vertex(234.447301,6.758765){1}
\Vertex(235.475578,6.750033){1}
\Vertex(236.503856,6.870201){1}
\Vertex(237.532134,6.813609){1}
\Vertex(238.560411,6.996963){1}
\Vertex(239.588689,7.064738){1}
\Vertex(240.616967,7.010617){1}
\Vertex(241.645244,7.230552){1}
\Vertex(242.673522,7.286878){1}
\Vertex(243.701799,7.222685){1}
\Vertex(244.730077,7.120219){1}
\Vertex(245.758355,7.244596){1}
\Vertex(246.786632,7.238667){1}
\Vertex(247.814910,7.267865){1}
\Vertex(248.843188,7.254953){1}
\Vertex(249.871465,7.318452){1}
\Vertex(250.899743,7.365258){1}
\Vertex(251.928021,7.348791){1}
\Vertex(252.956298,7.405697){1}
\Vertex(253.984576,7.492409){1}
\Vertex(255.012853,7.455096){1}
\Vertex(256.041131,7.500524){1}
\Vertex(257.069409,7.629110){1}
\Vertex(258.097686,7.615722){1}
\Vertex(259.125964,7.609755){1}
\Vertex(260.154242,7.795370){1}
\Vertex(261.182519,7.845663){1}
\Vertex(262.210797,7.796605){1}
\Vertex(263.239075,7.865721){1}
\Vertex(264.267352,7.846746){1}
\Vertex(265.295630,7.944784){1}
\Vertex(266.323907,8.139854){1}
\Vertex(267.352185,8.080221){1}
\Vertex(268.380463,7.979475){1}
\Vertex(269.408740,8.038471){1}
\Vertex(270.437018,8.268402){1}
\Vertex(271.465296,8.294370){1}
\Vertex(272.493573,8.236305){1}
\Vertex(273.521851,8.140623){1}
\Vertex(274.550129,8.267727){1}
\Vertex(275.578406,8.216722){1}
\Vertex(276.606684,8.303055){1}
\Vertex(277.634961,8.320965){1}
\Vertex(278.663239,8.387857){1}
\Vertex(279.691517,8.371182){1}
\Vertex(280.719794,8.481791){1}
\Vertex(281.748072,8.424239){1}
\Vertex(282.776350,8.552863){1}
\Vertex(283.804627,8.539428){1}
\Vertex(284.832905,8.532606){1}
\Vertex(285.861183,8.609712){1}
\Vertex(286.889460,8.595706){1}
\Vertex(287.917738,8.653287){1}
\Vertex(288.946015,8.718658){1}
\Vertex(289.974293,8.700567){1}
\Vertex(291.002571,8.768171){1}
\Vertex(292.030848,8.781236){1}
\Vertex(293.059126,8.849012){1}
\Vertex(294.087404,8.839757){1}
\Vertex(295.115681,8.893252){1}
\Vertex(296.143959,8.947934){1}
\Vertex(297.172237,8.933634){1}
\Vertex(298.200514,8.993912){1}
\Vertex(299.228792,9.006730){1}
\Vertex(300.257069,9.031321){1}
\Vertex(301.285347,9.058837){1}
\Vertex(302.313625,9.092673){1}
\Vertex(303.341902,9.137806){1}
\Vertex(304.370180,9.215986){1}
\Vertex(305.398458,9.220917){1}
\Vertex(306.426735,9.270706){1}
\Vertex(307.455013,9.311107){1}
\Vertex(308.483290,9.351518){1}
\Vertex(309.511568,9.394418){1}
\Vertex(310.539846,9.433926){1}
\Vertex(311.568123,9.532563){1}
\Vertex(312.596401,9.556450){1}
\Vertex(313.624679,9.595825){1}
\Vertex(314.652956,9.636226){1}
\Vertex(315.681234,9.654089){1}
\Vertex(316.709512,9.785545){1}
\Vertex(317.737789,9.735234){1}
\Vertex(318.766067,9.775720){1}
\Vertex(319.794344,9.841425){1}
\Vertex(320.822622,9.971227){1}
\Vertex(321.850900,9.923994){1}
\Vertex(322.879177,10.051887){1}
\Vertex(323.907455,10.007010){1}
\Vertex(324.935733,10.062462){1}
\Vertex(325.964010,10.101533){1}
\Vertex(326.992288,10.142333){1}
\Vertex(328.020566,10.182734){1}
\Vertex(329.048843,10.223126){1}
\Vertex(330.077121,10.246500){1}
\Vertex(331.105398,10.309810){1}
\Vertex(332.133676,10.337108){1}
\Vertex(333.161954,10.386184){1}
\Vertex(334.190231,10.426576){1}
\Vertex(335.218509,10.467072){1}
\Vertex(336.246787,10.499283){1}
\Vertex(337.275064,10.559229){1}
\Vertex(338.303342,10.599507){1}
\Vertex(339.331620,10.641134){1}
\Vertex(340.359897,10.682390){1}
\Vertex(341.388175,10.690533){1}
\Vertex(342.416452,10.727390){1}
\Vertex(343.444730,10.763943){1}
\Vertex(344.473008,10.813684){1}
\Vertex(345.501285,10.863083){1}
\Vertex(346.529563,10.906905){1}
\Vertex(347.557841,10.958936){1}
\Vertex(348.586118,11.006264){1}
\Vertex(349.614396,11.055483){1}
\Vertex(350.642674,11.101015){1}
\Vertex(351.670951,11.148476){1}
\Vertex(352.699229,11.196231){1}
\Vertex(353.727506,11.225012){1}
\Vertex(354.755784,25.791110){1}
\Vertex(355.784062,55.805651){1}
\Vertex(356.812339,79.599413){1}
\Vertex(357.840617,101.591522){1}
\Vertex(358.868895,118.538552){1}
\Vertex(359.897172,132.050068){1}
\Vertex(360.925450,149.715938){1}
\Vertex(361.953728,165.042892){1}
\Vertex(362.982005,180.043293){1}
\Vertex(364.010283,199.265672){1}
\Vertex(365.038560,213.517616){1}
\Vertex(366.066838,224.620683){1}
\Vertex(367.095116,235.316813){1}
\Vertex(368.123393,266.649133){1}
\Vertex(369.151671,263.277689){1}
\Vertex(370.179949,289.703293){1}
\Vertex(371.208226,303.318862){1}
\Vertex(372.236504,313.349142){1}
\Vertex(373.264781,326.725249){1}
\Vertex(374.293059,335.757130){1}
\Vertex(375.321337,355.322443){1}
\Vertex(376.349614,363.636396){1}
\Vertex(377.377892,359.524667){1}
\Vertex(378.406170,358.669943){1}
\Vertex(379.434447,193.263507){1}
\Vertex(380.462725,76.290528){1}
\Vertex(381.491003,72.718808){1}
\Vertex(382.519280,62.053882){1}
\Vertex(383.547558,48.028724){1}
\Vertex(384.575835,44.268927){1}
\Vertex(385.604113,38.563747){1}
\Vertex(386.632391,35.513022){1}
\Vertex(387.660668,34.184706){1}
\Vertex(388.688946,33.585198){1}
\Vertex(389.717224,33.495682){1}
\Vertex(390.745501,33.363314){1}
\Vertex(391.773779,33.310371){1}
\Vertex(392.802057,33.268554){1}
\Vertex(393.830334,12.617389){1}
\Vertex(394.858612,11.257109){1}
\Vertex(395.886889,11.019510){1}
\Vertex(396.915167,20.619990){1}
\Vertex(397.943445,18.394563){1}
\end{picture}
\caption{Performance of the program. On the $x$-axis we have the number of 
the module in which one group of equations is substituted and on the $y$-axis 
the size of the expression at the end of the module (arbitrary units). The 
spikes are due to the shuffles.}
\label{fig:weight21}
\end{center}
\end{figure}
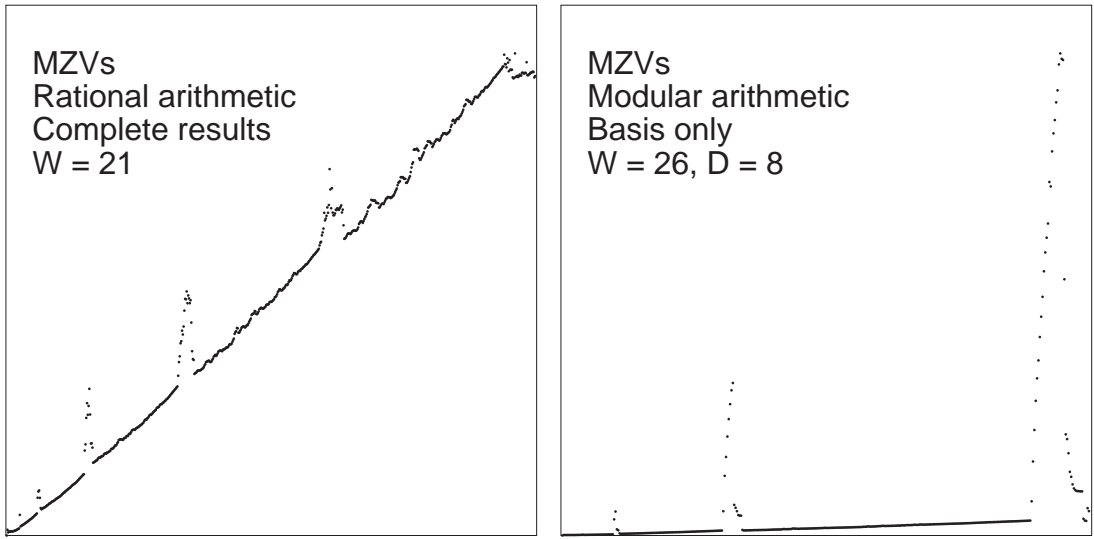

For constructing tables of all sums at a given weight we run the full 
program.
The performance of the program is shown in Figure~\ref{fig:weight21} for a 
complete run at weight ${\sf w=21}$ and a run to depth ${\sf d=8}$ at weight 
${\sf w=26}$.
We see that the stuffles give a steady growth of the master expression but 
that the shuffles cause intermediate expression swell which is worse when 
the depth is much less than half the weight. The result is that when we run 
the complete system most time is spent with the stuffle relations while for 
the limited depth runs by far most time is spent with the shuffle 
relations.

In the case of Euler sums the master expression is created with a 
three letter alphabet $(-1,0,1)$ rather than the two letter alphabet (0,1) 
for the MZVs. In addition there are many more equations to 
consider because the number of lower weight objects that we can multiply 
either by shuffles or stuffles is correspondingly greater. Of course also 
for the Euler sums it is possible to just study the basis.

In addition it is possible to study sums to a limited depth. This way we 
can go to much greater values of the weight. This is of course only 
possible if we use a basis in which the concept of depth is relevant, like 
the basis of the odd negative indices that form a Lyndon word. Without such 
a basis the calculations become much harder.

When we are constructing tables we cannot go quite as far in weight as when 
we are determining rank deficiency. When we use a Lyndon basis, the 
majority of terms consists of products of basis elements of lower weights. 
This means that we have many more terms to carry around. We observe, in 
addition, that the coefficients containing the most digits are in the terms 
with powers of $\zeta_2$. This is to be expected since $\zeta_2^N$ is our 
repository for all terms of the form $\zeta_{2a}^m\zeta_{2b}^n$ with 
$N=ma+nb$.

The representation we have selected, together with the modular arithmetic, 
makes for a very fast treatment of the terms. This is reflected in the 
number of terms that can be processed. In one run, which took more than 30 
days the program generated a total of more than $7\ \cdot 10^{12}$ terms. This 
seems to be a new record.


\section{The Running of the Programs}
\renewcommand{\theequation}{\thesection.\arabic{equation}}
\setcounter{equation}{0}
\label{Runningprograms}

\vspace{1mm}
\noindent
We have used the programs of the previous Section to obtain results to as 
high a weight and depth as possible, both for MZVs and Euler sums. Before we 
start discussing these results we show the parameters of 
these runs to give the reader an impression of what is available and why 
there are limitations to obtain more.

We start with the Euler sums. We have first run the complete system for the 
given weights, see Table~\ref{run12}. This means that for ${\sf w = 12}$ there 
are expressions for all 236196 Euler sums with that weight, all expressed 
in terms of the basis of Lyndon words of the negative odd integers, see 
Appendix~A, which 
is the basis we use for all Euler sums, unless mentioned differently.

The columns marked `variables' mentions how many variables there are at the 
start of the program. `Remaining' tells how many basis elements remain in the 
end. Under 
`output' we give the size of the output expression in text format. The column 
`size' refers to the largest size of the master expression during the 
calculation. Time refers to real time to run the program. If the column `CPU 
time' is present it refers to the total CPU time by all processors.
\begin{table}\centering
\begin{tabular}{|c|c|c|c|c|c|c|} \hline
 {\sf w}  & variables & eqns   & remaining &  size  & output  & time [sec] \\ 
\hline
  4 &       36  &     57 &      1    &    4.3K&    2.0K &   0.06 \\ %
  5 &      108  &    192 &      2    &   21K  &    8.9K &   0.12 \\ %
  6 &      324  &    665 &      2    &   98K  &    42K  &   0.37 \\ %
  7 &      972  &   2205 &      4    &  472K  &   219K  &   1.71 \\ %
  8 &     2916  &   7313 &      5    &  2.25M &   1.15M &   7.78 \\ %
  9 &     8748  &  23909 &      8    &   11M  &    6.3M &    50  \\ %
 10 &    26244  &  77853 &     11    &   58M  &    36M  &   353  \\ %
 11 &    78732  & 251565 &     18    &  360M  &   213M  &  3266  \\ %
 12 &   236196  & 809177 &     25    &  3.1G  &  1.29G  & 47311  \\ %
\hline
\end{tabular}
\caption{
Runs on an 8-core Xeon computer at 3 GHz and with 32 Gbytes of memory. The 
column `eqns' gives the number of equations that was considered.}
\label{run12}
\end{table}
We notice that computer time is not the 
issue here, see Table~\ref{run12}\footnote{The first time we ran the ${\sf w = 
12}$ case on an 8-core 
Xeon machine at 2.33 GHz the run took two full weeks. It just shows how 
good a test case this problem is. Both {\tt (T)FORM} and the MZV program 
have been improved greatly during this project.}. The size of the results 
becomes the major problem. This is one of the reasons why we stopped at ${\sf 
w = 12}$. Technically the 
run at ${\sf w = 13}$ is feasible as it should take of the order of 10~days. 
The output is, however, projected at almost 8 Gbytes which we considered 
excessive.

We have also run programs that go to a maximum value of the depth. This 
involves only a subset of the Euler sums of that weight and hence such 
programs are much faster. As a consequence we can go to much greater values 
of the weight. 

In Table~\ref{depth4} we show the statistics of the runs up to depth ${\sf 
d=4}$. 
These are full runs in the sense that they are over the rational numbers and we 
have kept all terms, including the products of lower weight objects.

\begin{table}\centering
\begin{tabular}[htb]{|c|c|c|c|} \hline
  weight &  constants & running time [sec] & output [Mbyte] \\ \hline
    9    &     956    &      7             &   0.26         \\  %
   10    &    1412    &     13             &   0.64         \\  %
   11    &    1996    &     24             &   1.25         \\  %
   12    &    2724    &     39             &   3.18         \\  %
   13    &    3612    &     68             &   5.04         \\  %
   14    &    4676    &    108             &  17.1          \\  %
   15    &    5932    &    199             &  17.1          \\  %
   16    &    7396    &    436             &  71.1          \\  %
   17    &    9084    &    602             &  54.9          \\  %
   18    &   11012    &   1323             & 275.9          \\  %
   19    &   13196    &   2761             & 157.1          \\  %
   20    &   15652    &   5424             & 877            \\  %
   21    &   18396    &  14090             & 395            \\  %
   22    &   21444    &  21875             & 2559           \\  %
      \hline
\end{tabular}
\caption{Summary of the runs at {\sf d = 4}. The runs were  performed on a 
computer with 8 Xeons at 3 GHz, using {\tt TFORM}.}
\label{depth4}
\end{table}

The dependence on the parity of the weight
for the higher values is 
due to the fact that we run up to an even depth and the independent 
variables we use have an even depth for even weights and an odd depth for 
odd weights. This means for instance that the depth 4 objects for weight 
${\sf w=17}$ 
can all be expressed in terms of depth ${\sf d=3}$ objects.
The results for the depth 5 runs are summarized in Table~\ref{depth5}.

\begin{table}\centering
\begin{tabular}[htb]{|c|c|c|c|c|} \hline
  weight &  constants &  remaining & running time [sec] & output [Mbyte] \\ 
\hline
    9    &    3394    &      7     &       27     &    1.15  \\ %
   10    &    5702    &      7     &       72     &    3.11  \\ %
   11    &    9042    &     13     &      172     &    8.5   \\ %
   12    &   13686    &     11     &      478     &   20.9   \\ %
   13    &   19938    &     22     &     1330     &   68.9   \\ %
   14    &   28134    &     17     &     4306     &  133     \\ %
   15    &   38642    &     35     &    27607     &  473     \\ %
   16    &   51862    &     24     &   110336     &  688     \\ %
   17    &   68226    &     55     &   450462     & 2767     \\ \hline
%
%
\end{tabular}
\caption{Summary of the runs at {\sf d = 5}. Same computer as used in 
Table~\ref{depth4}.}
\label{depth5}
\end{table}

We have a nice example here of what happens if we change the order in which 
we deal with the shuffles and the stuffles. We reran the program of 
Table~\ref{depth5} for the weights ${\sf w=14}$ and ${\sf w=15}$ under 
these conditions, obtaining running times of 100973 and 493489 sec 
respectively. This is more than an order of magnitude slower than the order 
we select in the regular programs.

Because we like to compare results of the MZV runs with those of the Euler 
runs to as high a weight as possible we made also runs in which we do all 
calculus modulus a 31-bit prime number. The number we selected is 
2147479273. We never ran into a case in which this seemed to cause 
problems. In the programs in which we used this modulus we also dropped all 
terms that are products of lower weight objects. This means that in the end 
all sums are expressed into elements from the same-weight Lyndon part of 
the basis only. Such programs are much faster. This can be seen in 
Tables~\ref{depth4m},~\ref{depth5m} and \ref{depth6m} which are for depth 
${\sf d \leq 
4}$, depth ${\sf d \leq 5}$ and depth ${\sf d \leq 6}$, respectively.

\begin{table}\centering
\begin{tabular}[htb]{|c|c|c|c|} \hline
  weight &  constants & running time [sec] & output [Mbyte] \\ \hline
   14    &    4676    &           35       &        1.3     \\  %
   16    &    7396    &          105       &        2.9     \\  %
   18    &   11012    &          323       &        6.0     \\  %
   20    &   15652    &          939       &       11.3     \\  %
   22    &   21444    &         2211       &       20.5     \\  %
   24    &   28516    &         5335       &       35       \\  %
   26    &   36996    &        13127       &       57       \\  %
   28    &   47012    &        47056       &       89       \\  %
   30    &   58692    &       100813       &      137       \\  %
      \hline
\end{tabular}
\caption{Summary of the runs at ${\sf d = 4}$ in modular arithmetic, dropping 
all terms that are products of lower weight objects.}
\label{depth4m}
\end{table}

\begin{table}\centering
\begin{tabular}[htb]{|c|c|c|c|} \hline
  weight &  constants & running time [sec] & output [Mbyte] \\ 
\hline
   13    &   16812    &       388    &     5.5  \\
   15    &   33388    &      2932    &    18    \\
   17    &   60044    &     18836    &    53    \\
   19    &  100236    &    118874    &   131    \\
   21    &  157932    &    554870    &   299    \\
\hline
\end{tabular}
\caption{Summary of the runs at ${\sf d = 5}$ in modular arithmetic, dropping 
all terms that are products of lower weight objects.}
\label{depth5m}
\end{table}

\begin{table}\centering
\begin{tabular}{|c|c|c|c|c|} \hline
   weight &  constants &  remaining & running time [sec] & output [Mbyte] \\ 
\hline
    13    &     56940  &       22   &      2611    &          \\ %
    14    &     90564  &       37   &     12716    &    51    \\ %
    15    &    138636  &       35   &     55204    &    87    \\ %
    16    &    205412  &       66   &    206951    &   214    \\ %
    17    &    295916  &       55   &    789540    &   288    \\ %
    18    &    416004  &      109   &   2622157    &   711    \\ %
\hline
\end{tabular}
\caption{Summary of the runs at ${\sf d = 6}$ in modular arithmetic, dropping 
all terms that are products of lower weight objects. Times refer to an 8 
Xeon core machine at 3 GHz and 32 GBytes of memory.}
\label{depth6m}
\end{table}

The run at ${\sf w = 18, d = 6}$ deserves some special attention. It was our 
most costly run and during the running {\tt TFORM} processed more than $7 
\cdot 10^{12}$ terms.

We come now to our runs for the Multiple Zeta Values. Those runs look more 
spectacular because there is much more literature on them. First we present 
the `complete' runs in which all calculus is over the rational numbers and all 
terms are kept, cf.  Table~\ref{MZVall}.

\begin{table}\centering
\begin{tabular}{|c|c|c|c|c|c|c|c|c|c|} \hline
{\sf w} & {\sf d} &  $G$   &  size & output &  num &   CPU[sec]   &  
real[sec]  & Eff. & 
Rat. \\ \hline
 16 &  8 &  128 &   11M &     7M &  22  &     289 &     56 & 5.16 & 0.99 \\
 17 &  8 &  256 &   30M &    21M &  19  &     677 &    129 & 5.25 & 0.96 \\
 18 &  9 &  256 &   88M &    64M &  29  &    3071 &    517 & 5.94 & 1.11 \\
 19 &  9 &  512 &  224M &   182M &  28  &    6848 &   1206 & 5.68 & 1.00 \\
 20 & 10 &  512 &  790M &   558M &  36  &   44883 &   6834 & 6.57 & 1.42 \\
 21 & 10 & 1024 & 1766M &  1821M &  40  &   86318 &  13851 & 6.23 & 1.12 \\
 22 & 11 & 1024 & 8856M &  5927M &  46  & 1572605 & 208972 & 7.53 & 3.18 \\
\hline                                                            
\end{tabular}
\caption{
Runs on an 8-core Xeon computer at 3 GHz and with 32 Gbytes of memory.
`Num' indicates, for the final expressions, the maximum number of decimal 
digits in either a numerator or a denominator. `Eff.' is the ratio of CPU time 
versus real time indicating how well the processors are used. The meaning 
of the column labeled `Rat.' is explained in the text. The anomaly between 
size and output for ${\sf w=21}$ is due to the fact that the output is in text 
and size is in {\tt FORM} binary notation.}
\label{MZVall}
\end{table}

`Rat' is the real time of this run divided by the real time of a run with a 
31-bit prime number dropping also products of lower weight objects. 
Together with the numbers in the `num' column it shows that making several 
runs modulus a 31-bit prime and then using the Chinese remainder theorem~\cite{CRT}, 
will not be efficient. We would need at least 12 runs for the 
${\sf w = 22}$ case and even then we have to account for dropping the lower 
weight terms.

We indicate the maximum value of the depth which, due to the duality 
relation for MZVs, is sufficient to obtain all MZVs at the given weight.

The basis in which these results are presented is described in
Appendix~\ref{ap:bases}. If we let the program select the basis, the 
outputs are shorter but from the viewpoint of basis elements selected there 
is less structure.

The next sequence of runs is performed using
in modular arithmetic in which we refer to  the same 
31-bit prime number as before. Again we run the full range of depths 
needed to obtain all sums. As usual in modular runs, we drop the 
products of lower weight objects. The results are given in 
Table~\ref{MZVrun24}.

\begin{table}\centering
\begin{tabular}{|c|c|c|c|c|c|c|} \hline
{\sf  w}  &  $G$   &  size & output &   CPU[sec]   &  real[sec]  & Eff. \\ \hline
 16 &  128 &  1.7M &   1.2M &     300 &     57 & 5.25 \\ %
 17 &  256 &  5.6M &   3.2M &     713 &    134 & 5.32 \\ %
 18 &  256 & 14.4M &   7.2M &    2706 &    465 & 5.82 \\ 
 19 &  512 &   39M &   19M  &    6901 &   1206 & 5.72 \\ 
 20 &  512 &  104M &   45M  &   30097 &   4819 & 6.25 \\ 
 21 & 1024 &  239M &  114M  &   75302 &  12379 & 6.08 \\ 
 22 & 1024 &  767M &  280M  &  449202 &  65644 & 6.84 \\ 
 23 & 2048 & 2.17G &  734M  &  992431 & 151337 & 6.56 \\ 
 24 & 2048 & 8.04G & 1.77G  & 9251325 &1268247 & 7.29 \\ 
\hline
\end{tabular}
\caption{
Runs on an 8-core Xeon computer at 3 GHz and with 32 Gbytes of memory. G is 
the size of the group used in the Gaussian elimination, `size' is the maximum 
size of the master expression during the run, `output' is the size of the 
master expression in the end, CPU is the total CPU time of all processors 
together in seconds, `real' denotes the elapsed time in seconds and `Eff.' is 
the 
pseudo efficiency, defined by the CPU time divided by the real time.}
\label{MZVrun24}
\end{table}
The output of the run at ${\sf w = 23}$ gives the results for $2^{20}$ MZVs 
expressed in terms of the 28 same-weight elements of a Lyndon basis selected 
by the program.

In Table~\ref{MZVrun30} we give the statistics of runs to a more restricted 
depth. If the conjecture \cite{BK1} is correct the runs at ${\sf w = 25, 26}$ 
should 
still give us a complete basis. In the higher runs some elements will be 
missing. 

\begin{table}\centering
\begin{tabular}{|c|c|c|c|c|c|c|c|} \hline
{\sf  w}  & {\sf D} &  $G$  &  size & output &   CPU[sec]   &  real[sec]  & Eff. \\ 
\hline
 23 & 7  & 2048 & 1.55G &    89M &   61447 &   9579 & 6.41 \\
 24 & 8  & 2048 &  673M &   380M &  536921 &  72991 & 7.36 \\
 25 & 7  & 4096 & 6.37G &   244M &  369961 &  50197 & 7.37 \\
 26 & 8  & 4096 & 38.3G &  1160M & 4786841 & 651539 & 7.35 \\ 
 27 & 7  & 6144 & 12.7G &   914M & 2152321 & 277135 & 7.77 \\
 28 & 6  & 6144 & 2.88G &   314M &  235972 &  30960 & 7.62 \\
 29 & 7  & 6144 & 41.0G &  3007M & 8580364 &1112836 & 7.71 \\
 30 & 6  & 6144 & 6.27G &   658M &  829701 & 106353 & 7.80 \\
\hline
\end{tabular}
\caption{
Runs on an 8-core Xeon computer at 3 GHz and with 32 Gbytes of memory. ${\sf D}$ 
indicates the maximum depth (see text). We reran at ${\sf w = 23}$ and ${\sf 
w = 24}$ to have information for extrapolation purposes.}
\label{MZVrun30}
\end{table}
We would have liked to have a run for depth ${\sf d \leq 9}$ at ${\sf w = 
27}$, but it would probably take more than a year with current technology. 
A run for depth ${\sf d \leq 8}$ at ${\sf w = 28}$ will require a smaller CPU 
time. 
The reason why these runs are interesting is explained in 
Section~\ref{sec:pushdown} on 
pushdowns. They may give us a new type of basis element that would indicate a 
double pushdown.

The outputs of all of the above runs are collected in the data mine, together 
with 
some files in which the results have been processed to make them more 
accessible.

At the end of this Section we would like to discuss the status of the general 
investigation of MZVs and Euler sums in the foregoing literature.
The relations between MZVs were studied both by mathematicians and 
physicists. An early study is due to Gastmans and Troost \cite{GT}, which 
gave a nearly complete list for the Euler sums of {\sf w = 4} and many 
relations for {\sf w = 5}, supplemented in~\cite{HSUM3} later. Various 
authors, among them D.~Broadhurst, to {\sf w = 9}, and D.~Zagier, performed 
precision numerical studies~\cite{BZ} using {\tt PARI}~\cite{PARI} during 
the 1990's for MZVs, which were not published. A very far-reaching 
investigation concerned the study of some of the MZVs at {\sf w = 23} and 
depth {\sf d = 7} by Broadhurst by numerical techniques (PSLQ). Double sums 
were studied in~\cite{BBG} using the {\tt PSLQ} method \cite{PSLQ}.
Vermaseren both studied the MZVs and the Euler sums to {\sf w = 
9}~\cite{Vermaseren:1} using a {\tt FORM} program~\cite{FORM}. This was 
the situation around the year 2000, when the Lille group presented their 
{\sf w = 12} results for the MZVs and {\sf w = 7} results for the Euler 
sums~\cite{LILLE0}. In Ref.~\cite{LILLE5} the solution of {\sf w = 8} for 
the Euler sums is mentioned by the Lille--group. However, the data-tables 
made available~\cite{LILLE0} only contain the relations to {\sf w = 7}. 
Moreover, the relations used in~\cite{LILLE5} do not cover the doubling 
relation, which is needed to reduce to the conjectured basis at this 
weight, as will be shown later. For the MZVs {\sf w = 10} had been solved 
in~\cite{LILLE1} and {\sf w = 13} in~\cite{LILLE2}, cf.~\cite{LILLE3}. 
Vermaseren could extend the MZVs to {\sf w = 16} \cite{VER1}. Studies for 
{\sf w = 16} were also performed at Lille \cite{LILLE4} without making the 
results public. In the studies by Vermaseren also the divergent harmonic 
sums $\zeta_{1,\vec{a}}$ were included, as this is sometimes necessary for 
physics applications, cf. also~\cite{HSUM3}.

The primary goal in this paper is to derive explicit representations of the 
MZVs over several bases suitable to the respective questions investigated. 
If one only wants to determine the size of the basis one may proceed 
differently, cf.~\cite{ENR}. Here for {\sf w = 19} in the MZV 
case it was shown, that the basis has the expected length, but modulo 
powers of $\pi^2$ at even weights. In \cite{KNT} the case {\sf w = 20} was 
studied determining the size of the basis calculating the rank of the 
associated matrix modulo a 15-bit prime. Although the computation times 
are not excessive, higher weights could not be investigated  yet because of 
memory limitations. Since these methods are based on the respective algebra 
only they can be extended to colored multiple zeta values by extending the 
underlying alphabet.


\section{The Data Mine}
\renewcommand{\theequation}{\thesection.\arabic{equation}}
\setcounter{equation}{0}
\label{TheDataMine}


\vspace{1mm}
\noindent
The results of our runs, together with a number of {\tt FORM} programs to 
manipulate them and clarifying text, are available on the internet in pages 
that we call the MZV data mine. It can be located as a link in the {\tt FORM} 
home page~\cite{FORMWEB}. Here we will describe the 
notations and how to use the programs.

The notations we use in the data mine are that the MZVs are represented 
either by a function {\tt Z} of which the variables are its indices or by a 
single symbol that consists of a string of objects of which the first 
character is the letter {\tt z} and the remaining characters are decimal 
digits. 
Each of these strings refers to an index of the MZV. Let us give an example~:
\begin{verbatim}
    z11z3z3 = Z(11,3,3)
\end{verbatim}
For the Euler sums we use mostly the function {\tt H}. It can have positive 
and negative indices, the negative ones indicating alternating or Euler 
sums. When we use basis elements a compact notation is the letter {\tt h} 
followed by a number of alphabetic characters or digits. Each character 
stands for a negative index. The digits $1,\cdots,9$ stand for the indices 
$-1,\cdots,-9$ and the upper case characters $A,\cdots,Z$ stand for the 
indices $-10,\cdots,-35$. We had no need to go further in this notation. 
The next example should illustrate this:
\begin{verbatim}
    hL33 = H(-21,-3,-3).
\end{verbatim}
If there is ever any doubt about which variable indicates which object one 
can look in the corresponding library file (always included as a file with 
the extension {\tt .h} in the directory in which the integrals reside) in 
the procedure `frombasis'.

For reasons of economy\footnote{It turns out that the number of digits in 
the fractions is somewhat smaller in $\eta$-notation than in 
$\zeta$-notation.} the {\tt H}-functions with a single negative index have 
a different notation. They are related to the constants $\eta_k$ defined by
\begin{equation}
\label{eq:eta}
\eta_k = \left(1 - \frac{1}{2^{k-1}}\right) \zeta_k~.
\end{equation}
In the program we call these constants \verb:e3,e5,...:.

In some cases we use a variable with a notation similar to the notation for 
the MZVs, except for that the character {\tt z} is replaced by the character 
{\tt a}. 
\begin{verbatim}
    aiajak = A(i,j,k)
           = Z(i,j,k)+Z(-i,j,-k)+Z(i,-j,-k)+Z(-i,-j,k)
           = H(i,j,k)-H(-i,j, k)-H(i,-j, k)+H(-i,-j,k)
\end{verbatim}
Here $A$ is the function defined in (\ref{eq:AFUN}).

In exceptional cases we refer to $Z$-functions with negative indices. The 
most common notation for this in the literature is to put a bar over the 
number. This is however a notation that cannot be used in programs like 
{\tt FORM}. Hence we use negative indices for the alternating sums there. 
For the symbolic variables we use the notation for the MZVs but with the 
character {\tt m} between {\tt z} and the number~:
\begin{verbatim}
    zm11zm3z3 = Z(-11,-3,3) = -H(-11,3,3)
\end{verbatim}

The programs run in what we call integral notation. This means that the 
master expression has the index fields of the functions {\tt E, H} and {\tt 
HH}\footnote{The function {\tt HH} is the same as the function {\tt H}. We 
need two different names because when we present the results the function 
{\tt H} marks the brackets and the function {\tt HH} marks the remaining 
basis elements.} in terms of the three letter alphabet $\{0,1,-1\}$ for 
Euler sums and the two letter alphabet $\{0,1\}$ for MZVs. This is then the 
way the outputs are presented. Actually, internally the whole string of 
indices is put together as one large ternary number for Euler sums and one 
large binary number for MZVs. This speeds up the calculation, but makes it 
virtually impossible to interpret intermediate results.

The outputs are presented in a method that one may consider unusual. In 
{\tt FORM} it is often more efficient to have one big expression, rather than 
$2^{20}$ expressions as would be the case for the MZVs at ${\sf w=23}$. Hence 
the output contains functions {\tt H} with the indices of the corresponding 
MZV 
and each {\tt H} is multiplied by what this MZV is equal to. In the case that 
we 
fixed a basis this can be an expression that consists of symbols like we 
defined above. In the case that we did the calculus modulus a prime and 
only wanted to determine a basis, it will be an expression that consists of 
terms that each contain a single function {\tt HH} with its indices in 
integral 
notation. These {\tt HH} functions form the basis. Often at the end of the 
program there is a list of the {\tt HH} functions used. Because {\tt FORM} 
will print 
the output in such a way that the functions {\tt H} are taken outside 
brackets, 
the contents of each bracket are what each {\tt H} function is equal to. With a 
decent editor it takes very few ($\leq 4$) edit commands to convert such 
output into the definition of $2^{20}$ table elements.

If this output should be used 
as input for other systems, this 
can be done, provided that the expressions do not cause memory problems. 
The format is in principle compatible with {\tt  Pari/GP, Reduce} and 
{\tt Maple}. There 
may be a problem with large coefficients. {\tt FORM} does not like to make output 
lines that are longer than a typical screen width. Hence they are usually 
broken up after some 75 characters. This holds also for long numbers. These 
are broken off by a backslash character and continued on the next line. The 
problem is usually that {\tt FORM} places some white space at the beginning of 
the line and some programs may have problems with that. Hence one can use 
an editor to remove all white space (blanks and tabs) at the beginning of 
the lines.

The data mine consists of several parts. The main part is formed by the 
different data sets. The remainder files give information about how to use 
the data mine and links to other useful information and/or programs. The 
data are divided over a number of directories, each containing the results 
of one type of runs for a range of values of the weight. In each directory 
there are several types of files again. The log--files of the runs are stored. 
These contain the run time statistics and the output of the runs in text 
format. Then there are the table files. They are in text format and contain 
table definitions for FORM programs. Their extension is {\tt .prc} as in 
{\tt mzv21.prc}. Some of these files have been split into several files 
because they become much to big to be handled conveniently. These tables can 
be 
read and compiled. Yet the case of the MZVs at ${\sf w=22}$ with its nearly 
6 Gbytes can be too large for a system with `only' 16 Gbytes. If one does 
not have a bigger machine to ones disposal, one should use either the 
binary {\tt .sav} file or the {\tt .tbl} file defined below.

The third type of files are the binary {\tt .sav} files. They can be used 
to read in the complete tables without having to go through the compiler 
and without having to load the complete table as table elements (which 
needs also big compiler buffers). Finally we have created so-called 
tablebases which allow very fast access to individual elements. A tablebase 
is a type of database for large tables. They are particular to {\tt FORM} 
and have been used with great success in a number of very large 
calculations. Their working is explained in Ref.~\cite{tablebase} and the 
{\tt FORM} manual. The tablebase files have traditionally the {\tt .tbl} 
extension.

In each directory we have also the programs that were used to create the 
various files and in some cases some example programs.

There is another section in the data mine that contains pages in which it 
is explained how to manipulate the information in the files. Although many 
files are in text format it is not easy to manipulate a 4 Gbyte text file 
and hence it might become necessary to either use {\tt FORM} and one of the 
binary files, or to use the {\tt STedi} editor which has been used to 
manipulate 
these files on a computer with 16 Gbytes of main memory. Links are set 
to these programs.
{\tt FORM} programs are provided for the most common manipulations of the data. 
They contain much commentary. This should make it easy for the user to 
customize the programs should the need arise.
The data mine is located at \hfill \\ 
{\tt http://www.nikhef.nl/$\sim$form/datamine/datamine.html}.
Its structure is given in Figure~2 :
\begin{center}
\begin{picture}(400,370)(0,0)
\SetPFont{Helvetica}{12}
\SetColor{Red}
\Line(20,150)(80,250)
\Line(20,150)(80,100)
\Line(20,150)(80,40)
\Line(20,150)(62,10)
\Line(100,250)(150,315)
\Line(100,250)(150,200)
\Line(170,200)(220,200)
\Line(240,200)(290,230)
\Line(240,200)(290,200)
\Line(240,200)(290,170)
\Line(170,315)(220,360)
\Line(170,315)(220,330)
\Line(170,315)(220,300)
\Line(170,315)(220,270)
\Line(100,100)(150,130)
\Line(100,100)(150,100)
\Line(100,100)(150,70)
              \CText(230,360){Black}{Yellow}{depth 3}
                        \CText(370,360){Blue}{Yellow}{W <= 29}
              \CText(230,330){Black}{Yellow}{depth 4}
                        \CText(370,330){Blue}{Yellow}{W <= 22}
         \CText(160,315){Black}{Yellow}{rational}
              \CText(230,300){Black}{Yellow}{depth 5}
                        \CText(370,300){Blue}{Yellow}{W <= 17}
              \CText(230,270){Black}{Yellow}{alldepth}
                        \CText(370,270){Blue}{Yellow}{W <= 12}
    \CText(90,250){Black}{Yellow}{Euler}
         \CText(160,200){Black}{Yellow}{modular}
                   \CText(300,230){Black}{Yellow}{depth 4}
                        \CText(370,230){Blue}{Yellow}{W <= 30}
              \CText(230,200){Black}{Yellow}{basis}
                   \CText(300,200){Black}{Yellow}{depth 5}
                        \CText(370,200){Blue}{Yellow}{W <= 21}
                   \CText(300,170){Black}{Yellow}{depth 6}
                        \CText(370,170){Blue}{Yellow}{W <= 18}
\CText(20,150){Black}{Yellow}{datamine}
         \CText(160,130){Black}{Yellow}{complete}
                        \CText(370,130){Blue}{Yellow}{W <= 22}
         \CText(160,100){Black}{Yellow}{modular}
                        \CText(370,100){Blue}{Yellow}{W <= 24}
    \CText(90,100){Black}{Yellow}{MZV}
         \CText(160,70){Black}{Yellow}{limited}
                        \CText(370,70){Blue}{Yellow}{W <= 30}
    \CText(90,40){Black}{Yellow}{programs}
    \CText(90,10){Black}{Yellow}{other things}
\end{picture}
\\
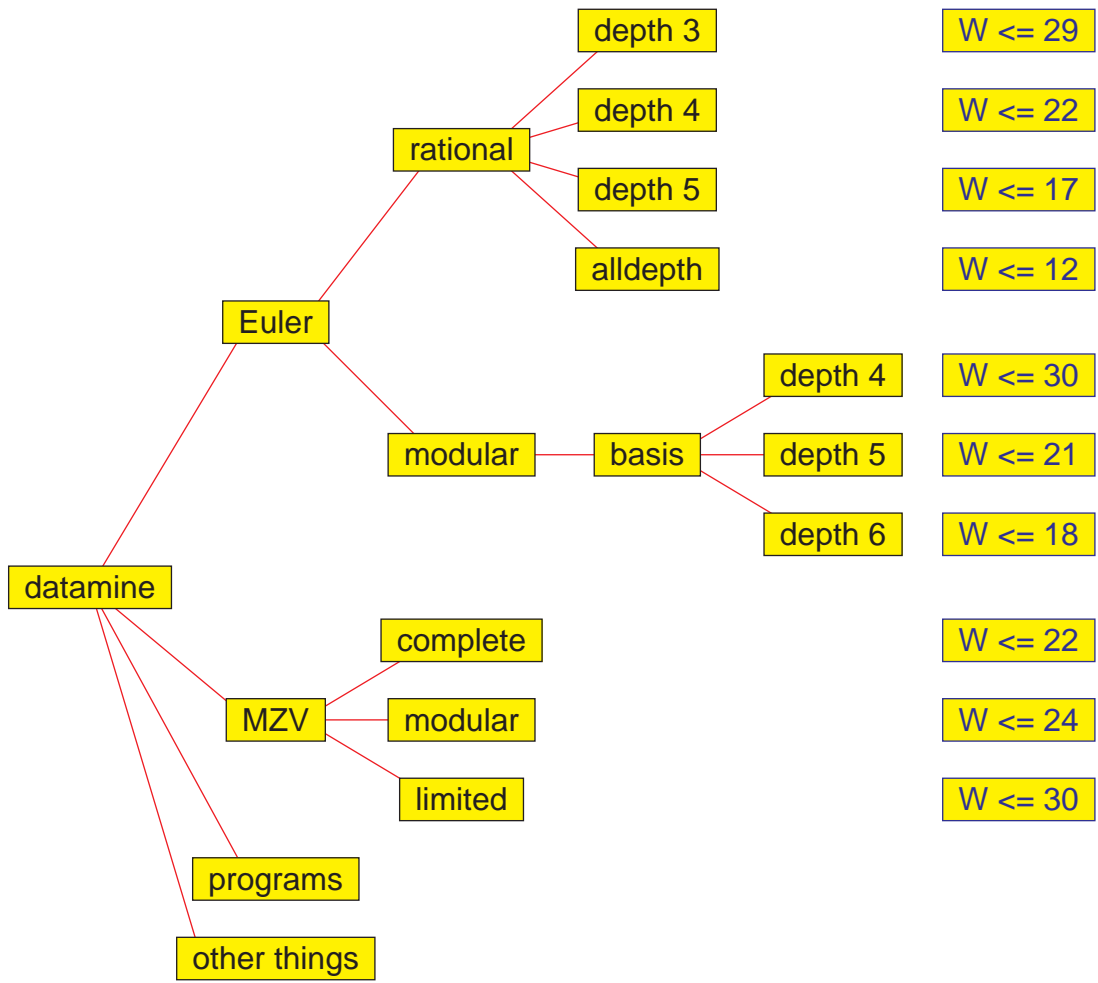
{Figure 2: Layout of the data part of the data mine.}
\end{center}

\noindent
In this figure we use the following names:
\begin{center}
\begin{tabular}{ll}
complete  & Complete expressions over the rational numbers. \\
modular   & Products of lower weight terms are dropped and the 
computation is \\ &
performed modulus a large prime. \\
limited   & As modular but incomplete bases. \\
rational  & Complete expressions over the rational numbers. \\
other things & Conventions, publications, help, links, etc. \\
\end{tabular}
\end{center}
The main problem with the data mine is its size. Many files are several 
Gbytes long. We have used {\tt bzip2} on most files, because it gives a better 
compression ratio than {\tt gzip}, even though it is much slower, both in 
compressing and decompressing. But even with {\tt bzip2} the combined files 
are larger than 30 Gbytes.

All programs are {\tt FORM} (or {\tt TFORM}) codes. They will run with 
the latest 
versions of {\tt FORM} (or {\tt TFORM}). The executables of {\tt FORM} can be obtained from 
the {\tt FORM} web site: {\tt http://www.nikhef.nl/$\sim$form}. Please 
remember the license condition: if you use {\tt FORM} (or {\tt TFORM}) for a 
publication, you 
should refer to Ref.~\cite{FORM}.


\section{FORM Aspects}
\renewcommand{\theequation}{\thesection.\arabic{equation}}
\setcounter{equation}{0}
\label{FORMaspects}

\vspace{1mm}
\noindent
As mentioned the running of the programs used posed great challenges for {\tt 
FORM} 
and {\tt TFORM}. This is not simply a matter of whether the system contains 
errors. It is much more a matter of whether the system deals with the 
problem in a sensible and efficient way. Where are the bottlenecks? What is 
inefficient? A clear example is the conversion between sum notation and 
integral notation. This can be programmed in one line:
\begin{verbatim}
    repeat id H(?a,n?!{-1,0,1},?b) = H(?a,0,n-sig_(n),?b);
\end{verbatim}
for going to integral notation and
\begin{verbatim}
    repeat id H(?a,0,n?!{0,0},?b) = H(?a,n+sig_(n),?b);
\end{verbatim}
for going to sum notation. It turns out that when one goes to large weights 
(for instance more than 20), this becomes very slow because it involves 
very much pattern matching. Considering also that the use of harmonic sums 
is becoming more and more popular it was decided to built two 
new commands in {\tt FORM} for this transformation:
\begin{verbatim}
    ArgImplode,H;
    ArgExplode,H;
\end{verbatim}
The first one converts {\tt H} to sum notation and the second one to integral 
notation. This made the program noticeably faster and easier to read.

Another addition to {\tt FORM} concerns  built-in shuffle and stuffle 
commands. One of 
the problems with shuffles is that the simple programming of it usually 
gives many identical terms. This means that the shuffle product of two 
MZVs can become very slow, which is illustrated by the following little 
program:
\begin{verbatim}
    S   n1,n2;
    CF  H,HH;
    L   F = H(3,5,3)*H(6,2,5);
    ArgExplode,H;
    Multiply HH;
    repeat;
      id  HH(?a)*H(n1?,?b)*H(n2?,?c) =
                +HH(?a,n1)*H(?b)*H(n2,?c)
                +HH(?a,n2)*H(n1,?b)*H(?c);
    endrepeat;
    id  HH(?a)*H(?b)*H(?c) = H(?a,?b,?c);
    .end

Time =      37.38 sec    Generated terms =    2496144
               F         Terms in output =       2146
                         Bytes used      =      63176
\end{verbatim}
By putting much combinatorics in the built-in 
shuffle statement we could solve most of these problems (although not all 
as the combinatorics can become very complicated). With the shuffle command 
the program becomes:
\begin{verbatim}
    S   n1,n2;
    CF  H,HH;
    L   F = H(3,5,3)*H(6,2,5);
    ArgExplode,H;
    Shuffle,H;
    .end

Time =       0.01 sec    Generated terms =       5163
               F         Terms in output =       2146
                         Bytes used      =      63176
\end{verbatim}
This is a great improvement of course.

For the stuffle product things are much easier. There we have the 
complication that there are two definitions. One is the product used for 
the $Z$-sums and the other is the product used for the $S$-sums. We have 
resolved that by appending a + for the $Z$-notation and a - for the 
$S$-notation:
\begin{verbatim}
    stuffle,Z+;
    stuffle,S-;
\end{verbatim}
Not only did this make the program significantly faster, it also made it 
more readable.

This way the stuffle product of two Euler sums in integral notation becomes 
in principle (assuming that we are in integral notation):
\begin{verbatim}
    ArgImplode,H;
    #call convertHtoZ(H,Z)
    Stuffle,Z+;
    #call convertZtoH(Z,H)
    ArgExplode,H;
\end{verbatim}
except for that in the actual program we substituted the contents of the 
two conversion procedures. Of course for MZVs the conversions are not 
needed and we can use just:
\begin{verbatim}
    ArgImplode,H;
    Stuffle,H+;
    ArgExplode,H;
\end{verbatim}

A third improvement concerns the parallelization. The original 
parallelization of {\tt TFORM}~\cite{Vermaseren:2} assumed the treatment of 
a single large expression of which the terms are distributed over the 
workers and later gathered in by the master. During the phase in which we 
execute a Gaussian elimination inside a group of identities, this is very 
inefficient, because we deal with many small expressions, each giving a 
certain amount of overhead when they are distributed over the worker 
threads. Hence it was decided to create a new form of parallelization in 
which the user tells the program that there are many small expressions 
coming. The reaction of the master thread is now to divide the expressions 
over the workers. It only has to tell each worker which expression to do 
next, after which the worker is responsible for obtaining its input and 
writing its output. The only remaining inefficiencies are that the writing 
of the output causes a traffic jam because that has to be done 
sequentially. The final results are kept in principle in a single file 
or its cached version. Additionally, there may be some load balancing 
problem in the end. This load balancing becomes rapidly less when the size 
of the groups of equations that is treated becomes bigger. The running of 
this phase of the program can give nearly ideal efficiencies.

A fourth improvement concerns the fact that very lengthy programs run a 
risk of discontinuity. This could be a power failure or a sudden urge of 
the service department to `update' the system, etc. For this a facility has 
been implemented inside {\tt FORM} that allows one to 
make `snapshots' of the current internal state, cf.~\cite{FORMsnapshot}.
At a later moment one can 
then restart from the point of the snapshot. The completion of this 
facility came however too late to have a practical impact for this paper.

The possibility to perform the calculus modulus a prime number has existed 
in {\tt FORM} since its first version. Much of it remained untested because these 
facilities had not been used extensively. It turned out to be necessary to 
redesign parts of it and add a few new features. 

Other aspects of {\tt TFORM} performed amazingly well. We have seen the 
program running with eight workers who all eight had to enter the fourth 
stage of the sorting simultaneously. This is rather rare even for single 
threads and only happens for very large expressions. It gives a bit of a 
slow down due to the great amount of disk accesses, but it all worked 
without any problems. The most impressive single module result 
observed was
\begin{verbatim}
    Time =   15720.03 sec    Generated terms =1202653196013
                  FF         Terms in output =   1508447974
     substitution(7-sh)-7621 Bytes used      =  36215474400
\end{verbatim}
The execution time is that of the master. Actually the master spent 1000 
CPU sec on this step and the eight workers each almost 200000 CPU sec.

One may wonder about the probability that calculations, done with a 
system under development, give correct answers. We have several remarks 
concerning this topic:
\begin{itemize}
\item Whenever {\tt FORM} failed, it was always in a very obvious way, like 
crashing because it couldn't interpret something.
\item The full all-depth outputs from the MZVs up to ${\sf w=22}$ and the 
Euler 
sums up to ${\sf w=12}$ have been tested numerically by completely 
independent 
programs, run under {\tt PARI-GP} \cite{PARI}.
\item Because of both {\tt TFORM} and the MZV programs being under development 
many programs have been run at least several times with different 
configurations and/or different orders of solving the equations.
\item {\tt TFORM} operates in a rather non-deterministic fashion. Terms are 
rarely distributed twice in the same way over the workers because the 
master serves the workers when they have finished a task and this is 
usually not in the same order. In the case of errors this would lead to 
different results in different runs.
\item There are effects that are expected on the basis of extrapolation, 
like the pushdowns and the construction of a basis. If anything goes 
wrong, such effects are absent.
\item If for instance a term gets lost in a calculation over the 
rational numbers, usually the output would have terms with fractions that are 
abnormally much more complicated than the others. This is due to the fact 
that in intermediate stages the coefficients are usually much more 
complicated than at the end. Such terms are spotted relatively easily.
\end{itemize}

\section{Results}
\renewcommand{\theequation}{\thesection.\arabic{equation}}
\setcounter{equation}{0}
\label{FirstResults}

\vspace{1mm}
\noindent
Armed with the vast amount of information contained in the data mine we 
start with having a look at a number of conjectures in this this field. 
They concern the number of basis elements, either 
just as a function of the weight or as a function of weight and depth.
We first check some conjectures made in the literature using the data mine and 
then describe the selection of the basis to represent the Euler sums and MZVs
in the data mine. 

\subsection{Checking some Conjectures with the Data Mine}

\vspace*{2mm}\noindent
{\bf Zagier conjecture \cite{ZAG1}:}\\
The number of elements in a Lyndon-basis for the MZVs at weight ${\sf w}$ is 
given by Eq.~(\ref{eq-WI1b}). $\Box$ \\
As far as we can check, the Zagier conjecture holds to weight 22. Assuming 
that in the modular calculus no terms were lost due to spurious zeroes, we 
can say that it holds to weight 24. With the additional assumption 
that all (Lyndon) basis elements have a depth of at most one third of the 
weight we can even say that it holds to weight 26. If we combine the 
findings in the thesis of Racinet~\cite{Racinet} that there may be 2 basis 
elements of depth 
9 for weight 27 with our runs to depth 7, the Zagier conjecture holds also 
at weight 27. This conjecture is in accordance with 
the  upper bound for the size of the basis being derived in Refs.~\cite{W1}. 

\vspace*{2mm}\noindent
{\bf Hoffman conjecture \cite{HOFC}:}\\
A Fibonacci-basis for the MZVs at a given weight ${\sf w}$ is formed out of 
MZVs the index set of which is formed out of all words over the 
alphabet $\{2,3\}$. $\Box$ \\
We could test the basis conjectured by Hoffman  up to weight ${\sf w=22}$. 
If we take the sub-variety in which we only look at the Lyndon words made 
from the indices 2 and 3, we can even verify this Lyndon basis to weight 
24. Because this basis is not centered around the concept of depth, we 
cannot use the partial runs at larger weights and limited depths for 
further validation.

\vspace*{2mm}\noindent
{\bf Broadhurst conjecture \cite{Broadhurst:1}:}\\
The number of basis elements of the
Euler sums  at fixed weight ${\sf w}$ and depth ${\sf d}$ 
is given by Eq.~(\ref{Enk}). $\Box$\\
All our runs for Euler sums are in complete agreement with the Broadhurst 
conjecture about the size and the form of a basis for these sums. This 
means complete verification up to weight 12, for depth 6 verification (in 
modular arithmetic) to weight 18, for depth 5 complete verification to 
weight 17 and modular verification to weight 21. For depth 4 these numbers 
are weight 22 and weight 30 respectively.

\vspace*{2mm}\noindent
{\bf Broadhurst-Kreimer conjectures \cite{BK1}:}\\
The number of basis elements of the
MZVs at fixed weight ${\sf w}$ and depth ${\sf d}$ 
is given by Eq.~(\ref{conj2}). The number of basis elements for MZVs when 
expressed in terms of Euler sums in a minimal depth representation is given 
by Eq.~(\ref{eq:BK2}) $\Box$\\
The runs for the MZVs confirm this conjecture over a large range, cf. 
Tables~\ref{BK1table}, \ref{BK2table}. The second part of the conjecture 
is harder to check than the first part, 
because for this we need the results for the corresponding Euler sums.

\vspace*{2mm}\noindent
{\bf Another conjecture by Hoffman \cite{Hof1}:}

\begin{eqnarray}
H_{2,1,2,3}-H_{2,2,2,2}-2 H_{2,3,3} &=& 0 \\
H_{2,1,2,2,3}-H_{2,2,2,2,2}-2 H_{2,2,3,3} &=& 0 \\
H_{2,1,2,2,2,3}-H_{2,2,2,2,2,2}-2 H_{2,2,2,3,3} &=& 0 \\
H_{2,1,2,2,2,2,3}-H_{2,2,2,2,2,2,2}-2 H_{2,2,2,2,3,3} &=& 0 \\
H_{2,1,\{2\}_k,3}-H_{\{2\}_{k+3}}- 2 H_{\{2\}_k,3,3} &=& 0
~~~~~~\Box 
\end{eqnarray}
We verified these relations up to weight ${\sf w=22}$. At ${\sf w=24}$  
we checked the weight-24 part, since we have only the modular representation 
at this level. \vspace{2mm}


\begin{table}\centering
\begin{tabular}{|c|c|c|c|c|c|c|c|c|c|c|}
\hline
  {\sf w /d}   &  1   &  2   &  3   &  4   &  5   &  6   &  7   &  8   &  9   
& 10 \\ \hline
   1    &      &      &      &      &      &      &      &      &      &    \\
   2    &\ul(1)&      &      &      &      &      &      &      &      &    \\
   3    &\ul(1)&      &      &      &      &      &      &      &      &    \\
   4    &      &      &      &      &      &      &      &      &      &    \\
   5    &\ul(1)&      &      &      &      &      &      &      &      &    \\
   6    &      &\ul(0)&      &      &      &      &      &      &      &    \\
   7    &\ul(1)&      &      &      &      &      &      &      &      &    \\
   8    &      &\ul(1)&      &      &      &      &      &      &      &    \\
   9    &\ul(1)&      &\ul(0)&      &      &      &      &      &      &    \\
  10    &      &\ul(1)&      &      &      &      &      &      &      &    \\
  11    &\ul(1)&      &\ul(1)&      &      &      &      &      &      &    \\
  12    &      &\ul(1)&      &\ul(1)&      &      &      &      &      &    \\
  13    &\ul(1)&      &\ul(2)&      &      &      &      &      &      &    \\
  14    &      &\ul(2)&      &\ul(1)&      &      &      &      &      &    \\
  15    &\ul(1)&      &\ul(2)&      &\ul(1)&      &      &      &      &    \\
  16    &      &\ul(2)&      &\ul(3)&      &      &      &      &      &    \\
  17    &\ul(1)&      &\ul(4)&      &\ul(2)&      &      &      &      &    \\
  18    &      &\ul(2)&      &\ul(5)&      &\ul(1)&      &      &      &    \\
  19    &\ul(1)&      &\ul(5)&      &\ul(5)&      &      &      &      &    \\
  20    &      &\ul(3)&      &\ul(7)&      &\ul(3)&      &      &      &    \\
  21    &\ul(1)&      &\ul(6)&      &\ul(9)&      &\ul(1)&      &      &    \\
  22    &      &\ul(3)&      &\ul(11)&     &\ul(7)&      &      &      &    \\
  23    &\ul(1)&      &\ul(8)&      &\ul(15)&     &\ul(4)&      &      &    \\
  24    &      &\ul(3)&      &\ul(16)&     &\ul(14)&     &\ul(1)&      &    \\
  25    &\ul(1)&      &\ul(10)&     &\ul(23)&     &\ul(11)&     &      &    \\
  26    &      &\ul(4)&      &\ul(20)&     &\ul(27)&     &\ul(5)&      &    \\
  27    &\ul(1)&      &\ul(11)&     &\ul(36)&     &\ul(23)&     &  2   &    \\
  28    &      &\ul(4)&      &\ul(27)&     &\ul(45)&     & 16   &      &    \\
  29    &\ul(1)&      &\ul(14)&     &\ul(50)&     &\ul(48)&     &  7   &    \\
  30    &      &\ul(4)&      &\ul(35)&     &\ul(73)&     & 37   &      &  2 \\
\hline
\end{tabular}
\caption{Number of basis elements for MZVs as a function of weight and depth
in a minimal depth representation. Underlined are the values we have 
verified with our programs.}
\label{BK1table}
\end{table}
\begin{table}\centering
\begin{tabular}{|c|c|c|c|c|c|c|c|c|c|c|}
\hline
  {\sf w/d }  &  1   &  2   &  3   &  4   &  5   &  6   &  7   &  8   &  9   & 
10 \\ \hline
   1    &      &      &      &      &      &      &      &      &      &    \\
   2    &\ul(1)&      &      &      &      &      &      &      &      &    \\
   3    &\ul(1)&      &      &      &      &      &      &      &      &    \\
   4    &      &      &      &      &      &      &      &      &      &    \\
   5    &\ul(1)&      &      &      &      &      &      &      &      &    \\
   6    &      &      &      &      &      &      &      &      &      &    \\
   7    &\ul(1)&      &      &      &      &      &      &      &      &    \\
   8    &      &\ul(1)&      &      &      &      &      &      &      &    \\
   9    &\ul(1)&      &      &      &      &      &      &      &      &    \\
  10    &      &\ul(1)&      &      &      &      &      &      &      &    \\
  11    &\ul(1)&      &\ul(1)&      &      &      &      &      &      &    \\
  12    &      &\ul(2)&      &      &      &      &      &      &      &    \\
  13    &\ul(1)&      &\ul(2)&      &      &      &      &      &      &    \\
  14    &      &\ul(2)&      &\ul(1)&      &      &      &      &      &    \\
  15    &\ul(1)&      &\ul(3)&      &      &      &      &      &      &    \\
  16    &      &\ul(3)&      &\ul(2)&      &      &      &      &      &    \\
  17    &\ul(1)&      &\ul(5)&      &\ul(1)&      &      &      &      &    \\
  18    &      &\ul(3)&      &\ul(5)&      &      &      &      &      &    \\
  19    &\ul(1)&      &\ul(7)&      &  3   &      &      &      &      &    \\
  20    &      &\ul(4)&      &  8   &      &  1   &      &      &      &    \\
  21    &\ul(1)&      &\ul(9)&      &  7   &      &      &      &      &    \\
  22    &      &\ul(4)&      & 14   &      &  3   &      &      &      &    \\
  23    &\ul(1)&      & 12   &      & 14   &      &  1   &      &      &    \\
  24    &      &\ul(5)&      & 20   &      &  9   &      &      &      &    \\
  25    &\ul(1)&      & 15   &      & 25   &      &  4   &      &      &    \\
  26    &      &\ul(5)&      & 30   &      & 20   &      &  1   &      &    \\
  27    &\ul(1)&      & 18   &      & 42   &      & 12   &      &      &    \\
  28    &      &\ul(6)&      & 40   &      & 42   &      &  4   &      &    \\
  29    &\ul(1)&      & 22   &      & 66   &      & 30   &      &  1   &    \\
  30    &      &  6   &      & 55   &      & 75   &      & 15   &      &    \\
\hline
\end{tabular}
\caption{Number of basis elements for MZVs as a function of weight and depth
when expressed as Euler sums in a minimal depth representation. Underlined 
are the values we have verified with our programs.}
\label{BK2table}
\end{table}



There are identities for special patterns of indices as
\begin{equation}
2 \zeta_{m,1} = m \zeta_{m+1} - \sum_{k=1}^{m-2} \zeta_{m-k} \zeta_{k+1},
~~2 \leq m~\in~{\bf Z}~, 
\end{equation}
cf.~\cite{Euler,OLD} or
\begin{equation}
\zeta_{\{3,1\}_{n}} = \frac{1}{2n+1} 
\zeta_{\{2\}_{n}} = \frac{1}{4^n} 
\zeta_{\{4\}_n} = \frac{2 \pi^{4n}}{(4n+2)!}~,
\end{equation}
conjectured in \cite{ZAG1} and proven in \cite{BBBL}. Another relation is
\begin{eqnarray}
\zeta_{2,\{1,3\}_{n}} &=&  
\frac{1}{4^n} \sum_{k=0}^n (-1)^k \zeta_{\{4\}_{n-k}}
\left\{(4k+1) \zeta_{4k+2}
- 4 \sum_{j=1}^k \zeta_{4j-1} \zeta_{4k-4j+3}\right\}
\nonumber\\
\end{eqnarray}
conjectured in \cite{BBB} and proven in \cite{BB}. For the Euler sums one 
finds, \cite{ZHAO2},
\begin{eqnarray}
\zeta_{\{3\}_n} = 8^n \zeta_{\{-2,1\}_n}~.
\end{eqnarray}
In Ref.~\cite{BBB} conjectures were given for special cases based on {\tt 
PSLQ},
\begin{eqnarray}
\label{rel1}
\zeta_{\{4,1,1\}_2} &=& \frac{3 \pi^4}{16} 
\left[\zeta_{6,2}-4\zeta_5\zeta_3\right] - 
\frac{41\pi^6}{5040}\left[\zeta_3^2 - \frac{77023 \pi^6}{14414400}\right]
+ \frac{397}{8} \zeta_9 \zeta_3 + \zeta^4_3
\nonumber\\ \\ 
\label{rel2}
\zeta_{2,2,1,2,3,2} &=& \frac{75 \pi^2}{32} \left[\zeta_{8,2} - 2 
\zeta_7 \zeta_3 + \frac{34}{225} \zeta_5^2 + \frac{4528801 
\pi^{10}}{61297236000}\right] - \frac{825}{8} \zeta_7 \zeta_5~,
\end{eqnarray}
which we verified.
A series of special relations for the Euler sums were conjectured in 
\cite{BBB} based on {\tt PSLQ}~:
\begin{eqnarray}
\label{rel3}
\zeta_{2,1,-2,-2} &=& \frac{39}{128} \zeta_4 \zeta_3 - 
\frac{193}{64} \zeta_5 \zeta_2 + \frac{593}{128} \zeta_7 \\
\label{rel4}
\zeta_{-2,-2,1,2} &=& \frac{9}{128} \zeta_4 \zeta_3 + \frac{447}{128} \zeta_5 
\zeta_2 - \frac{1537}{256} \zeta_7 
 \\
\zeta_{\{-3,1\}_2} &=& -7 \left[\alpha_5 - \frac{39}{64} \zeta_5 + \frac{1}{8} 
\zeta_4 \ln(2) \right] \zeta_3 + \left[2 \alpha_4 - \frac{1}{4} 
\zeta_4\right]^2 \nonumber\\ && + 2 \left[\alpha_4 - \frac{15}{16} \zeta_4 +
\frac{7}{8} \zeta_3 \ln(2) \right]^2 -\frac{1}{32} \zeta_8~. 
\end{eqnarray}
Here 
\begin{eqnarray}
\alpha_n = \Li_n(1/2) + (-1)^n \left[
\frac{\ln^n(2)}{n!} 
- \frac{\zeta_2}{2} 
\frac{\ln^{(n-2)}(2)}{(n-2)!} \right]~. 
\end{eqnarray}
These relations are verified analytically as well by our data base.
Relations (\ref{rel1}--\ref{rel4}) were also obtained in \cite{LILLE0}. 

In Ref.~\cite{Broadhurst:1} a series of relations was conjectured for weight {\sf w = 8 
...12} and {\sf d = 3,4} for Euler sums being related to values 
$\zeta_{-|a_1|,-|a_2|}$. 
\begin{eqnarray}
\zeta_{3,-3,-3} &=& 6 \zeta_{5,-1,-3} +6 \zeta_{3,-1,-5}
-\frac{315}{32} \ln(2) \zeta_3 \zeta_5
+6 \zeta_{-5,-1} \zeta_3 \nonumber
\end{eqnarray} \begin{eqnarray}
&+&\frac{40005}{128} \zeta_2 \zeta_7
-\frac{39}{64} \zeta^3_3
+\frac{1993}{256} \zeta_3 \zeta_6
+\frac{8295}{128} \zeta_4 \zeta_5
-\frac{226369}{384} \zeta_9,
\nonumber\\
\\
\zeta_{3,-5,-3} &=&
 \frac{1059}{80}\zeta_{5,3,3}
+15\zeta_{7,-1,-3}
+15\zeta_{3,-1,-7}
+\frac{701}{69} \zeta_{-5,-3} \zeta_3
\nonumber\\
&+&15 \zeta_{-7,-1} \zeta_3
-\frac{6615}{256} \ln(2) \zeta_3 \zeta_7
-\frac{11852967}{2560} \zeta_{11}
+\frac{301599}{128} \zeta_2 \zeta_9 \nonumber\\
&-&\frac{124943}{5888}\zeta^2_3 \zeta_5
+\frac{1753577}{35328} \zeta_3 \zeta_8
+\frac{2960103}{5120} \zeta_4 \zeta_7
+\frac{3405}{32} \zeta_5 \zeta_6,
\nonumber\\
\\
\zeta_{3,-1,3,-1}&=&
\frac{61}{27}\,\zeta_{-3,-3,-1,-1}
-\frac{14}{3}\,\zeta_{-5,-1,-1,-1}
-\frac{185}{27} \zeta_{-5,-1} \zeta_2
\nonumber\\
&-&\frac{163499}{22356} \zeta_{-5,-3}
+\frac{2051}{54} \zeta_{-7,-1}
+\frac{28}{9}\ln^2(2) \zeta_{-5,-1}
+\frac{35}{96} \ln^2(2) \zeta^2_3
\nonumber\\
&-&\frac{581}{64} \ln^2(2)\zeta_6
-\frac{8735}{576} \ln(2) \zeta_2 \zeta_5
-\frac{903}{64} \ln(2) \zeta_3 \zeta_4
\nonumber\\
&-&\frac{1441}{288} \zeta_2 \zeta^2_3
+\frac{10365875}{476928} \zeta_3 \zeta_5
+\frac{36916435}{1907712}\zeta_8.
\label{cnot}
\\
2^5\cdot 3^3 \zeta_{4,4,2,2} &=& 2^5 \cdot 3^2\zeta_3^4
+2^6 \cdot 3^3 \cdot 5 \cdot 13 \zeta_9 \zeta_3
+2^6 \cdot 3^3 \cdot 7 \cdot 13 \zeta_7 \zeta_5 
\nonumber\\ &&
+2^7 \cdot 3^5\zeta_7 \zeta_3 \zeta_2
+2^6 \cdot 3^5 \zeta^2_5 \zeta_2
-2^6 \cdot 3^3 \cdot 5 \cdot 7 \zeta_5 \zeta_4 \zeta_3
\nonumber\\ &&
-2^8 \cdot 3^2 \zeta_6 \zeta^2_3
-\frac{13177 \cdot 15991}{691} \zeta_{12}
\nonumber\\ &&
+2^4 \cdot 3^3 \cdot 5 \cdot 7 \zeta_{6,2} \zeta_4
-2^7 \cdot 3^3 \zeta_{8,2} \zeta_2
-2^6 \cdot 3^2 \cdot 11^2 \zeta_{10,2}
\nonumber\\ &&
+2^{14} \zeta_{-9,-3} \label{zu93}~.
\end{eqnarray}
These relations were verified using the current data base. Eq.~(\ref{zu93}) is 
particularly interesting since it implies a relation between 
MZVs mediated by one term of the kind $\zeta_{-|a_1|,-|a_2|}$.

There is a series of Theorems proven on the MZVs, which can be verified using 
the data base. We used already the duality theorem~\cite{ZAG1}.
For the MZVs a large variety of relations has been 
proven, which can be verified for specific examples using the data mine. 

The first of these general relations
is the {\bf Sum Theorem},~Ref.~\cite{Euler,SUMTH},
\begin{equation}
\sum_{i_1 + \ldots +i_k =n, i_1 > 1} \zeta_{i_1, \ldots, i_k} = \zeta_n~.	
\end{equation}
The sum-theorem was conjectured in \cite{HOF3}, cf.~\cite{HOF2}. For its derivation 
using the Euler connection formula for polylogarithms, cf.~\cite{OU}.

Further identities are given by the {\bf Derivation Theorem}, \cite{HOF3,HOF4}
Let $I=(i_1, \ldots, i_k)$ any sequence of positive integers with $i_1 > 1$. 
Its derivation $D(I)$ is given by
\begin{eqnarray}
D(I) &=& 
	(i_1 + 1, i_2, \ldots, i_k) + 
(i_1,   i_2+1, \ldots, i_k) + \ldots 
(i_1,   i_2, \ldots, i_k+1) \nonumber\\
\zeta_{D(I)} &=& \zeta_{(i_1 + 1, i_2, \ldots, i_k)} + \ldots 
+ \zeta_{(i_1,   i_2, \ldots, i_k+1)}~.
\end{eqnarray}
The Derivation Theorem states
\begin{equation}
\zeta_{D(I)} = \zeta_{\tau(D(\tau(I)))}~.	
\end{equation}
Here $\tau$ denotes the duality-operation (\ref{DUAL}). 
We call an index-word 
$w$ admissible, if its first letter is not {\sf 1}. The words form the set
$\mathfrak{H}^0$. $|w|$ = {\sf w} is the weight and $d(w)$ the depth of $w$. 
For the MZVs the words $w$ are build in terms of concatenation 
products $x_0^{i_1 - 1} x_1 x_0^{i_2 - 1} x_0 ... x_0^{i_k - 1} x_1$.
The height of a word, ht$(w)$, counts the number of (non-commutative) factors  
$x_0^a x_1^b$ of $w$. The operator $D$ and 
its dual $\overline{D}$ act as follows \cite{ZUD},
\begin{equation}
D x_0 =0,~~~~ D x_1 = x_0 x_1,~~~~ \overline{D} x_0 = x_0 x_1,~~~~ 
\overline{D} x_1 = 0~.
\nonumber
\end{equation}
Define an anti-symmetric derivation 
\begin{equation}
\partial_n x_0 = x_0 (x_0 + x_1)^{n-1} x_1~.
\nonumber
\end{equation}
A generalization of the Derivation Theorem was given in 
\cite{HOF4,KANEK1}~:\newline 
The identity
\begin{equation}
\zeta(\partial_n w) = 0
\end{equation}
holds for any $n \geq 1$ and any word $w \in \mathfrak{H}^0$.
Further theorems are the  {\bf Le--Murakami Theorem},~\cite{LEMURA},  
the {\bf Ohno Theorem},~\cite{OHNO}, which generalizes the sum- and
duality theorem, the {\bf Ohno--Zagier Theorem}, \cite{OHNZAG}, 
which covers the Le--Murakami theorem and the sum theorem,
and generalizes a theorem by Hoffman~\cite{HOF3,HOF2}, and the {\bf cyclic sum 
theorem},~\cite{OHNO1}.

Finally, we mention a main conjecture for the MZVs. Consider tuples ${\bf 
k} = (k_1, \ldots, 
k_r)~\in~\mathbb{N}^r, k_1 \geq 1$. One defines
\begin{eqnarray}
{\mathcal Z}_0 &:=& \mathbb{Q} \nonumber\\
{\mathcal Z}_1 &:=& \{0\} \nonumber\\
{\mathcal Z}_w &:=& \sum_{|{\bf k}| = w} \mathbb{Q}
 \cdot \zeta({\bf k}) \subset \mathbb{R}~.
\end{eqnarray}
If further
\begin{eqnarray}
{\mathcal Z}^{\rm Go} &:=& \sum_{w=0}^\infty {\mathcal Z}_w \subset 
\mathbb{R}~~~~~~{\rm (Goncharov)} \\
{\mathcal Z}^{\rm Ca} &:=& \overset{\infty}{\underset{w=0}{ 
\bigoplus}} {\mathcal Z}_w ~~~~~~~~~~~~~~~{\rm (Cartier)}
\end{eqnarray}
the conjecture states 

\vspace*{1mm} \noindent
(a)~~${\mathcal Z}^{\rm Go} \cong {\mathcal Z}^{\rm Ca}$. There are no 
relations over $\mathbb{Q}$ between the MZVs of different weight {\sf w}.\\
(b)~~dim${\mathcal Z}_w  = d_w$, with $d_0 = 1, d_1=0, d_2=1, d_w = d_{w-2}
+d_{w-3}$.\\
(c)~~All relations between MZVs are given by the extended double-shuffle 
relations~\cite{IKZ}, cf. also \cite{ECALLE}. If this conjecture turns out to 
be true all MZVs are irrational numbers.

\subsection{Selection of a Basis}

\vspace*{1mm}\noindent
Thus far we have not specified which basis we have been using for the MZVs. 
In first instance, we actually let the program select the basis. The result 
was the collection of remaining elements after elimination of as many 
elements as possible. The ordering in the elimination process was such that 
the remaining elements would be minimal in depth and maximal in their sum 
notation. Hence $Z_{20,2,1,1}$ would be preferred over $Z_{18,4,1,1}$. As 
it turned out, all remaining elements had an index field which formed a 
Lyndon word. This is not really surprising due to the ordering. 
Unfortunately there was not much systematics found in these elements.

Next came the idea that if the Euler sums have a basis made out of Lyndon 
words of only negative odd indices, maybe one should investigate to 
which extent one can write a basis for the MZVs in terms of Lyndon words 
with positive odd indices only. It turns out that a number of elements 
can be selected with 
odd-only indices, but it is not possible for the whole basis. A number 
of basis elements needs at least two even indices.

\vspace*{2mm}\noindent
{\bf Definition.}\\
$L_w$ is the set of Lyndon words made out of positive odd-integer indices, 
with no index $i=1$ at given weight ${\sf w}$. $\Box$

\vspace*{2mm}\noindent
We observed that Table~\ref{BK2table} can be reproduced by basis elements
with indices in $L_w$.
As mentioned, this is not a basis for the MZVs, but if we write as many 
elements of the basis as possible as elements of the set $L_w$, the 
remaining elements of the basis have a depth that is at least two greater 
than the elements that are remaining in the $L_w$ set and need at least two 
even indices. Additionally, it looks like that they can be written as an 
extended version of these remaining elements by adding two indices 1 at the 
end and subtracting one from the first two indices as in
\begin{eqnarray}
	Z_{7,5,3} & \rightarrow & Z_{6,4,3,1,1}~.
\end{eqnarray}
We have been able to construct bases with these properties all the way up 
to weight ${\sf w=26}$. The complete (non-unique) recipe for such bases is:
\begin{enumerate}
\item Construct the set $L_w$ of all Lyndon words of positive odd integers 
excluding one that add up to ${\sf w}$.
\item Starting at lowest depth, write as many basis elements of the basis as 
possible in terms of elements of $L_w$. Call the remaining elements in 
$L_w$ at this depth $R_W^{(D)}$.
\item At the next depth, two units larger than the previous one, write 
again 
as many basis elements of the basis as possible in terms of elements of 
$L_w$ and construct $R_W^{(D+2)}$.
\item Write the elements of the basis with depth $D+2$ that could not be 
written as elements of $L_w$ as 1-fold extended elements of $R_W^{(D)}$.
\item Write the elements of the basis with depth $D+2$ that could not be 
written as elements of $L_w$ or 1-fold extended elements of $R_W^{(D)}$ as 
2-fold extended elements of what remains of $R_W^{(D-2)}$, etc.
\item If we are not done yet, raise $D$ by two and go back to step 3.
\end{enumerate}
The concept of $n$-fold extension is defined by subtracting one from the 
first $2n$ indices and adding $2n$ indices with the value one at the end of 
the index set.

%
%
%

To illustrate this we give two examples. First the basis at weight ${\sf 
w=12}$:
\begin{center}
\begin{tabular}{rcc}
	$L_{12}:$ & $H_{9,3}$ & $H_{7,5}$     \\
	$P_{12}:$ & $H_{9,3}$ & $H_{6,4,1,1}$ \\
\end{tabular}
\end{center}
and next the basis at weight ${\sf w=18}$ :
\begin{center}
\begin{tabular}{rcccccccc}
 $\!L_{18}:$ & $\!H_{15,3}$ & $\!H_{13,5}$ & $\!H_{11,7}$ & $\!H_{9,3,3,3}$ &
	$\!H_{7,5,3,3}$ & $\!H_{7,3,5,3}$ & $\!H_{7,3,3,5}$ & $\!H_{5,5,5,3}$ \\
 $\!P_{18}:$ & $\!H_{15,3}$ & $\!H_{13,5}$ & $\!H_{10,6,1,1}$ & $\!H_{9,3,3,3}$ &
	$\!H_{6,4,3,3,1,1}$ & $\!H_{7,3,5,3}$ & $\!H_{7,3,3,5}$ & 
$\!H_{5,5,5,3}$ \\
\end{tabular}
\end{center}
From the basis at weight 18 it should be clear why we put so much 
effort in obtaining the results for the Euler sums at weight 18, depth 6.

Because the construction does not tell which elements of $L_w$ to select 
the results are not unique. In fact quite a few selections are not possible 
because of dependencies between the elements of $L_w$. Hence the whole 
procedure requires a certain amount of experimenting before a good basis is 
found. In Appendix~\ref{ap:bases} we have tried to find a 
basis in which the elements that are taken from $L_w$ have the highest 
values when their index set is seen as a multi-digit number.
Because of reasons being explained in the next Section we call these bases 
`pushdown bases'.

We do not have complete runs for the weights ${\sf w=27}$ and ${\sf w=28}$. In 
these cases the 
elements with the greatest depth are missing. But we can go through the 
construction as far as possible and make predictions about the missing 
elements. It turns out that for both these weights a 2-fold extension is 
needed. For weight ${\sf w=27}$ this would be for depth 5 to depth 9 and for 
weight ${\sf w = 28}$ for depth 4 to depth 8. This concept was not taken into 
account in the 
conjectures in Ref.~\cite{BK1}. 
Hence we formulate a new conjecture 
that not only specifies the number of elements for each weight and depth 
but also how many elements need how many extensions. 

\vspace*{2mm}\noindent
{\bf Conjecture~2.}\\
The number of basis elements $D(w,d,p)$ of MZVs with weight ${\sf w}$, 
depth ${\sf d}$, and pushdown ${\sf p}$ is obtained from the generating 
function 
\begin{eqnarray}
\prod_{w=3}^\infty\prod_{d=1}^\infty\prod_{p=0}^\infty
	(1-x^w y^d z^p)^{D(w,d,p)} & = &
       1-\frac{x^3 y}{1-x^2}
       +\frac{x^{12} y^2 (1-y^2 z)}{(1-x^4)(1-x^6)}
\end{eqnarray}
solving for the coefficients of the monomials $x^w y^d z^p$. $\Box$ 

\vspace*{2mm}\noindent
This formula predicts 
the first $n$-fold extension ($n>1$) at weight $w=12n+3$ and it will be to 
depth $d=4n+1$. The exception is the first extension at weight 12. We show 
this in Table~\ref{BBVtable}.
\begin{table}\centering
\begin{tabular}{|c|c|c|c|c|c|c|c|c|c|c|}
\hline
  {\sf w/d}& 1& 2 &  3 &  4   &  5   &  6    &  7    &  8      &  9    & 10 \\ 
\hline
  1 &   &   &    &      &      &       &       &         &       &  \\
  2 & 1 &   &    &      &      &       &       &         &       &  \\
  3 & 1 &   &    &      &      &       &       &         &       &  \\
  4 &   &   &    &      &      &       &       &         &       &  \\
  5 & 1 &   &    &      &      &       &       &         &       &  \\
  6 &   & 0 &    &      &      &       &       &         &       &  \\
  7 & 1 &   &    &      &      &       &       &         &       &  \\
  8 &   & 1 &    &      &      &       &       &         &       &  \\
  9 & 1 &   & 0  &      &      &       &       &         &       &  \\
 10 &   & 1 &    &      &      &       &       &         &       &  \\
 11 & 1 &   & 1  &      &      &       &       &         &       &  \\
 12 &   & 1 &    &  0,1 &      &       &       &         &       &  \\
 13 & 1 &   & 2  &      &      &       &       &         &       &  \\
 14 &   & 2 &    &    1 &      &       &       &         &       &  \\
 15 & 1 &   & 2  &      &  0,1 &       &       &         &       &  \\
 16 &   & 2 &    &  2,1 &      &       &       &         &       &  \\
 17 & 1 &   & 4  &      &  1,1 &       &       &         &       &  \\
 18 &   & 2 &    &  4,1 &      &   0,1 &       &         &       &  \\
 19 & 1 &   & 5  &      &  3,2 &       &       &         &       &  \\
 20 &   & 3 &    &  6,1 &      &   1,2 &       &         &       &  \\
 21 & 1 &   & 6  &      &  6,3 &       &   0,1 &         &       &  \\
 22 &   & 3 &    & 10,1 &      &   3,4 &       &         &       &  \\
 23 & 1 &   & 8  &      & 11,4 &       &   1,3 &         &       &  \\
 24 &   & 3 &    & 14,2 &      &   8,6 &       &     0,1 &       &  \\
 25 & 1 &   & 10 &      & 18,5 &       &   4,7 &         &       &  \\
 26 &   & 4 &    & 19,1 &      & 16,11 &       &     1,4 &       &  \\
 27 & 1 &   & 11 &      & 29,7 &       & 11,12 &         & 0,1,1 &  \\
 28 &   & 4 &    & 25,2 &      & 31,14 &       &  4,11,1 &       &  \\
 29 & 1 &   & 14 &      & 42,8 &       & 25,23 &         & 1,5,1 &  \\
 30 &   & 4 &    & 33,2 &      & 52,21 &       & 14,22,1 &       & 0,1,1 \\
\hline
\end{tabular}
\caption{Number of basis elements for MZVs as a function of weight, depth 
and extension(or pushdown). If there are several numbers, separated by 
commas, the first indicates the number of elements that came from $L_w$, 
the second the number of 1-fold extensions from depth $d-2$, the third the 
number of 2-fold extensions from depth $d-4$, etc. A single number refers 
to the elements of $L_w$.}
\label{BBVtable}
\end{table}

It is a great pity that with the resources that were at our disposal we 
just could not get direct access to a double extension or pushdown. 
Extrapolating from the numbers in Table~\ref{MZVrun30} indicates computer 
times of the order of half a year (for weight 28, depth 8) to more than a 
year (for weight 27, depth 9).


\section{Pushdowns}
\label{sec:pushdown}
\renewcommand{\theequation}{\thesection.\arabic{equation}}
\setcounter{equation}{0}

\vspace{1mm}
\noindent
As mentioned in the previous Section, there are elements that as MZVs can 
only be written with a certain depth, while, when written in terms of Euler 
sums, can be written with a smaller depth. This phenomenon is called 
pushdown. The simplest example occurs at weight ${\sf w=12}$ and can be looked 
up in the Tables for the Euler sums. It is
\begin{eqnarray}
\label{eq:pushdownraw}
	Z_{6,4,1,1} & = &
       - \frac{2107648}{15825} H_{-11,-1}
       + \frac{50048}{9495} H_{-9,-3}
       - \frac{117568}{237375} H_{-7,-5}
       + \frac{100352}{1583} \zeta_2 H_{-9,-1}
			\nonumber \\ &&
       - \frac{3584}{1583} \zeta_2 H_{-7,-3}
       + \frac{320}{57} \zeta_2^2 H_{-7,-1}
       - \frac{64}{171} \zeta_2^2 H_{-5,-3}
       - \frac{2535128220786914}{481025690578125} \zeta_2^6
			\nonumber \\ &&
       + \frac{69528448}{427275} \eta_3 \eta_9
       - \frac{32}{35} \eta_3^2 \zeta_2^3
       + \frac{64}{243} \eta_3^4
       - \frac{21236224}{299187} \eta_7 \eta_3 \zeta_2
			\nonumber \\ &&
       - \frac{11072}{1425} \eta_5 \eta_3 \zeta_2^2
       + \frac{696654848}{4984875} \eta_5 \eta_7
       - \frac{11690624}{356175} \eta_5^2 \zeta_2~,
\end{eqnarray}
in which we remind the reader that $\eta_n = H_{-n}$. The next equation is 
at weight ${\sf w=15}$ and is already considerably lengthier. The rhs of 
$Z_{6,4,3,1,1}$ contains 49 terms when written in this form and some of the 
fractions consist of more than 100 decimal digits. The phenomenon of these 
pushdowns seems to be intimately connected with the doubling and 
generalized doubling relations. We have investigated this at the weight 
${\sf w=12}$ system. This is the only system over which we have complete 
control, because we have the full results for all depths for all Euler sums 
up to this weight. If we run this system without the use of the doubling and 
generalized doubling relations there are three more elements left in the 
`basis', see Table~\ref{nodoubling}. Two are of depth 4 and one is of depth 
2. And additionally there 
is no pushdown. The element $Z_{6,4,1,1} = H_{6,4,1,1}$ needs one of these 
extra elements at depth 4. If we use the doubling relations, but we do not 
use the GDRs, there is only one extra element of depth 4, but the pushdown 
does take place. If we use only the GDRs, there are no remaining elements 
beyond the regular basis and the pushdown takes place.

Unfortunately we cannot run this test for other weights. Not using the GDRs 
means that we cannot run at restricted depth, due to the phenomenon of 
leakage.
Of course it is rather adventurous to make the statement that doubling is 
at the origin of the pushdowns, when we have only a single case, but there 
is more supporting evidence as we will see below.

The way we have presented the pushdown in (\ref{eq:pushdownraw}), 
although correct, is not its most transparent form. One can rewrite it to 
as many MZVs as possible and obtain a much simpler representation. One can, 
for instance, write
\begin{eqnarray}
 H_{-9,-3} & = & \frac{1055}{1024} \Biggl[
       - Z_{9,3}
       - \frac{185874}{5275} \zeta_7 \zeta_5
       - \frac{37332}{1055} \zeta_9 \zeta_3
		\nonumber \\ &&
       + \frac{1024}{26375} H_{-7,-5}
       + \frac{187392}{5275} H_{-11,-1}
       + \frac{92649159488}{23101203125} \zeta_2^6 \Biggr]~.
\end{eqnarray}
Additionally, we introduce a new function $A$ as
\begin{eqnarray}
\label{eq:AFUN}
	A_{n_1,n_2,\cdots,n_{p-1},n_p} & = &
		\sum_{\pm} s H_{\pm n_1,\pm n_2,\cdots,\pm n_{p-1},n_p}
\end{eqnarray}
in which the sum is over the $2^{p-1}$ possible sign combinations and
$s = -1$ if the number of minus signs inside $H$ is odd and $s = + 1$ 
if this number is even as in
\begin{eqnarray}
	A_{7,5,3} & = & H_{7,5,3} - H_{-7,5,3} - H_{7,-5,3} + H_{-7,-5,3}~.
\end{eqnarray}
Notice that the last index is always positive. In terms of the $Z$-notation 
the function $A$ is the sum over all $Z$-sums with an even number of 
negative indices, but the absolute values of the indices are identical to 
the indices of the $A$-function. We rewrite then
\begin{eqnarray}
H_{-7,-5} & = & -\frac{25}{3} \Biggl[
       - A_{7,5}
       + \frac{1295}{2304} Z_{9,3}
       + \frac{461399}{15360} \zeta_7 \zeta_5
       + \frac{3213}{128} \zeta_9 \zeta_3
		\nonumber \\ &&
       - \frac{126}{5} H_{-11,-1}
       - \frac{39238805939}{12612600000} \zeta_2^6 \Biggr]~,
\end{eqnarray}
and finally the result for the pushdown becomes:
\begin{eqnarray}
\label{eq:pushdown12}
	Z_{6,4,1,1} & = &
       - \frac{64}{27} A_{7,5}
       - \frac{7967}{1944} Z_{9,3}
       + \frac{1}{12} \zeta_3^4
       + \frac{11431}{1296} \zeta_7 \zeta_5
		\nonumber \\ &&
       - \frac{799}{72} \zeta_9 \zeta_3
       + 3 \zeta_2 Z_{7,3}
       + \frac{7}{2} \zeta_2 \zeta_5^2
       + 10 \zeta_2 \zeta_7 \zeta_3
		\nonumber \\ &&
       + \frac{3}{5} \zeta_2^2 Z_{5,3}
       - \frac{1}{5} \zeta_2^2 \zeta_5 \zeta_3
       - \frac{18}{35} \zeta_2^3 \zeta_3^2
       - \frac{5607853}{6081075} \zeta_2^6~,
\end{eqnarray}
which is much simpler than equation (\ref{eq:pushdownraw}). We see the same 
happening in the expression for $Z_{6,4,3,1,1}$,
\begin{eqnarray}
\label{eq:pushdown15}
	Z_{6,4,3,1,1} & = &
       + \frac{1408}{81} A_{7,5,3}
       + \frac{16663}{11664} Z_{9,3,3}
       + \frac{150481}{68040} Z_{7,3,5}
       + 10 \zeta_3 Z_{6,4,1,1}
	\nonumber \\ &&
       + \frac{162823}{3888} \zeta_3 Z_{9,3}
       - \frac{17}{20} \zeta_3^5
       - \frac{101437}{38880} \zeta_5 Z_{7,3}
       - \frac{1520827}{38880} \zeta_5^3
	\nonumber \\ &&
       + \frac{1903}{120} \zeta_7 Z_{5,3}
       - \frac{93619}{1296} \zeta_7 \zeta_5 \zeta_3
       + \frac{3601}{48} \zeta_9 \zeta_3^2
       - \frac{20651486329}{4082400} \zeta_{15}
	\nonumber \\ &&
       + \frac{14}{5} \zeta_2 Z_{5,5,3}
       - 2 \zeta_2 Z_{7,3,3}
       - 27 \zeta_2 \zeta_3 Z_{7,3}
       - \frac{21}{2} \zeta_2 \zeta_5 Z_{5,3}
       - \frac{61}{2} \zeta_2 \zeta_5^2 \zeta_3
	\nonumber \\ &&
       - 84 \zeta_2 \zeta_7 \zeta_3^2
       + \frac{31753363}{12960} \zeta_2 \zeta_{13}
       - 4 \zeta_2^2 Z_{5,3,3}
       - 5 \zeta_2^2 \zeta_3 Z_{5,3}
	\nonumber \\ &&
       + \frac{9}{2} \zeta_2^2 \zeta_5 \zeta_3^2
       + \frac{979621}{1701} \zeta_2^2 \zeta_{11}
       + \frac{186}{35} \zeta_2^3 \zeta_3^3
       - \frac{490670609}{3572100} \zeta_2^3 \zeta_9
	\nonumber \\ &&
       - \frac{1455253}{283500} \zeta_2^4 \zeta_7
       + \frac{4049341}{311850} \zeta_2^5 \zeta_5
       + \frac{12073102}{1488375} \zeta_2^6 \zeta_3~.
\end{eqnarray}
In both relations there is only a single object in the 
equation that is not an MZV: the function $A$. This means that we can write 
this $A$-function alternatively as a combination of MZVs of which one has a 
depth ${\sf d' = d +2}$. We have done that with $A_{7,5}$ to 
obtain (\ref{eq:pushdown15}), see the fourth term in the right hand side. 
The intriguing part about it all is that this function $A$ 
contains half of the terms on the right hand side
of the doubling relation in equation (\ref{eq:inx2}). In terms of 
$H$-functions it are the terms in which the last index is positive and in 
terms of $Z$-functions it are all terms with an even number of negative 
indices.

We have been able to construct pushdown relations for all extended basis 
elements up to weight ${\sf w=21}$ and one for weight ${\sf w=22}$. Some of 
these could be 
constructed directly from the data mine. The more difficult ones are, 
however, outside the range of the files in the data mine. There we could use 
the data mine as an aid in limiting the search with numerical algorithms 
like {\tt LLL} or {\tt PSLQ}. More details are given in 
Appendix~\ref{ap:explicit}. This search for pushdowns is not always as simple 
as 
the two examples we gave above. Sometimes there is more than one pushdown 
at a given depth, and sometimes there are elements at a given depth that 
should be pushed down, but there are also elements that remain at that 
depth. In the last case it is usually a linear combination of the extended 
element(s) and the remaining element(s) that get(s) pushed down. But for 
all cases that we could check there is a single function $A$ associated 
with each pushed down element. If there are several pushdowns at a given 
weight and depth the right hand side may contain linear combinations of the 
corresponding $A$-functions. In all cases we could select the bases such that 
the index fields of the $A$-functions corresponded to the index fields of the 
elements of the set $L_w$ that had to be extended.

The above indicates that these $A$-functions have a special status within the 
Euler sums. They are quite similar to the MZVs.

It should be noted that not all $A$-functions can be written in terms of MZVs 
only. This holds only for a limited subset as we will see in the next 
Section. Additionally, not all $A$-functions that can be rewritten in terms of 
MZVs can be used for pushdowns, because a number of them can be rewritten 
in terms of MZVs that have at most the same depth as the $A$-function 
itself.

The above observations lead to the following conjecture:

\vspace*{2mm}
\noindent
{\bf Conjecture~3.}\\
At each weight $w$, there exists a set of Lyndon words $L_w$
from which one may construct a basis for MZVs as follows.
For each Lyndon word one chooses either the associated $Z$ value
or the associated $A$ value, with the number of $A$ values chosen
to agree with the Broadhurst-Kreimer conjectures. Linear
combinations of these $A$ values then provide the pushdowns
for the extensions of $Z$ values by a pair unit indices,
as exemplified in Appendix C. $\Box$

\vspace*{2mm}\noindent
What the above says is that we can find a good basis for the MZVs using the 
set $L_w$, provided we borrow some elements from the Euler sums. In such 
terms the basis for weight ${\sf w=18}$ would look like
\begin{center}
\begin{tabular}{rcccccccc}
 $\!L_{18}:$ & $\!Z_{15,3}$ & $\!Z_{13,5}$ & $\!Z_{11,7}$ & $\!Z_{9,3,3,3}$ &
	$\!Z_{7,5,3,3}$ & $\!Z_{7,3,5,3}$ & $\!Z_{7,3,3,5}$ & $\!Z_{5,5,5,3}$ \\
 $\!P_{18}:$ & $\!Z_{15,3}$ & $\!Z_{13,5}$ & $\!A_{11,7}$ & $\!Z_{9,3,3,3}$ &
	$\!A_{7,5,3,3}$ & $\!Z_{7,3,5,3}$ & $\!Z_{7,3,3,5}$ & $\!Z_{5,5,5,3}$ \\
\end{tabular}
\end{center}


\section{Special Euler Sums}
\renewcommand{\theequation}{\thesection.\arabic{equation}}
\setcounter{equation}{0}
\label{SpecialEulerSums}

\vspace{1mm}
\noindent
The discovery of the $A$-functions brings up a new point. Which Euler sums 
can be written as a linear combination of MZVs only? This is of course a 
perfect question for a system like the data mine in which exhaustive 
searches are relatively cheap. At the same time we ask of course the 
question which $A$-functions can be written in terms of MZVs only. We should 
distinguish two cases~:
\begin{itemize}
\item The object can be written in terms of MZVs that have at most the same 
depth as the object.
\item The object needs MZVs of a higher depth. This occurs when there is 
already an $A$-function that is used in a pushdown. In that case many other 
$A$-functions may be rewritten in terms of this $A$-function and MZVs of 
the same depth or lower depth.
\end{itemize}
We find that whenever the second case can occur, it will for a large 
fraction of the $A$-functions of that depth. The number of $H$-functions 
with at least one negative index that can be rewritten completely in terms 
of MZVs is given in Table~\ref{tab:HMZV}.
\begin{table}\centering
\begin{tabular}{|c|c|c|c|c|}
\hline
{\sf w/d}&   2 &   3 &   4 &   5 \\ \hline
  7 &  13 &   9 &   2 &   0 \\
  8 &   5 &  10 &   8 &   2 \\
  9 &  19 &  26 &   2 &   0 \\
 10 &   7 &  22 &  17 &   7 \\
 11 &  25 &  38 &   6 &   0 \\
 12 &   9 &  40 &  43 &  13 \\
 13 &  31 &  62 &   4 &   1 \\
 14 &  11 &  62 &  77 &  23 \\
 15 &  37 &  90 &   6 &   3 \\
 16 &  13 &  90 & 137 &  34 \\
 17 &  43 & 121 &   6 &   3 \\ \hline
\end{tabular}
\caption{Number of Euler sums with at least one negative index that can be 
rewritten in terms of MZVs only as a function of weight ${\sf(w)}$ and depth 
${\sf (d)}$.}
\label{tab:HMZV}
\end{table}
In Table~\ref{tab:AMZV} we show the same for the $A$-functions. Here there 
are clearly many more. Actually a sizable fraction of the $A$-functions can 
be rewritten like this. For example, there are 1365 finite $A$-functions of 
${\sf w=17, d=5}$ of which 449 can be rewritten in terms of MZVs only.

\begin{table}\centering
\begin{tabular}{|c|c|c|c|c|}
\hline
{\sf w/d}&   2 &   3 &   4 &   5 \\ \hline
  7 &   4 &   5 &   2 &   0 \\
  8 &   5 &   8 &   4 &   0 \\
  9 &   6 &  13 &   9 &   3 \\
 10 &   7 &  18 &  17 &   7 \\
 11 &   8 &  25 &  31 &  17 \\
 12 &   9 &  32 &  49 &  34 \\
 13 &  10 &  41 &  74 &  67 \\
 14 &  11 &  50 & 106 & 116 \\
 15 &  12 &  61 & 148 & 192 \\
 16 &  13 &  72 & 198 & 298 \\
 17 &  14 &  85 & 259 & 449 \\ \hline
\end{tabular}
\caption{Number of $A$-functions that can be 
rewritten in terms of MZVs only as a function of weight ${\sf (w)}$ and depth 
${\sf (d)}$.}
\label{tab:AMZV}
\end{table}

Considering that a number of the Euler sums can be rewritten in terms of 
MZVs only, one may raise the question whether the pushdowns can be 
rewritten in such a way that they do not have the $A$-functions, but rather 
have a single Euler sum in their right hand side. This turned out to be a 
difficult question to answer, because the pushdown at ${\sf w=21, d=7}$ was 
very time consuming and took several days for each trial. At first the 
number of candidates was rather large. We could make a list of candidates 
in a way, similar to that of Table~\ref{tab:HMZV} for ${\sf w=21, d=5}$ and 
see which Euler sums could be expressed in terms of MZVs and $A_{7,5,3,3,
3}$ which is the object that was used in the pushdown\footnote{Originally 
we worked with $A_{9,3,3,3,3}$ and it was only at a very late stage that we 
converted to $A_{7,5,3,3,3}$. Hence a number of the `raw' results still 
refer to $A_{9,3,3,3,3}$.}. Unfortunately the results for ${\sf w=21, d=5}$ 
are in modular arithmetic and without the products of lower weight objects. 
Trying several elements of the list gave negative results indicating that 
many objects that give only MZVs for the terms with the same weight may 
have terms that are products of Euler sums of a lower weight. Then, after 
constructing Table~\ref{tab:HMZV} we looked for patterns and we noticed 
that the only eligible elements for ${\sf w=13, w=15, w=17}$ are
\begin{eqnarray}
	Z_{3,-2,3,-2,3} & = & H_{3,-2,-3,2,3} \nonumber \\
	Z_{3,-4,3,-2,3} & = & H_{3,-4,-3,2,3} \nonumber \\
	Z_{3,-2,3,-4,3} & = & H_{3,-2,-3,4,3} \nonumber \\
	Z_{3,-6,3,-2,3} & = & H_{3,-6,-3,2,3} \nonumber \\
	Z_{3,-4,3,-4,3} & = & H_{3,-4,-3,4,3} \nonumber \\
	Z_{3,-2,3,-6,3} & = & H_{3,-2,-3,6,3}~.
\end{eqnarray}
Trying to rewrite $Z_{3,-6,3,-6,3}$ in terms of $A_{7,5,3,3,3}$ by means of 
{\tt LLL} (a 130 elements search) gave the desired result. Hence by now all 
pushdowns have been obtained as well in terms of MZVs as in terms of one 
single Euler sum only. Unfortunately the index field of these Euler sums 
seems to be completely unrelated to the index fields of our basis elements.

\section{Outlook}
\renewcommand{\theequation}{\thesection.\arabic{equation}}
\setcounter{equation}{0}
\label{Outlook}

The data mine has given us already much information and it may yield 
more yet. But the current results leave also many new questions. To name a 
few:
\begin{itemize}
\item Can the GDRs be derived and/or written in a simpler way?
\item Why can the GDRs resolve the problem of `leakage'?
\item Why do we need the doubling relations at all?
\item What is the relation between the doubling formula and the pushdowns?
\item Is it possible to see which $A$-functions can be used for pushdowns 
      without needing the Euler sums of the data mine?
\item Can a pushdown basis be constructed without needing the MZVs of 
      the data mine?
\end{itemize}

In addition there is some `unfinished business'. We did not get more than 
partial evidence for the double pushdowns at weight 27 and weight 28. 
Although we can guess the basis at weight 27, an {\tt LLL} search for the 
complete formula would involve more than 800 elements and probably more 
than 10 times the number of digits than what our current searches needed. 
Considering the asymptotic behaviour of the {\tt LLL} algorithm, this would 
mean 
at least $10^7$ times the computer time we needed for the current 
determinations. The data mine approach is also not very attractive. There 
we would need the Euler sums to weight 27, depth 9. This might need even 
more extra orders of magnitude in resources than for the {\tt LLL} algorithm. 
What would be very welcome is an algorithm by which we can determine a 
(small) subset of the Euler sums that includes the $A$-functions and combine 
this subset with the MZVs.
For the MZV part of these double pushdowns things look much brighter. In 
modular arithmetic the continuously improving hardware and software 
technology should place those runs within reach soon. With a better 
ordering of the processing of the equations, which unfortunately we do not 
have, the runs could already be attempted. Again, finding non-trivial 
subsets to which one might limit oneself, would immediately lead to great 
progress as well. We hope, that the empirical discoveries we made in this 
paper
for harmonic sums up to ${\sf w = 30}$ will stimulate mathematical research
and eventually lead to proofs of more far reaching theorems in the future.
Here we regard the consideration of the embedding of the MZVs into the Euler 
sums of importance. Likewise one may consider colored `MZVs' with even higher 
roots of unity \cite{ZHAO1} in the future, which have not been the objective 
of this paper.

The data mine will be extended whenever new and relevant results are 
obtained. there is a history page that shows additions and corrections. If 
others have interesting contributions, they should contact one of the 
authors.

\vspace*{5mm}
\noindent
{\bf Acknowledgments}.\\
\noindent
The work has been supported in part by the research program of the Dutch 
Foundation for Fundamental Research of Matter (FOM), by DFG
Sonderforschungsbereich Transregio 9, Computergest\"utzte Theoretische
Teilchenphysik, the European Commission MRTN HEPTOOLS under Contract No. 
MRTN-CT-2006-035505, J.V. would also like to thank the Humboldt foundation 
for its generous support, and DESY, Zeuthen and the University of Karlsruhe 
for its hospitality during this work. The runs for creating the data mine 
were done on the computer system of the Theoretical Particle Physics group 
(TTP) at the University of Karlsruhe and computers at DESY and Nikhef.


\newpage
\appendix
\section{Fibonacci and Lyndon Bases at Fixed Weight}
\label{gen:basis}
\renewcommand{\theequation}{\thesection.\arabic{equation}}
\setcounter{equation}{0}

\vspace{1mm}
\noindent
In the past several bases have been considered for both the MZVs and the 
Euler sums. In some of these the concept of depth is not relevant and hence 
for the counting rules we should sum over the depth. We will discuss those 
bases in this Appendix. For a number of these bases conjectures are 
formulated in the literature, which cannot be broken down fixing the depth. 
The counting relation for the MZVs was conjectured in \cite{ZAG1,BK1} and 
\cite{Broadhurst:1},~respectively. 

The vector space of MZVs can be constructed allowing basis elements, 
which contain besides the  $\zeta$--values the index of which is a 
Lyndon word products of this type of $\zeta$-values of lower weight. 
One basis of this kind is
\begin{eqnarray} 
{\sf w=~~2} & & \zeta_2 \\ 
{\sf w=~~3} & & \zeta_3 \\ 
{\sf w=~~4} & & \zeta_2^2 \\ 
{\sf w=~~5} & & \zeta_5, \zeta_2 \zeta_3 \\ 
{\sf w=~~6} & & \zeta_3^2, \zeta_2^3  \\ 
{\sf w=~~7} & & \zeta_7, \zeta_5 \zeta_2, \zeta_3 \zeta_2^2 \\
{\sf w=~~8} & & \zeta_{5,3}, \zeta_5 \zeta_3, \zeta_3^2 \zeta_2, \zeta_2^4 \\
{\sf w=~~9} & & \zeta_9, \zeta_7 \zeta_2, \zeta_5 
                 \zeta_2^2, \zeta_3^3, \zeta_3 \zeta_2^3 
\\
{\sf w=10}  & & \zeta_{7,3}, \zeta_{5,3} \zeta_2, \zeta_7 \zeta_3, \zeta_5^2, 
                \zeta_5 \zeta_3 \zeta_2,
                \zeta_3^2 \zeta_2^2, \zeta_2^5,~{\rm etc.}
\end{eqnarray}
The number of these basis elements is counted by the Padovan numbers, 
$\hat{P}_k$, \cite{PADO}, which have the same recursion as the Perrin 
numbers, but start from the initial values $\hat{P}_1 = \hat{P}_2 = 
\hat{P}_3 = 1$. Their generating function is
\begin{eqnarray}
G(\hat{P}_k,x) = \frac{1+x}{1-x^2-x^3} = \sum_{k=0}^\infty x^k \hat{P}_k~.
\end{eqnarray}
They also obey a Binet-like formula.
The first values are given in Table~\ref{tablepadovanb}.

{
\begin{table}[h]
\begin{center}
\begin{tabular}[htb]{|r||r|r|r|r|r|r|r|r|r|r|} 
\hline \hline
$w$   & 1 & 2 & 3 & 4 & 5 & 6 & 7 &  8 &  9 & 10 \\
\hline
$\hat{P}_w$ & 1 & 1 & 1 & 2 & 2 & 3 & 4 &  5 &  7 &  9 \\
\hline \hline 
$w$   & 11 & 12 & 13 & 14 & 15& 16 & 17 & 18 & 19 & 20\\
\hline
$\hat{P}_w$ & 12 & 16 & 21 & 28 & 37 &49  &65  &86  &114  &151  \\
\hline \hline 
$w$   & 21 & 22 & 23 & 24 & 25 & 26 & 27 & 28 & 29 & 30\\ \hline
$\hat{P}_w$ 
&200 &265  &351  &465  &616  &816  &1081 & 1432 &1897 &2513 \\
\hline \hline 
\end{tabular}
\end{center}
\caption{The first 30 Padovan numbers.}
\label{tablepadovanb}
\end{table}
\normalsize}

The above basis is of the Fibonacci type. Another basis of the Fibonacci 
type is the Hoffman basis~\cite{HOFC} which consists of all elements of 
which the index field is made up from 2's and 3's only. If one uses the 
following construction it is easy to see that the number of basis elements 
follows the Padovan sequence.
\begin{eqnarray}
{\sf w = 1} & &~~~~~{~\emptyset} \nonumber\\
{\sf w = 2} & &~~~~~(2)\nonumber\\
{\sf w = 3} & &~~~~~(3)~.
\end{eqnarray}
The index words at weight {\sf w} are given by
\begin{eqnarray}
I_w = 
\underset{{|a| = (w-2)}}{\cup} (2,I_{a})~~~\cup~~~ 
\underset{{|b| = (w-3)}}{\cup} (3,I_{b}) ~. 
\end{eqnarray}

Let us now turn to Lyndon bases for the MZVs.
Using a Witt-type relation~\cite{WITT} the size of the basis is conjectured 
to be given by
\begin{eqnarray}
\label{eq-WI1b}
l(w) &=& \frac{1}{w} \sum_{d|w} \mu\left(\frac{w}{d}\right) P_d,\nonumber\\
     & & P_1 = 0, P_2 = 2, P_3 =3, P_d = P_{d-2} + P_{d-3},~~d \geq 3~.
\end{eqnarray}
Here the sum runs over the divisors $d$ of the weight ${\sf w}$ and $P_d$ 
denotes 
the Perrin-numbers~\cite{LUCAS,Perrin}. They are given by the 
Binet-like formula
\begin{eqnarray}
P_n &=& \alpha^n + \beta^n + \gamma^n,~~~{\rm with~~\alpha,~\beta,~\gamma
~~the~roots~of} \nonumber\\
x^3 - x - 1    &=& 0 
\end{eqnarray}
and can be derived from the generating 
function
\begin{eqnarray}
G(P_k,x) = \frac{3-x^2}{1-x^2-x^3} = \sum_{k=0}^\infty x^k P_k~.
\end{eqnarray}
The first values are given in Table~\ref{tableperrinb}.

{
\begin{table}\centering
\begin{tabular}[htb]{|r||r|r|r|r|r|r|r|r|r|r|} 
\hline \hline
$w$   & 1 & 2 & 3 & 4 & 5 & 6 & 7 &  8 &  9 & 10 \\
\hline
$P_w$ & 0 & 2 & 3 & 2 & 5 & 5 & 7 & 10 & 12 & 17 \\
\hline \hline 
$w$   & 11 & 12 & 13 & 14 & 15& 16 & 17 & 18 & 19 & 20\\
\hline
$P_w$ & 22 & 29 & 39 & 51 & 68 & 90 & 119 & 158 & 209 & 277 \\
\hline \hline 
$w$   & 21 & 22 & 23 & 24 & 25 & 26 & 27 & 28 & 29 & 30\\ \hline
$P_w$ 
& 367 & 486 & 644 & 853 & 1130 & 1497 & 1983 & 2627 & 3480 & 4610 \\
\hline \hline 
\end{tabular}
\caption{The first 30 Perrin numbers.}
\label{tableperrinb}
\end{table}
\normalsize}
For the basis different choices are possible, which yield equivalent 
representations. Here we choose the basis in terms of $\zeta$--values, 
with an index field which forms a Lyndon word. Our first choice consists of 
indices, which contain as widely as possible odd integers. In case of even 
weights in a series of cases also indices with only even numbers occur from 
{\sf w = 12} onwards, as e.g. for
\begin{equation}
{\sf w=18~:} \hspace*{5mm}  \zeta_{15,3},~\zeta_{13,5},~\zeta_{9,3,3,3},~
               \zeta_{7,5,3,3},~\zeta_{5,5,5,3},~\zeta_{7,5,5,1},~
               \zeta_{8,2,2,2,2,2},~\zeta_{12,2,2,2}~.
\end{equation}

A second natural choice is to take the afore mentioned Hoffman basis and 
select from it only those elements of which the index field forms a Lyndon 
word. Because the algebraic relations for the product of basis elements of 
lower weight do not give objects that are closely related to the basis 
elements at the higher weight, this basis is not used very frequently.

As an example we consider the case {\sf w = 30} and calculate the 
size of the bases using the Witt formula (\ref{eq-WI1b}) resp. the number of 
Lyndon words made up by the letters 2 and 3 only with $2 < 3$. 30 
has the following decomposition
\begin{eqnarray}
30 \equiv k_i*3~+~l_i*2 = 2*3~+~12*2 = 4*3~+~9*2 = 6*3~+~6*2 = 8*3~+3*2~.
\nonumber\\
\end{eqnarray}
We now calculate the number of Lyndon words for each of these 
contributions, with $m_i = k_i + l_i$,
\begin{eqnarray}
n_i = \frac{1}{m_i} \sum_{d|m_i} \mu(d) \frac{(m_i/d)!}{(k_i/d)! (l_i/d)!}~. 
\end{eqnarray}
One obtains
\begin{eqnarray}
L_{\{2,3\}}(30) &=& \frac{1}{14} \left[\frac{14!}{12! 2!} - 
\frac{7!}{6!}\right] 
                + \frac{1}{13} \frac{13!}{9! 4!} 
                + \frac{1}{12} \left[\frac{12!}{6!^2} - \frac{6!}{3!^2} 
                - \frac{4!}{2!^2} + \frac{2!}{1!^2}\right]
                + \frac{1}{11} \frac{11!}{8! 3!} \nonumber\\ 
&=& 151~.
\end{eqnarray}
Using (\ref{eq-WI1b}) the result is
\begin{eqnarray}
l(30) &=& \frac{1}{30} \left[P_{30} - P_{15} - P_{10} - P_6 + P_5 + P_3 + P_2
                             -P_{0}\right] 
\nonumber\\
&=& \frac{1}{30} \left[4610 - 68 - 17 - 5 + 5 + 3 + 2 - 0\right] = 151~.
\end{eqnarray}
A basis up to weight ${\sf w=17}$ for the MZVs was also constructed in \cite{BAS1}.
{
\begin{table}\centering
\begin{tabular}[htb]{|r||r|r|r|r|r|r|r|r|r|r|} 
\hline \hline
$w$   & 1 & 2 & 3 & 4 & 5 & 6 & 7 &  8 &  9 & 10 \\
\hline
$l_w$ & 
0 &
1 &
1 &
0 &
1 &
0 &
1 &
1 &
1 &
1 \\
\hline \hline 
$w$   & 11 & 12 & 13 & 14 & 15& 16 & 17 & 18 & 19 & 20\\
\hline
$l_w$ &
2 &
2 &
3 &
3 &
4 &
5 &
7 &
8 &
11&
13\\
\hline \hline 
$w$   & 21 & 22 & 23 & 24 & 25 & 26 & 27 & 28 & 29 & 30\\ \hline
$l_w$ & 
17&
21&
28&
34&
45&
56&
73&
92&
120&
151\\
\hline \hline 
\end{tabular}
\caption{Number of basis elements of the Lyndon basis for the MZVs for fixed 
weight {\sf w}.}
\label{tablelyndMZV}
\end{table}
\normalsize}
 
For the Euler sums the Fibonacci basis is counted by the Fibonacci numbers.
When we consider also all divergent multiple zeta values the Fibonacci 
sequence is merely shifted. It is easily shown that the divergent Euler 
sums can be represented by the convergent sums and the element $\sigma_0$. 
As in the MZV case we may span the vector space of the Euler 
sums by forming a basis, which includes products of lower weight basis 
elements contained in a Lyndon-basis. One basis of this 
type, used in the {\tt summer} program~\cite{Vermaseren:1} reads
\begin{eqnarray} 
{\sf w=~~1} & & \ln(2) \\ 
{\sf w=~~2} & & \zeta_2, \ln^2(2) \\ 
{\sf w=~~3} & & \zeta_3, \zeta_2 \ln(2), \ln^3(2) \\ 
{\sf w=~~4} & & \Li_4(1/2), \zeta_3 \ln(2), \zeta_2^2, \zeta_2 \ln^2(2), \ln^4(2) \\ 
{\sf w=~~5} & & \Li_5(1/2), \zeta_5, \Li_4(1/2) \ln(2), \zeta_3 \zeta_2, 
\zeta_3 \ln^2(2),
                \zeta_2 \ln^3(2), \zeta_2^2 \ln(2),\ln^5(2) \nonumber\\
\\
{\sf w=~~6} & & \Li_6(1/2), \zeta_{-5,-1}, \Li_5(1/2) \ln(2), \zeta_5 \ln(2), \Li_4(1/2) 
                \zeta_2, \nonumber \\ &&
                \Li_4(1/2) \ln^2(2), \zeta_3^2, \zeta_3 \zeta_2 \ln(2), \zeta_3 \ln^3(2),
                \zeta_2^3, \zeta_2^2 \ln^2(2), \zeta_2 \ln^4(2), 
\nonumber\\ && \ln^6(2),
~{\rm etc.} 
\end{eqnarray}
These bases are counted by the Fibonacci-numbers~\cite{FIBO,HW}, $f_{w+1}$, 
which obey the same recursion relation as the Lucas numbers, but with the
initial conditions $f_0 = 0, f_1 = 1$.
They are represented by the formula given by J.P.M. Binet  (1843)\footnote{
The relation was known to Euler and Moivre.} 
\begin{eqnarray}
f_d = \frac{1}{\sqrt{5}}\left[\left( \frac{1+\sqrt{5}}{2}\right)^d
    - \left( \frac{1-\sqrt{5}}{2}\right)^d\right]~,
\end{eqnarray}
and result from the generating function
\begin{eqnarray}
G(f_k,x) = \frac{x}{1-x-x^2} = \sum_{k=0}^\infty x^k f_k~.
\end{eqnarray}
The first values are given in Table~\ref{tablefibonaccib}.
{
\begin{table}
\begin{center}
\begin{tabular}[htb]{|r||r|r|r|r|r|r|r|r|r|r|} 
\hline \hline
$w$   & 1 & 2 & 3 & 4 & 5 & 6 & 7 &  8 &  9 & 10 \\
\hline
$f_w$ & 1 & 1 & 2 & 3 &	5 & 8 & 13 & 21 & 34 & 55 \\
\hline \hline 
$w$   & 11 & 12 & 13 & 14 & 15& 16 & 17 & 18 & 19 & 20\\
\hline
$f_w$ & 89&144& 233& 377 &610 &	987& 1597& 2584& 4181& 	6765\\
\hline \hline 
\end{tabular}
\end{center}
\caption{The first 20 Fibonacci numbers.}
\label{tablefibonaccib}
\end{table}
\normalsize}

Another Fibonacci basis can be constructed as
\begin{eqnarray}
\label{startfibo}
{\sf w=0} & &~~~~~{~\emptyset} \nonumber\\
{\sf w=1} & &~~~~~(-1) \nonumber\\
{\sf w=2} & &~~~~~(0,-1)~.
\end{eqnarray}
$\HH_{-1}(1)$ and $\HH_{0,-1}(1) = \HH_{-2}(1)$ are chosen as basis
elements.  

\vspace*{2mm}
\noindent
{\bf Conjecture~4.}\\
With the above starting conditions, consider the index words at weight {\sf 
w} to be
\begin{eqnarray}
\label{Bm1m2}
I_w = 
\underset{{|a| = (w-1)}}{\cup} (-1,I_{a})~~~\cup~~~ 
\underset{{|b| = (w-2)}}{\cup} (-2,I_{b}) ~. 
\end{eqnarray}
The basis elements for the Euler sums are then given by the 
$\zeta$-values with indices out of $I_w$. The elements of which the index sets 
are a Lyndon word form a Lyndon basis. $\square$

The Fibonacci version of this basis seems to have been discovered 
independently by S. Zlobin, see Ref.~\cite{ZHAO2}.
 
\vspace*{2mm}\noindent
This construction is analogous to that by Hoffman in the case of MZVs. 
It also uses a 2-letter alphabet. The different decomposition of the 
weight {\sf w}, 
however, leads to a basis of different length. Again we may derive the length 
of the basis using the Witt-formula (\ref{WITT2b}) or counting the basis 
elements as Lyndon words of the index set (\ref{Bm1m2}). Let us give an 
example for {\sf w = 20}.
\begin{eqnarray}
20 &=& k_i*1+l_i*2 = 18*1~+~1*2  
                   = 16*1~+~2*2  
                   = 14*1~+~3*2 \nonumber\\  
                   &=& 12*1~+~4*2  
                   = 10*1~+~5*2  
                   =  8*1~+~6*2  
                   =  6*1~+~7*2 \nonumber\\  
                   &=&  4*1~+~8*2  
                   =  2*1~+~9*2  
\end{eqnarray}
Similar to the non-alternating case one obtains
\begin{eqnarray}
L_{\{-1,-2\}}(20) &=& \frac{1}{19} \frac{19!}{18! 1!} 
                     +\frac{1}{18} \left[\frac{18!}{16! 2!} - \frac{9!}{8! 1!}\right] 
                     +\frac{1}{17} \frac{17!}{14! 3!}                 
                     +\frac{1}{16} \left[\frac{16!}{12! 4!} - \frac{9!}{8! 1!}\right]
\nonumber\\ &+&                     
\frac{1}{15} \left[\frac{15!}{10! 5!} - \frac{3!}{2! 1!}\right]
                     +\frac{1}{14} \left[\frac{14!}{8! 6!} - \frac{7!}{4! 3!}\right]
                     +\frac{1}{13} \frac{13!}{7! 6!}                 
\nonumber \\ &+&
                     \frac{1}{12} \left[\frac{12!}{8! 4!} - \frac{6!}{4! 
2!}\right] 
                     +\frac{1}{11} \frac{11!}{9! 2!}                 
= 750~.
\end{eqnarray}
Likewise the Witt-formula (\ref{WITT2b}) yields
\begin{eqnarray}
l(20) &=& \frac{1}{20} \left[l_{20} - l_{10} - l_{4} + l_2 \right]
\nonumber\\
&=& \frac{1}{20} \left[15127 - 123 - 7 + 3 \right] = 750~.
\end{eqnarray}

The above basis suffers from the same shortcoming as the Hoffman basis in 
that the concept of depth lacks relevance. Hence we did not use it.

In a similar way we can construct yet another Fibonacci basis:
\vspace{2mm}

\noindent {\bf Conjecture~5.}\\
With the starting conditions of (\ref{startfibo}), consider the 
index words at weight {\sf w} to be
\begin{eqnarray}
I_w = 
\underset{{|a| = (w-1)}}{\cup} (-1,I_{a})~~~\cup~~~ 
\underset{{|b| = (w-2)}}{\cup} (0,0,I_{b}) ~. 
\end{eqnarray}
The basis elements for the Euler sums are then given by the $\zeta$-values 
of indices $I_w$. The elements of which the index fields are a Lyndon word 
and all indices are odd valued if $w > 2$ form a Lyndon basis. $\square$

The Lyndon basis of this construction happens to be the basis proposed in 
ref~\cite{Broadhurst:1}.
We can divide $I_w$ 
\begin{eqnarray}
I_w = I_w^{\sf odd} \oplus I_w^{\sf \neg odd}~,
\end{eqnarray}
with the indices in $I_w^{\sf odd}$ are all odd and the last index of 
$I_w^{\sf \neg odd}$ even, all others odd. The Lyndon words of $I_w^{\sf odd}$, $\Ly[I_w^{\sf 
odd}]$,
 form the basis elements at weight {\sf w} and they are counted
by (\ref{WITT2b}). Note, that the basis element at {\sf w = 2} is not odd, 
which is an exception.

As an illustration we consider the case {\sf w = 6}. The following words are 
generated, where we assume the ordering $0 < 1$ and let the digit 1 play 
the role of -1.
\begin{alignat}{1}
& \{000001,000011,001001,001101,001111\}; 
\nonumber\\
& \{100001,100101,100111,110001,110011,111001,111101,111111\}~.
\end{alignat}
The Lyndon words are
\begin{alignat}{3}
(000011) &\equiv (-5,-1);~~~~~~~&(001111) &\equiv (-3,-1,-1,-1);\nonumber\\
(000001) &\equiv (-6);          &(001101) &\equiv (-3,-1,-2)~.  
\end{alignat}
The Lyndon words with odd indices taken as index of an Euler sum are 
basis elements, which we express through the harmonic polylogarithms at 
argument $x=1$, $\HH_{-5,-1}(1)$ and $\HH_{-3,-1,-1,-1}(1)$. On the other 
hand,
\begin{eqnarray}
\HH_{-6} &=& \frac{62}{35} \HH_{-2}^3 \\
\HH_{-3,-1,-2} &=& \HH_{-5,-1} + \HH_{-2} \HH_{-3,-1} + 
\frac{452}{105} \HH_{-2}^3 - \frac{55}{18} \HH_{-3}^2
\end{eqnarray}
do not belong to the basis.

The last Lyndon basis is the one we actually use in the programs. It is 
depth oriented and no element can be written as a linear combination of 
elements of lower depth or products of elements with lower weight.
To weight ${\sf w=12}$ the complete basis for the finite elements is given by
\begin{eqnarray}
{\sf w=~~1} & & \HH_{-1}; \\
{\sf w=~~2} & & \HH_{-2}; \\
{\sf w=~~3} & & \HH_{-3}; \\
{\sf w=~~4} & & \HH_{-3,-1}; \\
{\sf w=~~5} & & \HH_{-5},~\HH_{-3,-1,-1}; \\
{\sf w=~~6} & & \HH_{-5,-1},~\HH_{-3,-1,-1,-1}; \\
{\sf w=~~7} & & \HH_{-7},~\HH_{-5,-1,-1},~\HH_{-3,-3,-1},
               ~\HH_{-3,-1,-1,-1}; \\
{\sf w=~~8} & & \HH_{-7,-1},~
\HH_{-5,-3},~
\HH_{-5,-1,-1,-1},~
\HH_{-3,-3,-1,-1},~
\nonumber\\ &&
\HH_{-3,-1,-1,-1,-1}; \\
{\sf w=~~9} 
& & \HH_{-9},~\HH_{-7,-1,-1},~\HH_{-5,-3,-1},~\HH_{-5,-1,-3},
   ~\HH_{-5,-1,-1,-1,-1},\nonumber\\ &&
    \HH_{-3,-3,-1,-1,-1},~\HH_{-3,-1,-3,-1,-1},
   ~\HH_{-3,-1,-1,-1,-1,-1,-1}; 
\\
{\sf w=10} & & \HH_{-9,-1},~
\HH_{-7,-3},~
\HH_{-7,-1,-1,-1},~
\HH_{-5,-3,-1,-1},~
\HH_{-5,-1,-3,-1}, \nonumber\\ &&
\HH_{-5,-1,-1,-3},~
\HH_{-3,-3,-1,-1},~
\HH_{-5,-1,-1,-1,-1,-1},~
\HH_{-3,-3,-1,-1,-1,-1}, \nonumber\\ &&
\HH_{-3,-1,-3,-1,-1,-1},~
\HH_{-3,-1,-1,-1,-1,-1,-1,-1};
\end{eqnarray}\begin{eqnarray}
{\sf w=11} & &         
\HH_{-11},~
\HH_{-9,-1,-1},~
\HH_{-7,-3,-1},~
\HH_{-7,-1,-3},~
\HH_{-5,-5,-1}, \nonumber\\ &&
\HH_{-5,-3,-3},~
\HH_{-3,-3,-1,-3,-1},~
\HH_{-3,-3,-3,-1,-1},~
\HH_{-5,-1,-1,-1,-3}, \nonumber\\ &&
\HH_{-5,-1,-1,-3,-1},~
\HH_{-5,-1,-3,-1,-1},~
\HH_{-5,-3,-1,-1,-1}, \nonumber\\ &&
\HH_{-7,-1,-1,-1,-1},~
\HH_{-3,-1,-1,-3,-1,-1,-1},~
\HH_{-3,-1,-3,-1,-1,-1,-1}, \nonumber\\ &&
\HH_{-3,-3,-1,-1,-1,-1,-1},~
\HH_{-5,-1,-1,-1,-1,-1,-1}, \nonumber\\ &&
\HH_{-3,-1,-1,-1,-1,-1,-1,-1,-1};
\\
{\sf w=12} & & 
\HH_{-7,-5},~
\HH_{-9,-3},~
\HH_{-11,-1},~
\HH_{-5,-1,-3,-3},~
\HH_{-5,-3,-1,-3}, \nonumber\\ &&
\HH_{-5,-3,-3,-1},~
\HH_{-5,-5,-1,-1},~
\HH_{-7,-1,-1,-3},~
\HH_{-7,-1,-3,-1}, \nonumber\\ &&
\HH_{-7,-3,-1,-1},~
\HH_{-9,-1,-1,-1},~
\HH_{-3,-3,-1,-1,-3,-1}, \nonumber\\ &&
\HH_{-3,-3,-1,-3,-1,-1},~
\HH_{-3,-3,-3,-1,-1,-1},~
\HH_{-5,-1,-1,-1,-1,-3}, \nonumber\\ &&
\HH_{-5,-1,-1,-1,-3,-1},~
\HH_{-5,-1,-1,-3,-1,-1},~
\HH_{-5,-1,-3,-1,-1,-1}, \nonumber\\ &&
\HH_{-5,-3,-1,-1,-1,-1},~
\HH_{-7,-1,-1,-1,-1,-1},~
\HH_{-3,-1,-1,-3,-1,-1,-1,-1}, 
\nonumber\\
&&
\HH_{-3,-1,-3,-1,-1,-1,-1,-1},~
\HH_{-3,-3,-1,-1,-1,-1,-1,-1}, \nonumber\\ &&
\HH_{-5,-1,-1,-1,-1,-1,-1,-1},~
\HH_{-3,-1,-1,-1,-1,-1,-1,-1,-1,-1};
\end{eqnarray}

For the Lyndon basis the conjectured length is 
\cite{Broadhurst:1}
\begin{eqnarray}
\label{WITT2b}
l(w) &=& \frac{1}{w} \sum_{d|w} \mu\left(\frac{w}{d}\right) l_d,~~w 
\geq 2 \nonumber\\
     & & l_1 = 1, l_2 = 3, l_3 =4, l_d = l_{d-1} + l_{d-2},~~d \geq 4~.
\nonumber\\
l(1) &=& 2
\end{eqnarray}
$l_d$ denote the Lucas-numbers~\cite{LUCAS,HW}. They are represented by
\begin{eqnarray}
l_d = \left( \frac{1+\sqrt{5}}{2}\right)^d
    + \left( \frac{1-\sqrt{5}}{2}\right)^d~,
\end{eqnarray}
and derive from the generating function
\begin{eqnarray}
G(l_k,x) = \frac{2-x}{1-x-x^2} = \sum_{k=0}^\infty x^k l_k~.
\end{eqnarray}
The first values are given in Table~\ref{tablelucasb}.
The case $w=1$ is special as two elements contribute. 

{
\begin{table}
\begin{center}
\begin{tabular}[htb]{|r||r|r|r|r|r|r|r|r|r|r|} 
\hline \hline
$w$   & 1 & 2 & 3 & 4 & 5 & 6 & 7 &  8 &  9 & 10 \\
\hline
$l_w$ & 
  1 &   3 &   4 &   7 &   11 &   18 &   29 &   47 &   76 &   123 \\
\hline \hline 
$w$   & 11 & 12 & 13 & 14 & 15& 16 & 17 & 18 & 19 & 20\\
\hline
$l_w$ & 199 & 322 & 521 & 843 & 1364 & 2207 & 3571 & 5778 & 9349 & 15127 \\
\hline \hline 
\end{tabular}
\end{center}
\caption{The first 20 Lucas numbers.}
\label{tablelucasb}
\end{table}
\normalsize}
\noindent
{
\begin{table}
\begin{center}
\begin{tabular}[htb]{|r||r|r|r|r|r|r|r|r|r|r|} 
\hline \hline
$w$   & 1 & 2 & 3 & 4 & 5 & 6 & 7 &  8 &  9 & 10 \\
\hline
$l_w$ &
1&
1&    
1&    
1&
2&
2&
4&
5&
8&
11\\
\hline \hline 
$w$   & 11 & 12 & 13 & 14 & 15& 16 & 17 & 18 & 19 & 20\\
\hline
$l_w$ & 
18&
25&
40&
58&
90&
135&
210&
316&
492&
750\\
\hline \hline 
\end{tabular}
\end{center}
\caption{Number of basis elements of the Lyndon basis 
for the Euler sums for fixed
weight {\sf w}.}\label{tablelyndonESb}
\end{table}
\normalsize}

\section{Pushdown Bases}
\label{ap:bases}
\renewcommand{\theequation}{\thesection.\arabic{equation}}
\setcounter{equation}{0}

\vspace{1mm}
\noindent
We have tried to select a basis in which the elements of the set $L_w$ are 
maximal and the extended elements are minimal. At the same time the 
extended elements should be Lyndon words. This means for instance that an 
element like $H_{5,5,5,3}$ cannot be extended and hence has to be part of 
the basis, even though it is the minimal element at weight ${\sf w=18}$. One 
could of 
course reverse the criteria. For the construction of the bases this does 
not really diminish the amount of work. In both cases there are elements 
that should be skipped because of linear dependencies. We call the basis 
below the `minimal pushdown basis'. In addition we have used the 
requirement that for the extended elements the corresponding $A$-function 
should be usable for a pushdown. This requirement we could enforce up to 
weight ${\sf w=22}$. For higher weights we do not have the information in the 
data 
mine, and hence we do not know whether this requirement can be achieved.

\begin{eqnarray}
	P_{2} & = & H_{2} \\
	P_{3} & = & H_{3} \\
	P_{5} & = & H_{5} \\
	P_{7} & = & H_{7} \\
	P_{8} & = & H_{5,3} \\
	P_{9} & = & H_{9} \\
	P_{10} & = & H_{7,3} \\
	P_{11} & = & H_{11}, H_{5,3,3} \\
	P_{12} & = & H_{9,3}, H_{6,4,1,1} \\
	P_{13} & = & H_{13}, H_{7,3,3}, H_{5,5,3} \\
	P_{14} & = & H_{11,3}, H_{9,5}, H_{5,3,3,3} \\
	P_{15} & = & H_{15}, H_{7,3,5}, H_{9,3,3}, H_{6,4,3,1,1} \\
	P_{16} & = & H_{11,5}, H_{13,3}, H_{5,5,3,3}, H_{7,3,3,3}, H_{8,6,1,1}
			\\
	P_{17} & = & H_{17}, H_{7,5,5}, H_{9,3,5}, H_{9,5,3}, H_{11,3,3},
		 H_{5,3,3,3,3}, H_{6,6,3,1,1} \\
	P_{18} & = & H_{13,5}, H_{15,3}, H_{5,5,5,3}, H_{7,3,3,5},, H_{7,3,5,3},
		 H_{9,3,3,3}, H_{10,6,1,1}, H_{6,4,3,3,1,1} \\
	P_{19} & = & H_{19}, H_{9,3,7}, H_{9,5,5}, H_{11,3,5}, H_{11,5,3}, H_{13,3,3},
			\nonumber \\ & &
		 H_{5,3,5,3,3}, H_{5,5,3,3,3}, H_{7,3,3,3,3},
		 H_{6,6,5,1,1}, H_{8,6,3,1,1}
\end{eqnarray}
\begin{eqnarray}
	P_{20} & = & H_{13,7}, H_{15,5}, H_{17,3},  H_{7,3,5,5},
			 H_{7,5,5,3}, H_{7,7,3,3}, H_{9,3,3,5}, \nonumber \\ &&
			 H_{9,3,5,3}, H_{11,3,3,3}, H_{10,8,1,1}, H_{5,3,3,3,3,3},
			 H_{6,4,3,5,1,1}, H_{8,4,3,3,1,1} \\
	P_{21} & = & H_{21}, H_{9,5,7}, H_{9,9,3}, H_{11,3,7}, H_{13,3,5},
			 H_{13,5,3}, H_{15,3,3}, \nonumber \\ &&
			 H_{5,5,3,5,3}, H_{5,5,5,3,3}, H_{7,3,3,3,5},
			 H_{7,3,3,5,3}, H_{7,3,5,3,3}, H_{9,3,3,3,3}, \nonumber \\ &&
			 H_{8,6,5,1,1}, H_{10,4,5,1,1}, H_{10,6,3,1,1}, 
			 H_{6,4,3,3,3,1,1} \\
	P_{22} & = & H_{15,7}, H_{17,5}, H_{19,3}, H_{7,5,7,3}, H_{7,7,3,5},
			 H_{9,3,5,5}, H_{9,3,7,3}, \nonumber \\ &&
			 H_{9,5,3,5}, H_{9,5,5,3}, H_{11,3,3,5},
			 H_{11,3,5,3}, H_{11,5,3,3}, H_{13,3,3,3}, H_{12,8,1,1}, \nonumber \\ &&
			 H_{5,3,5,3,3,3}, H_{5,5,3,3,3,3}, H_{7,3,3,3,3,3}
				\nonumber \\ &&
			 H_{6,4,5,5,1,1}, H_{6,6,5,3,1,1},
			 H_{8,2,3,7,1,1}, H_{8,6,3,3,1,1} \\
	P_{23} & = & H_{23},
		H_{11,7,5},
		H_{11,9,3},
		H_{13,3,7},
		H_{13,5,5},
		H_{13,7,3},
		H_{15,3,5},
		H_{15,5,3},
			\nonumber \\ &&
		H_{17,3,3},
		H_{5,5,5,5,3},
		H_{7,3,7,3,3},
		H_{7,3,5,5,3},
		H_{7,5,3,5,3},
		H_{7,5,5,3,3},
			\nonumber \\ &&
		H_{7,7,3,3,3},
		H_{9,3,3,3,5},
		H_{9,3,3,5,3},
		H_{9,3,5,3,3},
		H_{9,5,3,3,3},
		H_{11,3,3,3,3},
			\nonumber \\ &&
		H_{8,6,7,1,1},
		H_{8,8,5,1,1},
		H_{10,2,9,1,1},
		H_{10,4,7,1,1},
			\nonumber \\ &&
		H_{5,3,3,3,3,3,3}
		H_{6,2,3,5,5,1,1},
		H_{6,2,5,3,5,1,1},
		H_{6,4,3,3,5,1,1} \\
	P_{24} & = & H_{17,7},
		H_{19,5},
		H_{21,3},
		H_{7,7,7,3},
		H_{9,7,3,5},
		H_{9,7,5,3},
		H_{9,9,3,3},
			\nonumber \\ &&
		H_{11,3,3,7},
		H_{11,3,5,5},
		H_{11,3,7,3},
		H_{11,5,3,5},
		H_{11,5,5,3},
		H_{11,7,3,3},
		H_{13,3,3,5},
			\nonumber \\ &&
		H_{13,3,5,3},
		H_{13,5,3,3},
		H_{15,3,3,3},
		H_{12,10,1,1},
		H_{14,8,1,1},
		H_{5,5,3,3,5,3},
			\nonumber \\ &&
		H_{5,5,3,5,3,3},
		H_{5,5,5,3,3,3},
		H_{7,3,3,3,5,3},
		H_{7,3,3,5,3,3},
		H_{7,3,5,3,3,3},
		H_{7,5,3,3,3,3},
			\nonumber \\ &&
		H_{9,3,3,3,3,3},
		H_{6,6,5,5,1,1},
		H_{8,2,5,7,1,1},
		H_{8,2,7,5,1,1},
		H_{8,4,3,7,1,1},
		H_{8,4,5,5,1,1},
			\nonumber \\ &&
		H_{8,4,7,3,1,1},
		H_{6,2,3,3,3,5,1,1} \\
	P_{25} & = & H_{25},
		H_{11,11,3},
		H_{13,5,7},
		H_{13,7,5},
		H_{13,9,3},
		H_{15,3,7},
		H_{15,5,5},
		H_{15,7,3},
			\nonumber \\ &&
		H_{17,3,5},
		H_{17,5,3},
		H_{19,3,3},
		H_{7,3,7,3,5},
		H_{7,5,3,7,3},
		H_{7,5,7,3,3},
			\nonumber \\ &&
		H_{9,3,3,3,7},
		H_{9,3,3,5,5},
		H_{9,3,3,7,3},
		H_{9,3,5,3,5},
		H_{9,3,5,5,3},
		H_{9,3,7,3,3},
			\nonumber \\ &&
		H_{9,5,3,3,5},
		H_{9,5,3,5,3},
		H_{9,5,5,3,3},
		H_{9,7,3,3,3},
		H_{11,3,3,3,5},
		H_{11,3,3,5,3},
			\nonumber \\ &&
		H_{11,3,5,3,3},
		H_{11,5,3,3,3},
		H_{13,3,3,3,3},
		H_{8,8,7,1,1},
		H_{10,4,9,1,1},
			\nonumber \\ &&
		H_{10,6,7,1,1},
		H_{10,8,5,1,1},
		H_{12,2,9,1,1},
		H_{5,3,3,5,3,3,3},
		H_{5,3,5,3,3,3,3},
			\nonumber \\ &&
		H_{5,5,3,3,3,3,3},
		H_{7,3,3,3,3,3,3},
		H_{6,2,5,5,5,1,1},
		H_{6,4,3,5,5,1,1},
		H_{6,4,5,3,5,1,1},
			\nonumber \\ &&
		H_{6,4,5,5,3,1,1},
		H_{6,6,3,3,5,1,1},
		H_{6,6,3,5,3,1,1},
		H_{6,6,5,3,3,1,1} \\
	P_{26} & = & H_{17,9},
		H_{19,7},
		H_{21,5},
		H_{23,3},
		H_{7,7,7,5},
		H_{9,5,9,3},
		H_{11,3,9,3},
		H_{11,5,3,7},
			\nonumber \\ &&
		H_{11,5,5,5},
		H_{11,5,7,3},
		H_{11,7,3,5},
		H_{11,7,5,3},
		H_{11,9,3,3},
		H_{13,3,3,7},
		H_{13,3,5,5},
			\nonumber \\ &&
		H_{13,3,7,3},
		H_{13,5,3,5},
		H_{13,5,5,3},
		H_{13,7,3,3},
		H_{15,3,3,5},
		H_{15,3,5,3},
		H_{15,5,3,3},
			\nonumber \\ &&
		H_{17,3,3,3},
		H_{14,10,1,1},
		H_{5,5,5,3,5,3},
		H_{5,5,5,5,3,3},
		H_{7,3,3,5,5,3},
		H_{7,3,5,3,5,3},
			\nonumber \\ &&
		H_{7,3,5,5,3,3},
		H_{7,3,7,3,3,3},
		H_{7,5,3,3,5,3},
		H_{7,5,3,5,3,3},
		H_{7,5,5,3,3,3},
			\nonumber \\ &&
		H_{7,7,3,3,3,3},
		H_{9,3,3,3,3,5},
		H_{9,3,3,3,5,3},
		H_{9,3,3,5,3,3},
		H_{9,3,5,3,3,3},
			\nonumber \\ &&
		H_{9,5,3,3,3,3},
		H_{11,3,3,3,3,3},
		H_{8,2,7,7,1,1},
		H_{8,4,5,7,1,1},
		H_{8,4,7,5,1,1},
			\nonumber \\ &&
		H_{8,6,3,7,1,1},
		H_{8,6,5,5,1,1},
		H_{8,6,7,3,1,1},
		H_{8,8,3,5,1,1},
		H_{8,8,5,3,1,1},
			\nonumber \\ &&
		H_{10,2,3,9,1,1},
		H_{10,2,5,7,1,1},
		H_{10,2,7,5,1,1},
		H_{5,3,3,3,3,3,3,3},
			\nonumber \\ &&
		H_{6,2,3,3,5,5,1,1},
		H_{6,2,3,5,3,5,1,1},
		H_{6,2,5,3,3,5,1,1},
		H_{6,4,3,3,3,5,1,1}
\end{eqnarray}

The above bases are complete. For the following basis we miss the two 
elements at depth 9 due to limited computer resources. Yet the construction 
based on $L_{27}$ allows us to predict the last two elements:

\begin{eqnarray}
	  P_{27} & = &
		\!H_{27},
		\!H_{11,7,9},
		\!H_{13,11,3},
		\!H_{15,3,9},
		\!H_{15,5,7},
		\!H_{15,7,5},
		\!H_{15,9,3},
		\!H_{17,5,5},
		\!H_{17,7,3},
			\nonumber \\ &&
		\!H_{19,3,5},
		\!H_{19,5,3},
		\!H_{21,3,3},
		\!H_{7,5,5,7,3},
		\!H_{7,5,7,3,5},
		\!H_{7,7,3,7,3},
		\!H_{7,7,7,3,3},
			\nonumber \\ &&
		\!H_{9,3,9,3,3},
		\!H_{9,5,3,5,5},
		\!H_{9,5,3,7,3},
		\!H_{9,5,5,3,5},
		\!H_{9,5,5,5,3},
		\!H_{9,5,7,3,3},
		\!H_{9,7,3,3,5},
			\nonumber \\ &&
		\!H_{9,7,3,5,3},
		\!H_{9,7,5,3,3},
		\!H_{9,9,3,3,3},
		\!H_{11,3,3,3,7},
		\!H_{11,3,3,5,5},
		\!H_{11,3,3,7,3},
		\!H_{11,3,5,3,5},
			\nonumber \\ &&
		\!H_{11,3,5,5,3},
		\!H_{11,3,7,3,3},
		\!H_{11,5,3,3,5},
		\!H_{11,5,3,5,3},
		\!H_{11,5,5,3,3},
		\!H_{11,7,3,3,3},
		\!H_{13,3,3,3,5},
			\nonumber \\ &&
		\!H_{13,3,3,5,3},
		\!H_{13,3,5,3,3},
		\!H_{13,5,3,3,3},
		\!H_{15,3,3,3,3},
		\!H_{10,8,7,1,1},
		\!H_{10,10,5,1,1},
			\nonumber \\ &&
		\!H_{12,2,11,1,1},
		\!H_{12,4,9,1,1},
		\!H_{12,6,7,1,1},
		\!H_{12,8,5,1,1},
		\!H_{16,2,7,1,1},
			\nonumber \\ &&
		\!H_{5,3,5,3,5,3,3},
		\!H_{5,5,3,3,3,5,3},
		\!H_{5,5,3,3,5,3,3},
		\!H_{5,5,3,5,3,3,3},
		\!H_{5,5,5,3,3,3,3},
			\nonumber \\ &&
		\!H_{7,3,3,3,3,3,5},
		\!H_{7,3,3,3,3,5,3},
		\!H_{7,3,3,3,5,3,3},
		\!H_{7,3,3,5,3,3,3},
		\!H_{7,3,5,3,3,3,3},
			\nonumber \\ &&
		\!H_{9,3,3,3,3,3,3},
		\!H_{6,4,5,5,5,1,1},
		\!H_{6,6,3,5,5,1,1},
		\!H_{6,6,5,3,5,1,1},
		\!H_{6,6,5,5,3,1,1},
			\nonumber \\ &&
		\!H_{8,2,3,5,7,1,1},
		\!H_{8,2,3,7,5,1,1},
		\!H_{8,2,5,3,7,1,1},
		\!H_{8,2,5,5,5,1,1},
		\!H_{8,2,5,7,3,1,1},
			\nonumber \\ &&
		\!H_{8,2,7,3,5,1,1},
		\!H_{8,2,7,5,3,1,1},
		\!H_{8,4,3,3,7,1,1},
			\nonumber \\ &&
		\!H_{7,5,7,5,3} \rightarrow? H_{6,4,6,4,3,1,1,1,1},
		\!H_{7,5,3,3,3,3,3} \rightarrow ? H_{6,4,3,3,3,3,3,1,1}
 \nonumber
\end{eqnarray}
We have selected the last two elements for the necessary extension on the 
basis of the Appendix in the thesis by Racinet~\cite{Racinet} in which for 
these two elements the numbers 6 and 4 seem to play a special role.

Although we have also results for $P_{28}$ in which the leading depth is 
missing, there are too many elements missing to give a reliable list of the 
basis elements. It should be remarked though that also for $P_{28}$ we expect a 
2-fold pushdown from depth 8 to depth 4.


\section{Explicit pushdowns}
\label{ap:explicit}
\renewcommand{\theequation}{\thesection.\arabic{equation}}
\setcounter{equation}{0}

\vspace{1mm}
\noindent
Below we list all pushdowns up to ${\sf w=21}$ and one at ${\sf w=22}$ with 
the mixing 
with terms of equal weight and depth in the left hand side and all 
remaining Euler sums in the right hand side. The function $A$ is defined in 
(\ref{eq:AFUN}).

We only list that part of the pushdowns that we consider particularly 
interesting. The complete formulas can be found in the data mine in the 
programs part. The name of the file is {\tt pushdowns.h}.

\begin{eqnarray}
	Z_{6,4,1,1} &\!=\! & -\frac{64}{27} A_{7,5} + \cdots \\
	Z_{6,4,3,1,1} &\!=\!& \frac{1408}{81} A_{7,5,3} + \cdots \\
	Z_{8,6,1,1}+\frac{542}{175}Z_{5,5,3,3}-\frac{19}{7}Z_{7,3,3,3}
			 &\!=\!& -\frac{1024}{405}A_{9,7} + \cdots
\end{eqnarray}
\begin{eqnarray}
	Z_{6,6,3,1,1}-\frac{14}{5}Z_{5,3,3,3,3}
			 &\!=\!& \frac{5120}{243} A_{7,7,3} + \cdots \\
	Z_{10,6,1,1} -\frac{10}{3} Z_{9,3,3,3} \nonumber \\
		-\frac{124}{35} Z_{7,3,5,3}
		-\frac{124}{35} Z_{7,3,3,5} \nonumber \\
		-\frac{3282}{875} Z_{5,5,5,3}
			 &\!=\!& -\frac{8192}{3375} A_{11,7} + \cdots \\
	Z_{6,4,3,3,1,1} &\!=\!& -\frac{392}{27} A_{7,5,3,3} + \cdots \\
	Z_{8,6,3,1,1} - \frac{61}{7} Z_{7,3,3,3,3} \nonumber \\
        + \frac{1774}{175} Z_{5,5,3,3,3}
        + \frac{2}{5} Z_{5,3,5,3,3}
			 &\!=\!& \frac{647168}{34263} A_{7,7,5} 
			      + \frac{45056}{1215} A_{9,7,3} + \cdots \\
	Z_{6,6,5,1,1} + 13 Z_{7,3,3,3,3} \nonumber \\
			 - \frac{268}{25} Z_{5,5,3,3,3}
			+ \frac{6}{5} Z_{5,3,5,3,3}
			 &\!=\!& - \frac{3598336}{125631} A_{7,7,5} 
			      - \frac{759808}{4455} A_{9,7,3} + \cdots 
\nonumber\\ \\
	Z_{10,8,1,1}
       - \frac{13}{2} Z_{11,3,3,3}
       - \frac{304}{45} Z_{9,3,3,5} \nonumber \\
       - \frac{3601}{525} Z_{9,3,5,3}
       - \frac{3799}{525} Z_{7,3,5,5} \nonumber \\
       + \frac{1371}{196} Z_{7,7,3,3}
       + \frac{163}{2450} Z_{7,5,5,3}
			&\!=\!& -\frac{16384}{6615} A_{11,9} + \cdots \\
	Z_{6,4,3,5,1,1}-\frac{68}{5} Z_{5,3,3,3,3,3} &\!=\!&
		 - \frac{118784}{243} A_{9,5,3,3} 
		 - \frac{2560}{243} A_{7,5,3,5} 
			+ \cdots \nonumber\\ \\
	Z_{8,4,3,3,1,1} -\frac{28}{5} Z_{5,3,3,3,3,3} &\!=\!&
		\frac{32768}{81} A_{9,5,3,3} 
		 - \frac{10240}{2187} A_{7,5,3,5} 
			+ \cdots \\
	Z_{8,6,5,1,1}
       - \frac{68}{9} Z_{9,3,3,3,3}
       - \frac{832}{105} Z_{7,3,5,3,3}
			\nonumber \\
       - \frac{967}{105} Z_{7,3,3,5,3}
       - \frac{1042}{105} Z_{7,3,3,3,5}
			\nonumber \\
       - \frac{13182}{875} Z_{5,5,5,3,3}
       - \frac{6}{7} Z_{5,5,3,5,3}
       &\!=\!&
       - \frac{194240512}{9628875} A_{9,7,5}
			\nonumber \\ &&
       - \frac{229376}{1125} A_{11,7,3}
			\nonumber \\ &&
       - \frac{80972546048}{337010625} A_{11,5,5}
			+ \cdots \\
	Z_{10,4,5,1,1}
       - \frac{46}{9} Z_{9,3,3,3,3}
       - \frac{67}{21} Z_{7,3,5,3,3}
			\nonumber \\
       - \frac{73}{21} Z_{7,3,3,5,3}
       - \frac{79}{21} Z_{7,3,3,3,5}
			\nonumber \\
       - \frac{482}{175} Z_{5,5,5,3,3}
       - \frac{46}{175} Z_{5,5,3,5,3}
       &\!=\!&
       + \frac{15966208}{641925} A_{9,7,5}
       + \frac{32768}{2025} A_{11,7,3}
			\nonumber \\ &&
       - \frac{1691951104}{67402125} A_{11,5,5}
			+ \cdots
\end{eqnarray}

\begin{eqnarray}
%
	Z_{10,6,3,1,1}
       - \frac{46}{9} Z_{9,3,3,3,3}
       - \frac{632}{105} Z_{7,3,5,3,3}
			\nonumber \\
       - \frac{86}{15} Z_{7,3,3,5,3}
       - \frac{572}{105} Z_{7,3,3,3,5}
			\nonumber \\
       - \frac{4792}{875} Z_{5,5,5,3,3}
       + \frac{46}{175} Z_{5,5,3,5,3}
       &\!=\!&
       + \frac{124608512}{9628875} A_{9,7,5}
       + \frac{16384}{10125} A_{11,7,3}
			\nonumber \\ &&
       - \frac{758235136}{48144375} A_{11,5,5}
			+ \cdots  \\
\label{push21}
	Z_{6,4,3,3,3,1,1}
       &\!=\!&
        - \frac{5120}{81} A_{7,5,3,3,3} 
			+ \cdots\\
%
	Z_{12,8,1,1}
         +\frac{13598459235}{18816311591} Z_{7,5,7,3}  \nonumber \\
         -\frac{9790486696}{6109192075} Z_{9,3,5,5}
         -\frac{3021879830}{2688044513} Z_{9,3,7,3}  \nonumber \\
         -\frac{560739181022}{201603338475} Z_{9,5,3,5}
         -\frac{19968330538}{13440222565} Z_{9,5,5,3}  \nonumber \\
         -\frac{66543918797}{40320667695} Z_{9,7,3,3}
         +\frac{2598592817}{707380135} Z_{11,3,3,5}  \nonumber \\
         -\frac{3186058443}{2688044513} Z_{11,3,5,3}
         -\frac{20352278271}{13440222565} Z_{11,5,3,3}  \nonumber \\
         +\frac{7925677546}{1221838415} Z_{13,3,3,3}
      	&\!=\!& -\frac{524288}{212625} A_{13,9} + \cdots 
\end{eqnarray}
The $+\cdots$ indicates terms that are purely MZVs of lower depth or 
products of lower weight MZVs. The complete relations can have up to about 
150 terms. Hence we give them in a file in the data mine. The first 15 of 
these relations were derived with the help of {\tt PSLQ} and/or the {\tt LLL} 
algorithm. Seven of them could be derived with the data mine. Unfortunately 
for depth ${\sf d=5}$ objects we have only exact results up to weight ${\sf 
w=17}$ and for 
depth ${\sf d=4}$ we have only exact results up to weight ${\sf w=22}$.

The above results used the available resources to their limit. The formula 
in (\ref{push21}) needed 45 hours of running time using the {\tt LLL} 
algorithms as implemented in {\tt PARI} in a 152 parameter search at 8000 
digits and was checked afterwards at 10000 digits.
 
We have expressed the pushdowns in terms of the $A$-function that has the 
same indices as the element of $L_w$ that was extended. It is not clear 
whether this scheme can be maintained for pushdowns beyond the ones we 
present. Some $A$-functions cannot be used because they express directly in 
terms of equal or lower depth MZVs. This then has again influence on the 
selection of the basis. In the end it may be that we have to drop one or 
more requirements for the basis. A simple example of such an $A$-function 
exists already at weight ${\sf w=15}$ :
\begin{eqnarray}
	A_{7,3,5} & = &
       + \frac{7649}{143360} Z_{7,3,5}
       - \frac{7089}{143360} \zeta_5 Z_{7,3}
       - \frac{2097}{71680} \zeta_5^3
       - \frac{3429}{5120} \zeta_7 Z_{5,3}
			\nonumber \\ &&
       - \frac{116396017}{2867200} \zeta_{15}
       + \frac{1083797}{40960} \zeta_2 \zeta_{13}
       + \frac{81059}{71680} \zeta_2^2 \zeta_{11}
			\nonumber \\ &&
       - \frac{110993}{627200} \zeta_2^3 \zeta_9
       - \frac{43311}{448000} \zeta_2^4 \zeta_7
       - \frac{27831}{78400} \zeta_2^5 \zeta_5~.
\end{eqnarray}
 
It is also possible to express each pushdown in terms of a single Euler sum 
rather than an $A$-function. In a sense this is less telling. After all the 
$A$-function contains half of the terms of the doubling relation and the 
doubling relations seem to be at the origin of the pushdowns. Also we could 
not find much structure concerning which Euler sum(s) to select. There are 
often many possibilities. In the case of the $A$-functions one can make a 
unique selection: the $A$-function should have the same index field as the 
element of the set $L_w$ that represents the pushdown. Anyway, for 
completeness we give here a single Euler sum for each of the pushdowns. We 
have dropped all factors and terms which have MZVs of the same weight or 
products of MZVs with lower weight.

\begin{center}
\begin{tabular}{cccc}
   & & $H$-representation & $Z$-representation \\
   $A_{7,5}$      &$\rightarrow$&$H_{-9,3}$&$Z_{-9,-3}$\\
   $A_{7,5,3}$    &$\rightarrow$&$H_{-6,-3,6}$&$Z_{-6,3,-6}$\\
   $A_{9,7}$      &$\rightarrow$&$H_{-13,3}$&$Z_{-13,-3}$\\
   $A_{7,7,3}$    &$\rightarrow$&$H_{-6,-5,6}$&$Z_{-6,5,-6}$\\
   $A_{11,7}$     &$\rightarrow$&$H_{-15,3}$&$Z_{-15,-3}$\\
   $A_{7,5,3,3}$  &$\rightarrow$&$H_{6,-5,4,3}$&$Z_{6,-5,-4,3}$\\
   $A_{9,7,3}$    &$\rightarrow$&$H_{-8,-3,8},H_{-6,-7,6}$&
                                 $Z_{-8,3,-8},Z_{-6,7,-6}$\\
   $A_{7,7,5}$    &$\rightarrow$&$H_{-8,-3,8},H_{-6,-7,6}$&
                                 $Z_{-8,3,-8},Z_{-6,7,-6}$\\
   $A_{11,9}$     &$\rightarrow$&$H_{-17,3}$&$Z_{-17,-3}$\\
   $A_{7,5,3,5}$  &$\rightarrow$&$H_{8,-5,4,3},H_{6,-5,6,3}$&
                                $Z_{8,-5,-4,3},Z_{6,-5,-6,3}$\\
   $A_{9,5,3,3}$  &$\rightarrow$&$H_{8,-5,4,3},H_{6,-5,6,3}$&
                                $Z_{8,-5,-4,3},Z_{6,-5,-6,3}$\\
   $A_{9,7,5}$    &$\rightarrow$&$H_{-8,-5,8},H_{-6,-9,6},H_{-8,-3,10}$&
                                 $Z_{-8,5,-8},Z_{-6,9,-6},Z_{-8,3,-10}$\\
   $A_{11,5,5}$   &$\rightarrow$&$H_{-8,-5,8},H_{-6,-9,6},H_{-8,-3,10}$&
                                 $Z_{-8,5,-8},Z_{-6,9,-6},Z_{-8,3,-10}$\\
   $A_{11,7,3}$   &$\rightarrow$&$H_{-8,-5,8},H_{-6,-9,6},H_{-8,-3,10}$&
                                 $Z_{-8,5,-8},Z_{-6,9,-6},Z_{-8,3,-10}$\\
   $A_{7,5,3,3,3}$&$\rightarrow$&$H_{3,-6,-3,6,3}$&$Z_{3,-6,3,-6,3}$\\
   $A_{13,9}$     &$\rightarrow$&$H_{-19,3}$&$Z_{-19,-3}$ \\ 
\end{tabular}
\end{center}

Of course more complete results can be found in the data mine.



\newpage

\end{document}